%% file: clear_ionization_metallicity.tex
\documentclass[twocolumn,tighten,trackchanges]{aastex631}

\usepackage{amsmath,amstext}
\usepackage[T1]{fontenc}
\usepackage{apjfonts} 
\usepackage{natbib}
\usepackage{multirow}
\usepackage{microtype}
\usepackage{longtable}
\usepackage{verbatim}
\usepackage{hyperref}
\hypersetup{
    colorlinks=true,
    linkcolor=blue,
    filecolor=magenta,      
    urlcolor=cyan,
}

\include{defs}

\newcommand{\editone}[1]{{#1}}
\renewcommand{\added}[1]{{#1}}

\begin{document}

\title{\large \bf CLEAR: The Ionization and Chemical-Enrichment Properties of Galaxies at $1.1 < z < 2.3$}

\correspondingauthor{Casey Papovich}
\email{papovich@tamu.edu}

\author[0000-0001-7503-8482]{Casey Papovich}
\affiliation{Department of Physics and Astronomy, Texas A\&M University, College
Station, TX, 77843-4242 USA}
\affiliation{George P.\ and Cynthia Woods Mitchell Institute for
 Fundamental Physics and Astronomy, Texas A\&M University, College
 Station, TX, 77843-4242 USA}

\author[0000-0002-6386-7299]{Raymond C. Simons}
\affil{Space Telescope Science Institute, 3700 San Martin Drive,
  Baltimore, MD, 21218 USA}
\author[0000-0001-8489-2349]{Vicente Estrada-Carpenter}
\affiliation{Department of Physics and Astronomy, Texas A\&M University, College
Station, TX, 77843-4242 USA}
\affiliation{George P.\ and Cynthia Woods Mitchell Institute for
 Fundamental Physics and Astronomy, Texas A\&M University, College
 Station, TX, 77843-4242 USA}
\affiliation{Department of Astronomy \& Physics, Saint Mary's University, 923 Robie Street, Halifax, NS, B3H 3C3, Canada}
  \author[0000-0002-7547-3385]{Jasleen Matharu}
\affiliation{Department of Physics and Astronomy, Texas A\&M University, College
Station, TX, 77843-4242 USA}
\affiliation{George P.\ and Cynthia Woods Mitchell Institute for
Fundamental Physics and Astronomy, Texas A\&M University, College
Station, TX, 77843-4242 USA}

\author[0000-0003-1665-2073]{Ivelina Momcheva}
\affil{Space Telescope Science Institute, 3700 San Martin Drive,
  Baltimore, MD, 21218 USA}
\affil{Max-Planck-Institut für Astronomie, Königstuhl 17, D-69117 Heidelberg, Germany}

\author[0000-0002-1410-0470]{Jonathan R. Trump}
\affil{Department of Physics, 196A Auditorium Road Unit 3046, University of Connecticut, Storrs, CT 06269 USA}

\author[0000-0001-8534-7502]{Bren E. Backhaus }
\affil{Department of Physics, 196A Auditorium Road Unit 3046, University of Connecticut, Storrs, CT 06269 USA}

\author[0000-0003-2680-005X]{Gabriel Brammer}
\affil{Cosmic Dawn Centre, University of Copenhagen, Blegdamsvej 17, 2100 Copenhagen, Denmark}

\author[0000-0001-7151-009X]{Nikko J. Cleri} \affiliation{Department
of Physics and Astronomy, Texas A\&M University, College Station, TX,
77843-4242 USA} \affiliation{George P.\ and Cynthia Woods Mitchell
Institute for Fundamental Physics and Astronomy, Texas A\&M
University, College Station, TX, 77843-4242 USA}

\author[0000-0001-8519-1130]{Steven L. Finkelstein}
\affil{Department of Astronomy, The University of Texas at Austin, Austin, TX, 78759 USA}

\author[0000-0002-7831-8751]{Mauro Giavalisco}
\affil{Astronomy Department, University of Massachusetts, Amherst, MA, 01003 USA} 

\author[0000-0001-7673-2257]{Zhiyuan Ji}
\affil{Astronomy Department, University of Massachusetts, Amherst, MA, 01003 USA} 

\author[0000-0003-1187-4240]{Intae Jung}
\affil{Department of Physics, The Catholic University of America, Washington, DC, 20064 USA}
\affil{Astrophysics Science Division, Goddard Space Flight Center, Greenbelt, MD, 20771 USA}

\author[0000-0001-8152-3943]{Lisa J. Kewley}
\affiliation{Research School of Astronomy and Astrophysics, Australian National University, Canberra, ACT 2600, Australia}
\affiliation{ARC Centre of Excellence for All Sky Astrophysics in 3 Dimensions (ASTRO 3D), Australia}

\author[0000-0003-0892-5203]{David C. Nicholls}
\affiliation{Research School of Astronomy and Astrophysics, Australian National University, Canberra, ACT 2600, Australia}

\author[0000-0003-3382-5941]{Norbert Pirzkal}
\affil{Space Telescope Science Institute, 3700 San Martin Drive,
  Baltimore, MD, 21218 USA}

\author[0000-0002-9946-4731]{Marc Rafelski}
\affil{Space Telescope Science Institute, 3700 San Martin Drive,
  Baltimore, MD, 21218 USA}
\affil{Department of Physics \& Astronomy, Johns Hopkins University, Baltimore, MD 21218, USA}

\author[0000-0001-6065-7483]{Benjamin Weiner}
\affil{MMT/Steward Observatory, 933 N. Cherry St., University of Arizona, Tucson,
AZ 85721, USA}

%
\begin{abstract}
  We use deep spectroscopy from the \textit{Hubble Space Telescope}
\textit{Wide-Field-Camera 3} IR grisms combined with broad-band
photometry to study the stellar populations, gas ionization and chemical
abundances in star-forming galaxies at $z\sim~1.1-2.3$.  The data stem
from the \textit{CANDELS Lyman-$\alpha$ Emission At Reionization}
(CLEAR) survey.  At these redshifts the grism spectroscopy measure the
\oii\ $\lambda\lambda$3727, 3729, \oiii\ $\lambda\lambda$4959, 5008,
and \hb\ strong emission features, which constrain the ionization
parameter and oxygen abundance of the nebular gas.   We compare the
line flux measurements to predictions from updated photoionization
models (MAPPINGS V, \citealt{kewley19b}), which include an updated
treatment of nebular gas pressure, $\log P/k=n_eT_e$.   Compared to
low-redshift samples ($z\sim 0.2$) at fixed stellar mass, $\log
M_\ast/M_\odot = 9.4-9.8$, the CLEAR galaxies at $z=1.35$ (1.90) have
lower gas-phase metallicity, $\Delta(\log Z)$ = 0.25 (0.35) dex, and
higher ionization parameters, $\Delta(\log q)$ = 0.25 (0.35) dex,
where $U\equiv q/c$.  We provide updated analytic calibrations between
the \oiii, \oii, and \hb\ emission line ratios, metallicity, and
ionization parameter.  The CLEAR galaxies show that at fixed stellar mass, the gas
ionization parameter is correlated with the galaxy specific
star-formation rates (sSFRs), where $\Delta\log
q\simeq0.4\times\Delta(\log~\mathrm{sSFR})$, derived from changes in
the strength of galaxy \hb\ equivalent width.  We interpret this as a
consequence of higher gas densities, lower gas covering fractions,
combined with higher escape fraction of
H-ionizing photons.  We discuss both tests to confirm these assertions
and implications this has for future observations of galaxies at
higher redshifts.
\end{abstract}


\section{Introduction}\label{section:intro}

Two of the fundamental processes of galaxy evolution are
star-formation and chemical enrichment.  These determine nearly all
their physical and observable properties.  These processes are
diagnostics of the history of gas in galaxies (the ``cosmic baryon
cycle''):  accretion of gas, the conversion of the gas into stars, the
production of heavy elements (i.e., metals), and the distribution of
those metals into and around galaxies. Understanding the history of
these observables is paramount, and for this reason they are a major
focus of galaxy formation theory \cite[see, e.g., reviews
by][]{some15,tuml17,peroux20}.  Because star-formation and metal
production occurred most rapidly in the past at $z\sim 1-3$
\citep{madau14}, it is during this era where measurements of the
relation between star-formation and gas properties is so crucial to
test our theories.  

One of the most important ways to study the properties of gas
involved in star-formation is through the strength and intensity of
nebular emission lines.  These lines are produced from transitions of
ionized (or neutral) gas, where the emission depends on a balance
between heating from ionizing sources (e.g., star-formation) and gas
cooling (which depends on the physical conditions and elemental
abundances of the nebular gas).  The strongest emission lines
associated with these processes reside in the rest-frame optical
portion of the electromagnetic spectrum (e.g., \oii\ $\lambda\lambda
3727,3729$, \hb\ $\lambda4862$, \oiii $\lambda\lambda 4959,5008$, \ha\
$\lambda 6564$, \nii\ $\lambda 6548,6584$).  These lines specifically
contain important information about the instantaneous
flux of ionizing photons (which is related to the star-formation rate
[SFR] and properties of massive stars), the density ($n_e \approx n_H$
for ionized gas) and temperature ($T_e$) of the nebular gas, and
elemental abundances in the gas (specifically for the lines above, the oxygen
abundance (\OH) and nitrogen--to--oxygen abundance (N/O)).

At $z\sim 1-3$ the strong rest-frame optical lines are shifted to
near-IR wavelengths.  It is therefore necessary to study them with
near-IR spectroscopy.  The past decade has seen significant progress
in this area with improvements in multiplexed and
slitless near-IR spectrographs on ground-based and space-based
telescopes \citep[e.g.,][]{straughn11,stei14,kriek15,wisn15,momc16}.
One major findings from these studies is that emission--line ratios in
high-redshift galaxies are offset compared to low-redshift galaxies
\citep[e.g.,][]{shap15,strom17}.    The conclusion is that there are
evolutionary changes \added{ either} in the properties of nebular gas,
where higher redshift galaxies have higher gas densities, lower
metallicities, and higher ionization parameters
\citep[e.g.,][]{kewley13,sand20,strom22}, \added{or in the
  metallicities and abundance ratios (e.g., [$\alpha$/Fe]) of the
  stellar populations \citep[e.g.,][]{sand16a,stei16a,strom17,topp20},
  or combination of these}.     Multiple studies have
analyzed the emission line ratios and (using assumptions about the
physical state of the gas) have quantified the evolution in the well-known
mass-metallicity relation (MZR)
\citep[e.g.,][]{trem04} to $z\sim 3$
\citep[e.g.,][]{sava05,erb06b,maio08,henry13b,henry21,lychun15,lychun16,sand15,sand18,sand21,onod16,suzuki17}.
The interpretation of this evolution is that the chemical enrichment
is tied to star-formation.  This is additionally borne out through
observations that the MZR has a secondary dependence on the SFR such
that O/H decreases with increasing SFR at fixed stellar mass
\citep[e.g.,][]{elli08,mann10,curti20}, and this persists out to at
least $z\sim 2$ \cite[e.g.,][]{zahid14,sand18,henry21}.

Therefore, to interpret the nebular emission of distant galaxies
requires that we understand the evolution of the physical conditions
of the nebular/star-forming gas in galaxies.   The analysis of line
ratios (e.g., the classic \nii--based \citealt{baldwin81} [BPT]
diagram) favors both harder ionizing spectra, higher ionization
parameters ($U = n_\gamma / n_\mathrm{H}$ where $n_\gamma$ is the
density of H-ionizing photons),  and higher gas densities in higher
redshift galaxies
\citep[e.g.][]{hainline09,bian10,kewley13,shap15,sand16a,sand20,strom17,strom18,sand20,runco21}. \citet{kaas18}
studied this evolution using a sample of galaxies at $z\sim 1.5$ and
$z < 0.3$, matched in stellar mass, SFR, and specific SFR.  They
concluded that the higher ionization parameters in galaxies at $z\sim
1.5$ is driven by higher specific SFRs, consistent with higher gas
densities in high redshift galaxies.  

Nevertheless, several key questions remain about the connections
between galaxy nebular emission lines and
their star formation.  One connection that has been less explored is
the relation between star formation and ionization.  \citet{brin08}
show that in low-redshift galaxies (specifically those from the Sloan
Digital Sky Survey, [SDSS], e.g., \citealt{york00,sdss_dr14}) that the
emission line strength (i.e., the rest-frame equivalent width [EW]) of
H-recombination lines (e.g., \ha, \hb) mirrors changes in the
ionization parameter.   This has also recently been observed in
observations of resolved \ion{H}{2} regions of individual galaxies in
the CALIFA survey \citep{espi22}.   Through several lines of
reasoning, \added{\citet{brin08}} argue that this is primarily driven by
higher gas densities for the case of non-zero escape fractions of
H-ionizing photons. This is consistent with the findings of
\citet{kaas18} described above.   If this interpretation is correct,
then there should be a relationship between gas ionization parameter
and the SFR.  This should be particularly
important at high redshifts, where both gas densities and SFRs are
higher \citep[e.g.,][]{madau14,sand16a} and will be even more
important for galaxies pushing to the earliest epochs (into the Epoch
of Reionization [EoR]).  If there exists a correlation between the
ionization parameter and SFR then it would indicate a change in the
physical conditions and/or geometry of the nebular gas, or it could
indicate a change in the nature of the ionizing sources (i.e., the
stars), or a combination of these.  This has yet to be tested in the
distant Universe. 

Here, we use slitless spectroscopy taken with the \textit{Hubble Space
Telescope} (\hst) \textit{Wide Field Camera 3} (WFC3) grisms to study
these questions.  The WFC3 grisms have several advantages over
ground-based spectrographs.  These data have no ``preselection'' (we
take spectra of all galaxies in the field) and the data have
continuous wavelength coverage (where ground-based data are littered
with atmospheric emission lines and limited by atmospheric
absorption).   The WFC3 data therefore provide a complementary picture
of galaxies at high redshift.  In this \textit{Paper}, we use these
data to diagnose the star-formation properties for galaxies at $z\sim
1-2.3$. The data probe observed-frame near-IR wavelengths covering
0.8-1.6 micron, and cover strong emission lines for galaxies at $z\sim
1-2$ that trace both gas ionization-parameter ($q$) and metallicity
(i.e., the oxygen abundance, \OH).
%
%
This allows us to study the evolution of the gas metallicity and
ionization, and compare it to other galaxy properties.  Importantly,
this work also demonstrates the capabilities of space-based slitless
spectroscopy to address this science.  This will be an important
capability of future telescopes (including both the \textit{James Webb
Space Telescope} [\jwst], and the \textit{Nancy Grace Roman Space
Telescope} [\textit{NGRST}]).    

The outline for this \textit{Paper} is as follows.  In
Section~\ref{section:data} we describe the datasets, sample selection,
and methods to derive stellar-population properties using broad-band
data and spectroscopy.  In Section~\ref{section:spectra_stack}, we
describe the grism spectra for the galaxies in our sample, including
the properties of stacked (average) spectra.  In
Section~\ref{section:gasproperties} we discuss the emission-line
ratios of galaxies in the sample, we describe the method to derive gas
metallicities and ionization parameters, and we discuss relations
between the line ratios and the measured parameters.   In
Section~\ref{section:results} we measure the mass--metallicity
relation (MZR) and the mass--ionization-parameter relation (MQR) for
the CLEAR samples.  In Section~\ref{section:discussion} we discuss the
implications for the evolution of gas metallicity,
ionization-parameter, and specific SFRs (sSFR $\equiv$ \mstar/SFR).
In Section~\ref{section:summary} we summarize our findings.
Appendix~\ref{section:appendix} compares the constraints on gas
metallicity and ionization parameter used here (derived from \oii,
\hb, and \oiii\ line emission) to those that also include \ha+\nii\
(and in some cases \sii).

Throughout we use a cosmology with $\Omega_{m,0}=0.3$,
$\Omega_{\Lambda,0}=0.7$, and $H_0 = 70$~km s$^{-1}$ Mpc$^{-1}$,
consistent with results from Planck \citep{planck18} and the local
distance scale \citep{riess22}.  We adopt 
Solar abundances from \citet{asplund09}, where $\OH_\odot = 8.69$, or
alternatively, $\OH = \log Z_\mathrm{gas}/Z_\odot + 8.69$.     All
magnitudes reported here are on the \textit{Absolute Bolometric} (AB)
system \citep{oke83}.

\section{Data and Sample}\label{section:data}

The primary datasets for this study include broadband photometric
catalogs for the GOODS-N and GOODS-S fields \citep[and see
below]{skel14} combined with WFC3 slitless spectroscopy from CLEAR
(GO-14227, PI: Papovich, see \citealt{estr19} and \citealt{simons21})
and 3D-HST \citep{momc16}.  We describe these datasets below
(Sections~\ref{section:phot} and \ref{section:reduction}), and our
sample selection for star-forming galaxies at $0.7 < z < 2.3$
(Section~\ref{section:selection}).  

\subsection{SDSS Comparison Catalog}\label{section:sdss}

As a low-redshift comparison sample, we make use of data from the SDSS
Data Release 14 \citep[DR14,][]{sdss_dr14} which includes emission
line fluxes and value-added catalogs.  This catalog includes emission
line fluxes corrected for Balmer absorption and dust attenuation for
SDSS III (including a reanalysis of galaxies from SDSS II;
\citealt{thomas13}).  We opt to use the stellar masses derived in the
value-added catalog of \citet{chen12} using the \citet{bruz03} stellar
populations (and a Kroupa IMF) as these more closely match those
derived for our CLEAR sample.    For consistency in the comparison, we
rederive the gas-phase oxygen abundances and ionization parameters of
the SDSS galaxies using the same emission lines (\oii, \oiii, \hb) and
method applied to the CLEAR sample (discussed below,
Section~\ref{section:izi}).  

\subsection{CLEAR Photometric Catalog}\label{section:phot}

The CLEAR \hst/WFC3 pointings all lie within the CANDELS
\citep{grogin11,koek11} GOODS-N and GOODS-S fields.   The foundation
of the CLEAR photometric catalog is the 3D-HST catalog from
\citet{skel14}, which provides multiwavelength catalogs with
photometric coverage from 0.3--8~\micron.  We have added to these
\hst\ F098M and/or F105W (i.e., $Y$-band) imaging as described in
\citet[see also Simons et al., in prep]{estr19}.    We then re-derived
photometric redshifts, rest-frame colors ($U-V$ and $V-J$) and derived
stellar masses using an updated version of \texttt{EAZY}
\citep{bram08, eazy-py}.
We refer to this catalog as 3D-HST+.   We use the 3D-HST+ catalog for
preliminary selection of the samples used here.  We subsequently
performed more sophisticated fits to the spectral energy distributions
(SEDs) including both the 3D-HST+ catalog broad-band photometry and
WFC3 G102 and G141 grism spectra, and use the quantities derived from
these latter fits for the analysis here (see
Section~\ref{section:sedfit}). 

\subsection{CLEAR WFC3 Slitless Spectroscopy, Data Reduction, and
  Line-Flux Measurements} \label{section:reduction}

The CLEAR program provides deep WFC3/G102 slitless spectroscopy in 12
pointings in the GOODS-N and GOODS-S fields.  These data use
observations of 10 or 12-orbit depth with WFC3/G102, which observe
wavelengths 0.80--1.15~\micron\ with $R\sim 210$.   We combined these
data with all other available G102 data that overlap the CLEAR fields,
including those data from programs GO-13420 (PI: Barro; see
\citealt{barro19}), GO/DD-11359 PI: O'Connell; see
\citealt{straughn11}) and GO-13779 (PI: Malhotra; see
\citealt{pirz18}).  These data are described fully in a forthcoming
paper (see \citealt{estr20,simons21}, and in prep).

We augment the CLEAR data with \hst\ WFC3 slitless spectroscopy with
the G141 grism from 3D-HST \citep{momc16} that cover the CLEAR fields.
The G141 data cover observed wavelengths 1.08--1.70~\micron\ with
$R\sim 130$ and achieve flux limits for emission lines of $2.1\times
10^{-17}$~erg s$^{-1}$ cm$^{-2}$ (3$\sigma$ for point sources,
\citealt{momc16}).  

We processed both the CLEAR and all ancillary WFC3 grism data in the
CLEAR fields (see, \citealt{simons21} and R.~Simons et al., in prep)
using the grism redshift line and analysis software
\grizli\ \citep{grizli}.  The full process is described elsewhere
(\citealt{simons21}, and see also see also \citealt
{estr19,estr20,matharu22}). In brief, we first reprocess the WFC3
G102 data, applying steps to correct for variable backgrounds and the
flat-field, and we perform a sky-subtraction using the ``Master Sky''
provided in \citet{bram15}.  We derive relative astrometric
corrections to the processed data by aligning to the WFC3 F140W mosaic
from \citet{skel14}.

We then use \grizli\ to model the G102 and G141 spectra of each object
using the F105W and F140W direct images, respectively, and a coarse
model fit to each galaxy's SED.  We correct for galaxy
contamination by subtracting the models for the spectra from nearby
objects.  This process is iterative.  On the first pass we model the
spectra of all objects with $m_\mathrm{F105W} < 25$~AB mag.   We
repeat the steps above. On the second pass we apply a finer model
correction to all objects with $m_\mathrm{F105W} < 24$~AB mag.   The
adopted magnitudes for these steps are similar to those applied in the
processing of the 3D-HST data \citep{bram12,momc16}.  Because the
CLEAR G102 data are similar in depth to the 3D-HST G141 data, we
achieve similar results (and visual inspection of the spectra and
their residuals shows this accounts for the majority of contamination).

Finally, we use \grizli\ to extract two-dimensional (2D) and
one-dimensional (1D) spectra for all galaxies in the 3D-HST+ catalog
that fall in the CLEAR fields with brightness, $m_\mathrm{F105W} \leq
25$~AB mag, including the corrections for contamination described
above.  We used \grizli\ to measure spectroscopic  redshifts, emission
line fluxes, and stellar-population parameters from the spectral fits
to the continua and emission lines from the G102 and G141 data and the
available multiwavelength broad-band photometry from the 3D-HST+
catalog  (see Section~\ref{section:phot}).  For this process, \grizli\
uses a set of template basis functions derived from the Flexible
Stellar Populations Synthesis models (FSPS; \citealt{conroy10b}) that
include a range of stellar populations and nebular emission lines.
\grizli\ integrates each model with the transmission functions of the
broad-band filters (including the system throughput of the telescope
and detectors), and projects each stellar population model to match
the G102 and G141 2D spectral ``beams'' using the observed direct
image (F105W for G102; F140W for G141), matching the role angle
(ORIENT) of \hst\ and object morphology as closely as possible.  This
approach is required to model the unique morphological broadening of
the spectral resolution (required for slitless spectroscopy).
\grizli\ performs a non-negative linear combination of the template
spectra and determines a redshift through $\chi^2$ minimization and a
marginalization over redshift.     \grizli\  fits emission line fluxes
using the best-fit redshift.  It subtracts the continua (correcting
for absorption features, e.g., from Balmer lines) using the best-fit
stellar population model.   The depth of the G102 data varies slightly
in some of the fields (which contain different numbers of orbits from
ancillary data) and the sensitivity depends somewhat on wavelength.
Nevertheless,  the bulk of the data (assuming the nominal 12 orbit
depth) are sensitive to emission line fluxes for point sources of
$\approx 2 \times 10^{17}$~erg s$^{-1}$ cm$^{-2}$ (3~$\sigma$),
comparable to the G141 data (Simons et al.\ in prep).
%

Here, we use the CLEAR v3.0 catalogs, which are an internal team
release. These include emission line fluxes, spectroscopic redshifts,
and other derived quantities and their respective uncertainties for
6048 objects from \grizli\ run on the combination of the G102 and G141
grism data and broad-band photometry using the 3DHST+ catalogs.   Of
these galaxies, 4707 galaxies have coverage with both G102 and G141.
These will be described fully in the forthcoming paper on the data
release (Simons et al., in prep.) and have been discussed in other
papers using these data
\citep[e.g.,][]{estr19,estr20,simons21,jung21,back21,cleri22,matharu22}.

\subsection{Galaxy Sample Selection}\label{section:selection}

%
Here we use the CLEAR spectroscopy to study galaxies with coverage of
strong emission lines that are tracers of the gas-phase oxygen
abundance and nebular ionization, namely \oii\ $\lambda\lambda
3727,3729$, \hb\ $\lambda4862$, and \oiii $\lambda\lambda 4959,5008$
\citep[e.g][and many others (see references therein)]{maio08, sand15,
sand21, curti17, strom18, kewley19b, maio19, henry21}.  Our  G102 and
G141 data cover all of these lines for galaxies at redshifts $1.1 < z
< 2.3$.  In addition, for galaxies in the redshift range $1.1 < z <
1.6$ our data include coverage of \ha\ $\lambda 6564$ + \nii\ $\lambda
6548,6584$ (which are blended at the grism data, see below). The
spectra also provide coverage of \neiii\ $\lambda 3869$,  which is not
detected in the majority of galaxies (but see \citealt{back21}), but
is observed in galaxy stacks (see below in
Section~\ref{section:spectra_stack}).
%

We selected galaxies from CLEAR for this study using the following
criteria:  
\begin{itemize}
    \item Spectroscopic redshift derived from the grism data in the
range, $1.1 < z_\mathrm{grism} < 2.3$.  This redshift range ensures
that all three of the lines \oii, \oiii, and \hb\ are all contained by
the G102 and/or G141 data.
    \item Detection of \oii, \hb, or \oiii\ with SNR $\ge$ 3 in at
least one line in the total (combined) 1D spectra.
    \item Galaxies are un-detected in X-ray catalogs based on the
\textit{Chandra} X-ray catalogs for CDF--N \citep{xue16} and CDF--S
\citep{luo17}; we rejected objects within 1\arcsec\ of sources flagged
as \texttt{Type = AGN}.  This step excludes strong AGN, and removes
5\% of sources in the GOODS-N and 6\% of sources in the GOODS--S
3DHST+ parent catalogs.  As an additional test, we checked if any
additional objects in our sample are flagged as potential AGN using
the ``Mass-Excitation'' (MEx) diagnostic of \citet{juneau14}, modified
to account for redshift \citep{coil15,henry21}.  This removed no
additional objects (which we interpret as evidence that all candidate
AGN in our sample are identified as such in the ultra-deep CDF--N and
CDF--S X-ray data). 
\end{itemize}
 In addition, we remind the reader that all galaxies have
$m_\mathrm{F105W} \le 25$~AB mag as they are drawn from our CLEAR
3DHST+ catalog (see Section~\ref{section:reduction}).

\begin{figure}[t]
\centering
\includegraphics[width=0.48\textwidth]{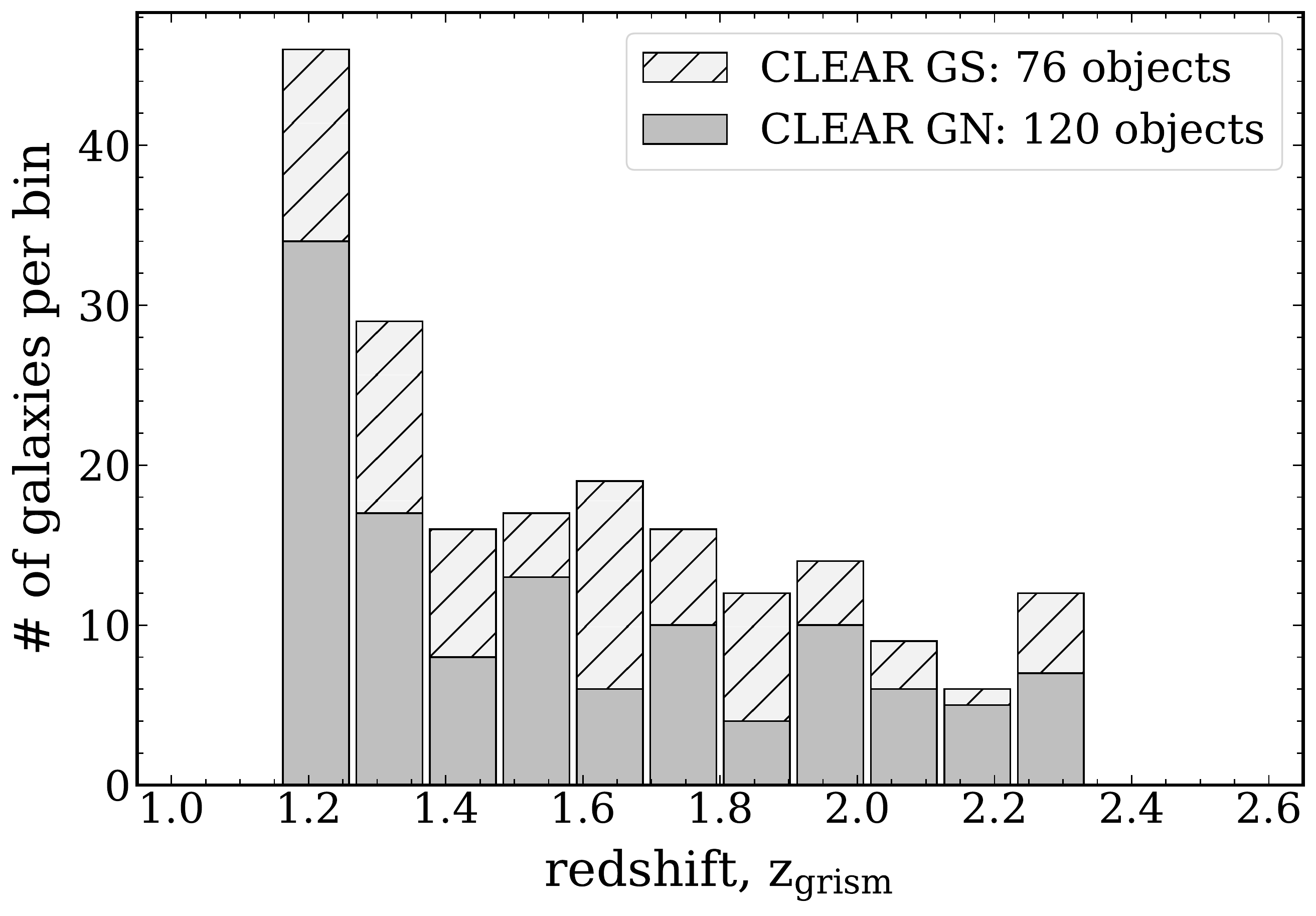}
\caption{Redshift distribution of galaxies in the CLEAR $1.1 < z <
2.3$ sample in this study.  This sample is selected from the 3D-HST+
parent catalog with spectroscopic redshifts from the grism data in
this redshift range and by requiring that all galaxies have SNR $>$3
in \oii, \hb, or \oiii.     \label{fig:zhist}}
\end{figure}

The selection produces a sample containing 196 galaxies.
Figure~\ref{fig:zhist} shows their redshift distribution.  The
redshifts span $1.1 < z < 2.3$ with a median of 1.5.  The distribution
is highly peaked at the first redshift bin, with $z\sim 1.25$.  This
is largely a result of galaxies in GOODS-N (GN), which makes up a
larger number of sources (120 galaxies) in our sample compared to
GOODS-S (GS, 76 galaxies).  We consider two bins in redshift, each
containing roughly 50\% of the sample, with one bin defined with $1.1
< z < 1.5$ (median $z = 1.3$) and the other with $1.5 < z < 2.3$
(median $z = 1.8$).  This allows us to test for redshift evolution in
the properties of the sample.

\begin{figure}[t]
\centering
\includegraphics[width=0.48\textwidth]{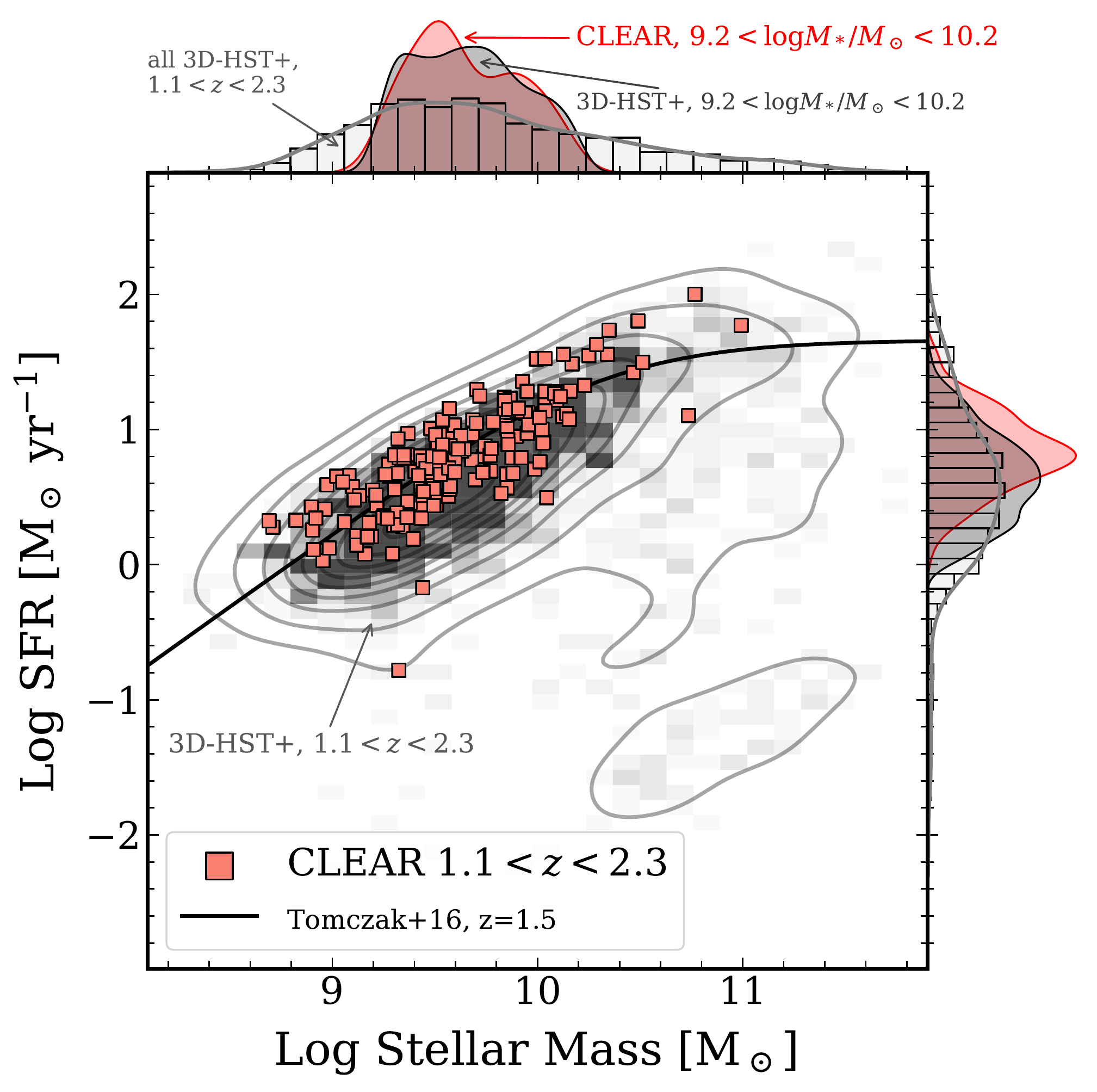}
\caption{SFR--Stellar-Mass distributions for galaxies in our CLEAR
sample.  The shaded regions and contours show the distribution of
sources with $1.1 < z < 2.3$ from the  3D--HST+ parent catalog.   The
red-colored squares show the CLEAR sample studied here.
The plots at the top and to the right show the distributions of the
samples in $\log M_\ast$ and $\log$ SFR using a histogram and  kernel
density estimator.
%
%
\label{fig:sfms}}
\end{figure}

Figure~\ref{fig:sfms} shows the stellar-mass--SFR distribution for our
CLEAR sample of $1.1 < z < 2.3$ galaxies using the selection criteria
above, compared to the 3D--HST+ parent sample in the same redshift
range.   The stellar-mass distribution of the CLEAR sample is
consistent with the 3D--HST+ sample when we restrict the stellar--mass
range to $9.2 < \log M_\ast/M_\odot < 10.2$ (where both samples are
reasonably complete).    The SFR distributions show that the CLEAR
sample here is biased toward higher SFRs, by 0.18~dex (a factor of
1.5) compared to the 3D--HST+ parent sample.  The bias can be
explained as a result of the emission line selection:  we require
galaxies to have SNR $>$3 in \hb, \oii, and/or \oiii.  The CLEAR
line-flux detection limit is $2\times 10^{-17}$ erg s$^{-1}$ cm$^{-2}$
(3$\sigma$), which for \hb\ corresponds to SFR$\simeq$3--7 $M_\odot$
yr$^{-1}$ (with no dust attenuation) at $z=1.5-2.0$ (assuming the
calibration of \citealt{kenn98} for a Chabrier IMF).  Comparing this
to Figure~\ref{fig:sfms} we see that this effectively limits our study
to objects with higher SFRs than the median. This bias in SFR is
similar to other studies of emission-line selected studies of galaxies
\citep[cf.,][]{shivaei15,sand18}.  We expect this bias to have only a
minor impact on our results as previous studies
have shown that a change in SFR of 1 dex
corresponds to a change in metallicity of $\simeq$0.3 dex
\citep[e.g.,][]{henry21}.  Based on this argument the bias in SFR
between the emission-line-selected sample and the parent sample,
$\Delta(\log~\mathrm{SFR}) \simeq 0.18$~dex (Figure~\ref{fig:sfms}),
corresponds to $\Delta(\log Z) = 0.05$~dex. 

In Appendix~\ref{section:appendix}
we also consider a subset of 87 galaxies from this sample with $1.2 < z
< 1.5$ for which \ha+\nii\ are covered by the data.   These galaxies
have a median redshift $z = 1.30$.  We use this subsample to test how
incorporating additional lines impacts the constraints on the
gas-phase metallicity and ionization \citep[cf.][]{henry21}.

\subsection{Estimating Galaxy Stellar Masses, Dust Attenuation, and
SFRs}\label{section:sedfit}

In what follows we compare the galaxy emission-line properties
(including derived quantities such as gas-phase metallicity and
ionization parameter) to galaxy stellar population parameters,
including stellar masses, SFRs, and sSFRs.    To derive these latter
quantities we use a custom-designed method that fits stellar
population synthesis models to the broad-band photometry (from our
3D-HST+ catalog, see Section~\ref{section:phot}) and the WFC3 G102 and
G141 1D spectra (see Section~\ref{section:reduction}).   The method is
discussed in detail elsewhere \citep{estr20,estr21}, and we summarize
it here.

\begin{figure*}
   \gridline{
    \fig{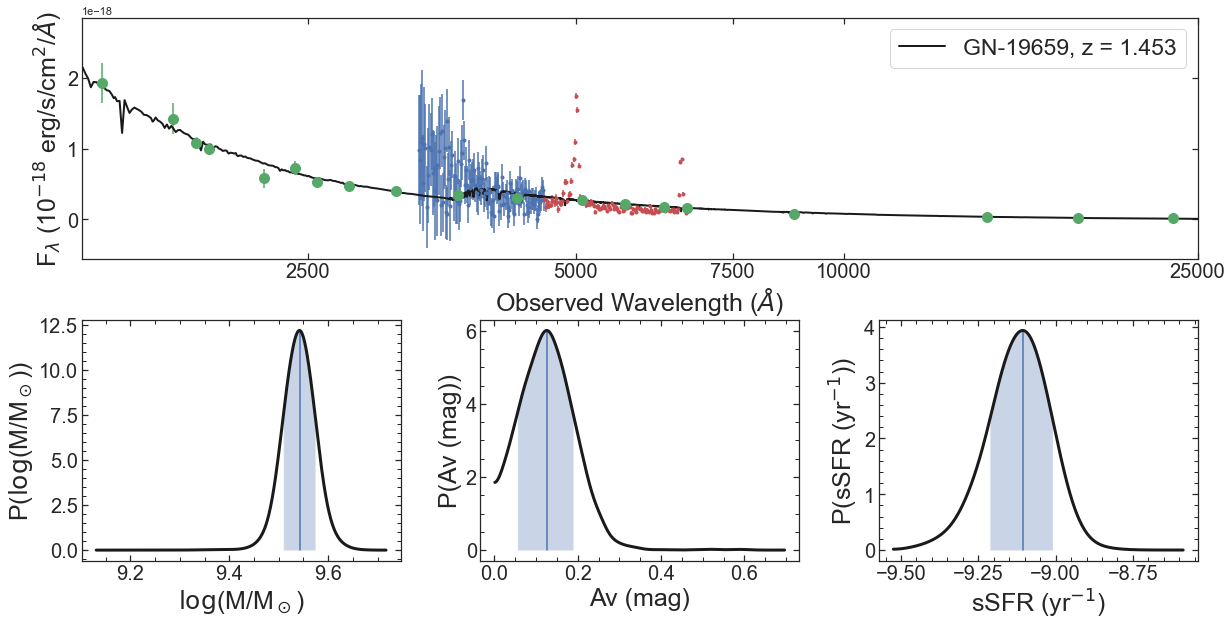}{0.7\textwidth}{}
  }
  \vspace{-12pt}
  \gridline{
    \fig{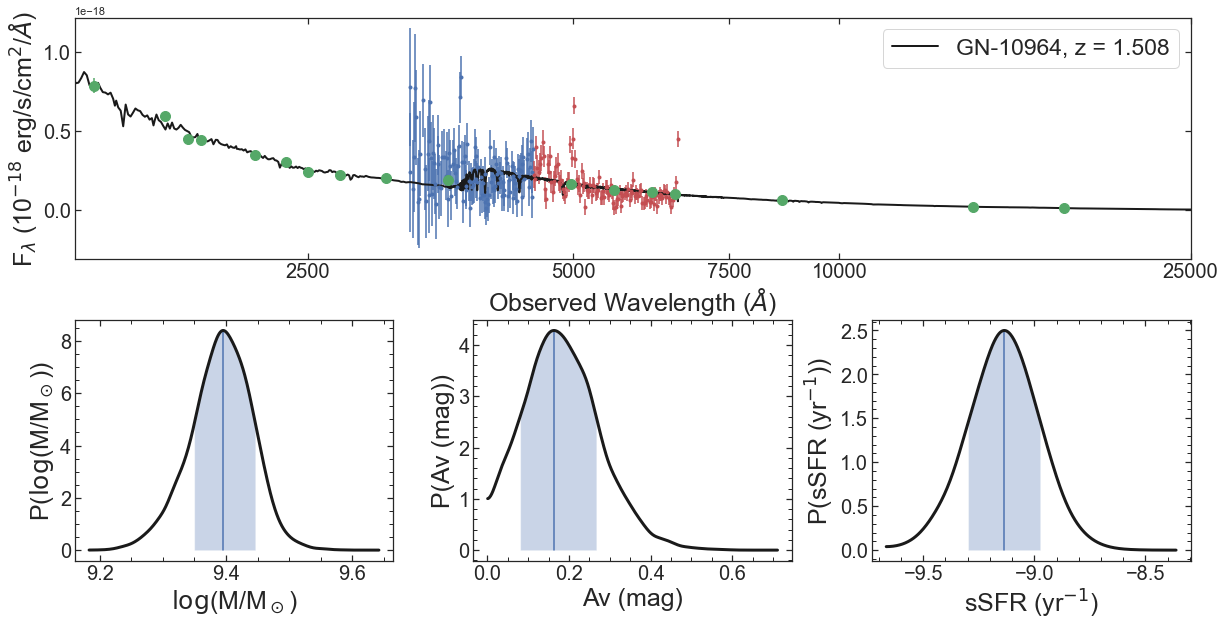}{0.7\textwidth}{}
}
  \caption{Examples of model fits to the broad-band photometry and
\hst\ grism data for two example galaxies at $z~\sim 1.5$ in the GOODS-N
CLEAR fields (GN 19659 is also shown in
Figure~\ref{fig:spectra_izi_results}).   The fitting procedure uses
the method of \citet[and in prep]{estr20})   The top row of each set of panels shows the
broad-band photometry (green dots) from the CLEAR 3DHST+ catalog, the
G102 (blue) and G141 (red) spectra along with a best-fit stellar
population model (black line).    The model fits currently exclude
emission lines (though the emission features are prominent in the data
for these galaxies).  The lower set of panels for each galaxy show the
posteriors for the stellar mass ($\log M/M_\odot$), dust attenuation
($A(V)$/mag), and specific SFR ($\log$ sSFR/yr$^{-1}$), where the
vertical line shows the median and the shaded region shows the 16--84
percentile range of the HDI. \label{fig:sed_results}}
  \end{figure*}


\begin{figure*}[th]
\centering
\includegraphics[width=0.9\textwidth]{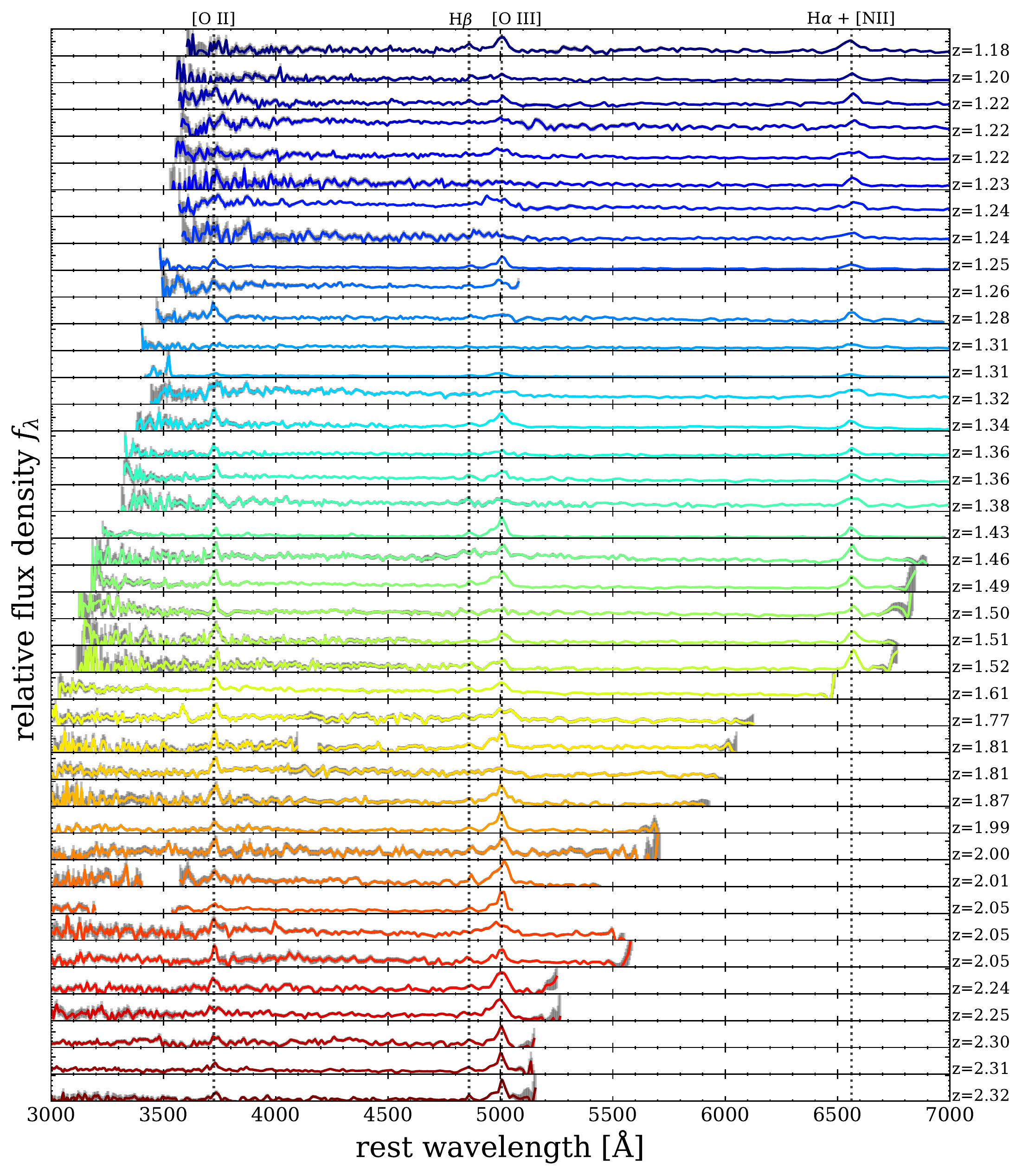}
\caption{Gallery of individual one-dimensional G102 + G141 spectra for the sample
of 40 galaxies in CLEAR with $1.1 < z < 2.3$ in the stellar mass
range $9.6 < \log M_\ast / M_\odot < 9.9$ with SNR $>$3 in at least
one of \oii\ $\lambda$3726,3729, \oiii\ $\lambda\lambda$4959, 5008,
and \hb\ $\lambda$4861.   The spectra have been shifted to the
rest-frame to illustrate common spectral features.  The color scale
changes with spectroscopic redshift (the gray shading of each spectrum
indicates the uncertainty).  For galaxies with $z \lsim 1.6$ the data
cover \ha\ $\lambda$6563 (which is blended with \nii\ $\lambda\lambda$
6584, 6584 at the resolution of the G141
grism). \label{fig:examplespec}}
\end{figure*}


We use the FSPS models 
\citep{conroy10} with a \citet{kroupa01} IMF.  We
fit a total of 23 parameters, including metallicity (of the stellar
population, $Z_\ast$), age, dust attenuation ($A_V$, assuming the
\citealt{calz00} model), and a flexible star-formation
history (allowing for 10 bins of SFR dynamically-spaced in time,
following \citealt{leja19}).  We also include 8 additional nuisance
parameters to allow offsets in the normalization/calibration between
the spectra and the photometry, and to allow for correlated noise
between spectral data points (see \citealt[and in prep]{estr20} for
more details).  We then fit to the broad-band data and grism
spectroscopy (for this modeling we currently exclude regions of strong line
emission) using a Bayesian formalism with a nested sampling to
predict the posteriors.  We  marginalize the posterior probability
distribution functions to derive constrains on the stellar population
parameters.

\added{We find that excluding the emission lines from the SED fitting
causes a small bias in the SFRs (and the specific SFRs) of the galaxies in our
sample.  We compared the SED-derived SFRs for objects in our sample to
SFRs estimated from dust-corrected \hb\ emission measured in the grism
data (assuming Case-B recombination and the calibration of
\citealt{kenn12}).  For galaxies with SFRs $\lesssim$ 5 $M_\odot$
yr$^{-1}$ the  \hb-derived SFRs estimated are higher by about
0.25~dex (for galaxies with higher SFRs the bias is negligible).
Because we use SED-derived specific SFRs, we ensure we are not biased
toward galaxies with strong emission lines only.   Furthermore, this
potential bias in SFR (and specific SFR) has negligible impact on our
conclusions related to the specific SFR as this bias is smaller than
the trends seen in the data and remains  present if we replace the
specific SFR with alternative measures (such as \hb\ equivalent width,
see \citealt{reddy18}, and Sections~\ref{section:logq_ssfr} and
\ref{section:ZQR}).}

Using this method we fit the stellar population parameters for all the
galaxies in our sample.  Here we focus on the stellar
population constraints for stellar mass ($M_\ast$), SFR, and dust
attenuation ($A_V$) for the study here.  We will present results
derived from these parameters elsewhere (V. Estrada-Carpenter et al.\
in prep).  Figure~\ref{fig:sed_results} shows results from the fitting
for two galaxies in our sample.  The figure includes the best-fit SED
(with parameters that maximize the likelihood) along with the
broad-band photometry and grism data.  The figure also shows the
marginalized posteriors for the three parameters above.  We take the
mode and highest density interval \citep[HDI,][]{bailer18} from the posteriors
as the measurement and inter-68 percentile range (e.g., the inter--16th-to-84th
percentile range) for each parameter, respectively.   In what follows,
we refer to these as ``SED--derived'' values as they were derived from
fitting models to the galaxy SEDs. 

\section{Characteristics of Grism Spectra of Emission-Line Galaxies at
  ${1.1 < {\lowercase{\mathrm{z}}} < 2.3}$}\label{section:spectra_stack}

Figure~\ref{fig:examplespec} shows a gallery of the G102 + G141
1D spectra for the 40 individual galaxies in our sample in the stellar mass range $9.6 <
\log M/M_\odot < 9.9$, ordered by increasing redshift.   The spectra
are shifted to the rest-frame to illustrate common features.  The most
prominent lines are \oii\ $\lambda$3726,3729, \oiii\
$\lambda\lambda$4959, 5008, and \hb\ $\lambda$4861.   For galaxies
with $z \lsim 1.6$, \ha\ $\lambda$6563 is also present.  At the
resolution of the G141 grism, this line is blended with the  \nii\
$\lambda\lambda$ 6584, 6584 lines.

\begin{figure*}[th]
\centering
\includegraphics[width=\textwidth]{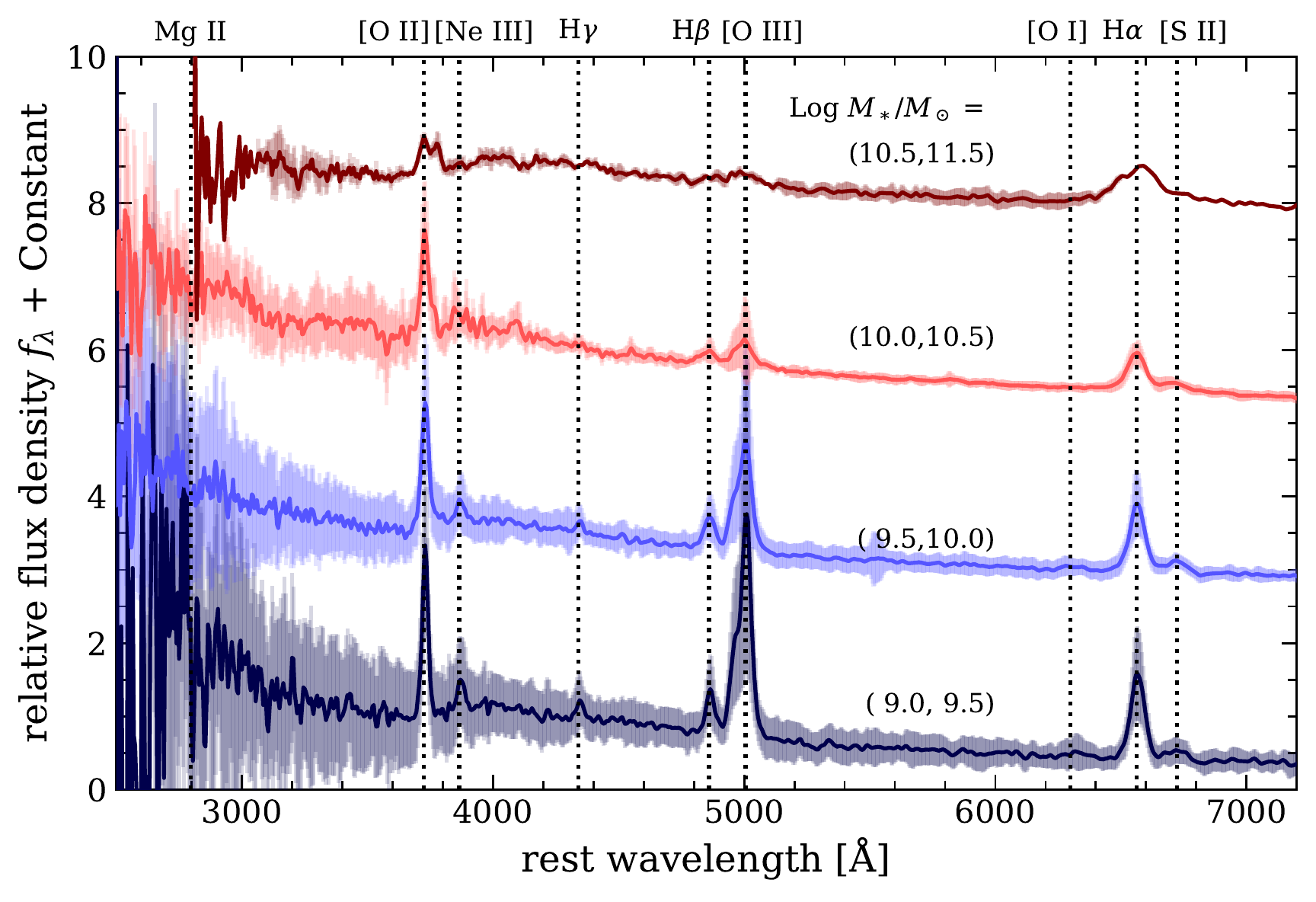}
\caption{One-dimensional G102 + G141 spectral stacks of galaxies in
the sample ($1.1 < z < 2.3$), divided in four bins of stellar mass.
The spectra have been shifted to the rest-frame, and stacked as
described in the text.  The solid lines show the weighted mean of each
stack and the shaded region shows standard deviation of the population
(NB:  this the dispersion of the objects in the stack, and
\textit{not} the error on the mean).  The locations of prominent
emission features are labeled. The stacked spectra illustrate the
effects of the instrumental resolution on the emission features.  The
data also show that the relative strength of emission features depends
on stellar mass, with galaxies with lower stellar masses having
stronger emission lines, and stronger \oiii\ compared to \oii\ and
\hbeta. \label{fig:allstack}}
\end{figure*}

\subsection{Stacked (Average) Spectra of CLEAR
Galaxies}\label{section:stacking}

To facilitate with the interpretation of the \hst\ spectra, we
constructed stacked spectra for galaxies in our sample in bins of
stellar mass.   We first divided the galaxies into subsamples of
stellar mass, $9.0 < \log M_\ast/M_\odot < 9.5$, $9.5 < \log
M_\ast/M_\odot < 10$, $10 < \log M_\ast/ M_\odot < 10.5$, and $10.5 <
\log M_\ast/\msol < 11.5$.   To create the stacks we corrected the
spectra for dust attenuation assuming the \citet{calz00} model and the
$A(V)$ values derived from the SED fits (see
Section~\ref{section:sedfit}).  We then shifted all 1D spectra for the
galaxies to the rest-frame using the measured redshift from the grism
data.  We linearly interpolated the data to a wavelength grid over
$2500-7500$~\AA\ at a resolution of $\delta \lambda$=5~\AA.   We
normalized each galaxy in the rest-wavelength range $4300-4500$~\AA\
(a window that avoids strong emission features, following
\citealt{zahid17}) and co-added the spectra, weighting by the inverse
variance of the flux density.  We then divided the spectra by the
total weights to obtain a mean spectrum.  We also created a weighted
sum of the variance of the flux density in the same way to study the
variation of the spectra among galaxies in the sample.   

Figure~\ref{fig:allstack} shows the stacked spectra of the CLEAR
galaxies in the bins of stellar mass.    The spectra show common
features, most prominently strong emission from \oii, \oiii, \hb, and
\ha.  In addition, weaker lines are also evident, including
[\ion{Ne}{3}], H$\gamma$, [\ion{S}{2}], and [\ion{O}{1}].  The shaded
region of the stacked spectra in the figure shows the scatter of the
population in each stack (i.e., this is \textit{not} the uncertainty
on the mean). The shading indicates the scatter is  generally larger
at shorter wavelengths, which we attribute to variations in
star-formation histories (although some of this may be caused by the
lower sensitivity of the G102 grism at bluer wavelengths). 

In general, the strength of the spectral features increases with
decreasing stellar mass.  In particular, it is in the lowest mass
galaxies ($9.0 < \log M_\ast/M_\odot < 9.5$) where the strongest
emission is seen, and where weaker lines such as [\ion{Ne}{3}] and
H$\gamma$ become prominent.  The relative strength of \oiii/\oii\ also
increases with decreasing stellar mass.  This is an indication of
increasing gas ionization and/or decreasing gas-phase oxygen
abundance.  We will explore this quantitatively below.

\begin{deluxetable*}{lcccccccc}
\tabletypesize{\footnotesize}
\tablecolumns{9}
\tablewidth{0pt}
\tablecaption{Relative Emission Line Flux Ratios and Line Equivalent Widths from
  Stacked Spectra of CLEAR Galaxies at $1.1 < z < 2.3$}. \label{table:ppxf}
\tablehead{
& \colhead{Number} &
 &  &
 &  & 
\colhead{EW \hb} & \colhead{EW H$\gamma$}  & 
 \colhead{EW H$\epsilon$\tablenotemark{\ddag}} \\ [-8pt]
\colhead{sample}  &  \colhead{of galaxies} &
\colhead{$\langle \OH \rangle$\tablenotemark{$\ast$}} &
\colhead{\editone{log} R$_{23}$}  &
\colhead{\editone{log} O$_{32}$} &
\colhead{\editone{log} \neiii/\oii} & 
\colhead{ [\AA]} &
\colhead{ [\AA]} &
\colhead{[\AA]}
}
\startdata
\cutinhead{Stacked Spectra of Galaxies in bins of Stellar Mass}
$\log M_\ast / M_\odot > 10.5$ & 9 & 8.83 $\pm$ 0.05 & 0.55
$\pm$ 0.06 & $-$0.40 $\pm$ 0.12 & $<  -1.33$ & 21.9
$\pm$ 1.0  & \phn 4.3 $\pm$
1.1\tablenotemark{$\dag$}  & 
2.5 $\pm$ 1.1\tablenotemark{$\dag$} \\[4pt]
$\log M_\ast / M_\odot = [10, 10.5)$ & 36 & 8.65 $\pm$ 0.03 & 0.70
$\pm$ 0.06 & $-$0.15 $\pm$ 0.11 & \editone{$-$}1.18 $\pm$ 0.12 & 23.6 $\pm$ 1.2
& \phn 9.7 $\pm$ 1.0\tablenotemark{$\dag$}   &
4.7 $\pm$ 1.1\tablenotemark{$\dag$} \\[4pt]
$\log M_\ast / M_\odot = [9.5, 10)$ & 84 & 8.55 $\pm$ 0.02 & 0.80
$\pm$ 0.03 & \phs 0.07 $\pm$  0.11 & $-$1.03 $\pm$ 0.17 & 44.2
$\pm$ 1.9\phs & 16.3 $\pm$ 2.0\phs  &
4.7 $\pm$ 1.7\phs    \\[4pt]
$\log M_\ast / M_\odot = [9.0, 9.5)$ & 66 & 8.42 $\pm$ 0.03 & 0.92
$\pm$ 0.03 & \phs 0.27 $\pm$ 0.11 & $-$0.87 $\pm$ 0.19 &  49.6 $\pm$
3.8\phs  & 18.5 $\pm$ 3.2\phs   &
8.0 $\pm$ 3.5\phs   \\[4pt] 
\cutinhead{Stacked Spectra of Galaxies with High- and Low-Ionization
  Parameters, all with $\log M_\ast /
  M_\odot = [9.3, 9.7]$ and $\OH > 8.3$} 
 High-ionization, & \multirow{2}{*}{31} & \multirow{2}{*}{8.49 $\pm$ 0.04} & \multirow{2}{*}{0.89 $\pm$ 0.01} & \multirow{2}{*}{\phs 0.35
$\pm$ 0.11} & \multirow{2}{*}{$-0.96$ $\pm$ 0.11} &  \multirow{2}{*}{61.3 $\pm$ 0.6} & \multirow{2}{*}{23.7 $\pm$ 0.5} & \multirow{2}{*}{8.2 $\pm$ 0.5} \\
$\log q > 7.8$ \\[6pt]
Low-ionization, & \multirow{2}{*}{32} & \multirow{2}{*}{8.52 $\pm$
  0.04} & \multirow{2}{*}{0.86 $\pm$ 0.03} & \multirow{2}{*}{$-$0.03
  $\pm$ 0.11} & \multirow{2}{*}{$-1.12$ $\pm$ 0.11} &
\multirow{2}{*}{35.4 $\pm$ 0.5} & \multirow{2}{*}{13.0 $\pm$ 0.4} &
\multirow{2}{*}{4.4 $\pm$ 0.5} \\
$\log q < 7.8$ \\ [4pt] 
\enddata
\tablenotetext{\ast}{Mean metallicity, and uncertainty on the mean,
derived from the measurements of the individual galaxies in the sample
(see text).}\vspace{-6pt}
\tablenotetext{\dag}{H$\gamma$ and/or H$\epsilon$ weakly detected;
flux ratios forced to their theoretical values in
model fitting, H$\gamma$/\hb\ $=$ 0.468 and H$\epsilon$/\hb\ $=$ 0.159
\citep{oste89} }\vspace{-6pt}
\tablenotetext{\ddag}{\added{Blended with [\ion{Ne}{3}] 3968.}}
\tablecomments{All line ratios and emission line equivalent widths
are measured from the stacked spectra using \ppxf\  as described in
the text (Sections~\ref{section:ppxf} and \ref{section:ZQR}).
Equivalent widths are in measured in the rest-frame.  }
\vspace{-6pt}
\end{deluxetable*}

\subsection{Measuring Emission Line Ratios from Stacked
Spectra}\label{section:ppxf}

Because of this sizable variation in the spectral properties of the
galaxies even at fixed stellar mass, we opt to study the individual
spectra in most of the analysis that follows.  However, we also use
emission line ratios and equivalent width measurements from the stacks
to interpret the average evolution of galaxies as a function of
stellar mass and gas-phase metallicity.  This complements work that
analyzes the average properties of galaxies from stacked \hst\ WFC3
grism spectra (including, e.g., \citealt{henry21}, which in part uses
stacks that include the same data used here).

To measure the emission line fluxes from stacked spectra in
Figure~\ref{fig:allstack} we adapted the Penalized Pixel-Fitting
method (\ppxf, \citealt{capp17}) for the CLEAR data.  The primary
difference between our use of \ppxf\ and other datasets is that the
CLEAR WFC3 grism data are lower resolution ($R\sim 100-200$) than
other spectroscopic studies \citep[see,][]{capp17}.  Nevertheless,
\ppxf\ fits simultaneously the stellar components and nebular emission
(i.e., correcting the nebular emission for stellar absorption),
fitting for the line width (which is important here as our stacks
include individual spectra with different spectral resolution owing to
the morphological broadening).  And, \ppxf\ can separate the Balmer
emission from the metal emission features.  For our purposes, we added
to \ppxf\ the \neiii\ $\lambda$3868 emission line as this is prominent
in our data (Figure~\ref{fig:allstack}).  We then ran \ppxf\  on the
stacked data in Figure~\ref{fig:allstack} (using stacks where the
input spectra have been corrected for dust attenuation).   We ran
\ppxf\  in two modes: one where each Balmer emission line is fit
separately and one where we tie the Balmer emission lines to their
theoretical Case-B values \citep[e.g.,][]{oste89}, and we use the
latter for cases where H$\gamma$ is too weak to be visible in the
spectra.   We report the results in Table~\ref{table:ppxf}.  Because
the flux density in the stacked spectra have been normalized we report
emission line flux ratios and equivalent widths (EWs) of prominent
features.  We use these results in the Section~\ref{section:ZQR} to
interpret the bulk trends between emission-line ratios and galaxy
properties.  

\section{Gas-Phase Metallicity and Ionization from Nebular Emission
Lines}\label{section:gasproperties}

The WFC3 grism data cover emission lines in the rest-frame optical
(for galaxies at $1.1 < z < 2.3$) that are sensitive to nebular
ionization and gas-phase metallicity.   The lower spectral resolution
of the \hst/WFC3 G102 and G141 data ($R\sim 100-200$) cause the
\ion{O}{2}~$\lambda\lambda 3726,~3729$ and \ion{O}{3}~$\lambda\lambda
4959,~5008$ lines to be blended  (see Figure~\ref{fig:allstack}).
Rather than attempting to de-confuse these lines, we adopt line ratios
that make use of the sums of these lines.  Specifically we define
ratios of these lines as:
\begin{eqnarray}
  \mathrm{O}_{32}&\equiv&\frac{ \oiii~\lambda\lambda 4959,~5008 } {
  \oii~\lambda\lambda 3726,~3729} \\
  \mathrm{R}_{23}&\equiv&\frac{ \oiii~\lambda\lambda 4959,~5008 +
  \oii~\lambda\lambda 3726,~3729}{\hb \lambda
  4861},
\end{eqnarray}
where $\oiii~\lambda\lambda 4959,~5008$ and $\oii~\lambda\lambda
3726,~3729$ are the sum of the emission from both lines in the
doublet, where we have corrected the line fluxes for dust attenuation
using the $A(V)$ values from the SED fits with the \citet{calz00}
model.  The line ratio $\mathrm{O}_{32}$ is sensitive to the
ionization of the gas (as it measures the relative amount of emission
from double-ionized oxygen to singly ionized oxygen, see
\citealt{strom18,kewley19}).   The line ratio $\mathrm{R}_{23}$ is
sensitive to the gas-phase metallicity (specifically the oxygen
abundance, $12 + \log(\mathrm{O/H})$) as for the typical conditions in
\ion{H}{2} regions the majority of the gas-phase oxygen is in the
singly or doubly ionized states \citep[e.g.,][]{delg14,kewley19}.  In
Appendix~\ref{section:appendix}, we test how including  measurements
of $\ha + \nii$ (which is the sum of $\ha~\lambda 6564 +
\nii~\lambda\lambda 6548,~6583$) and $\sii$ (which is unresolved in
the grism data, therefore we use the sum of the lines in the
\sii$\lambda\lambda 6716,~6731$ doublet) impact our results, and we
find there is no substantive change to our conclusions.

\begin{figure*}[th]
\centering
\includegraphics[width=0.67\textwidth]{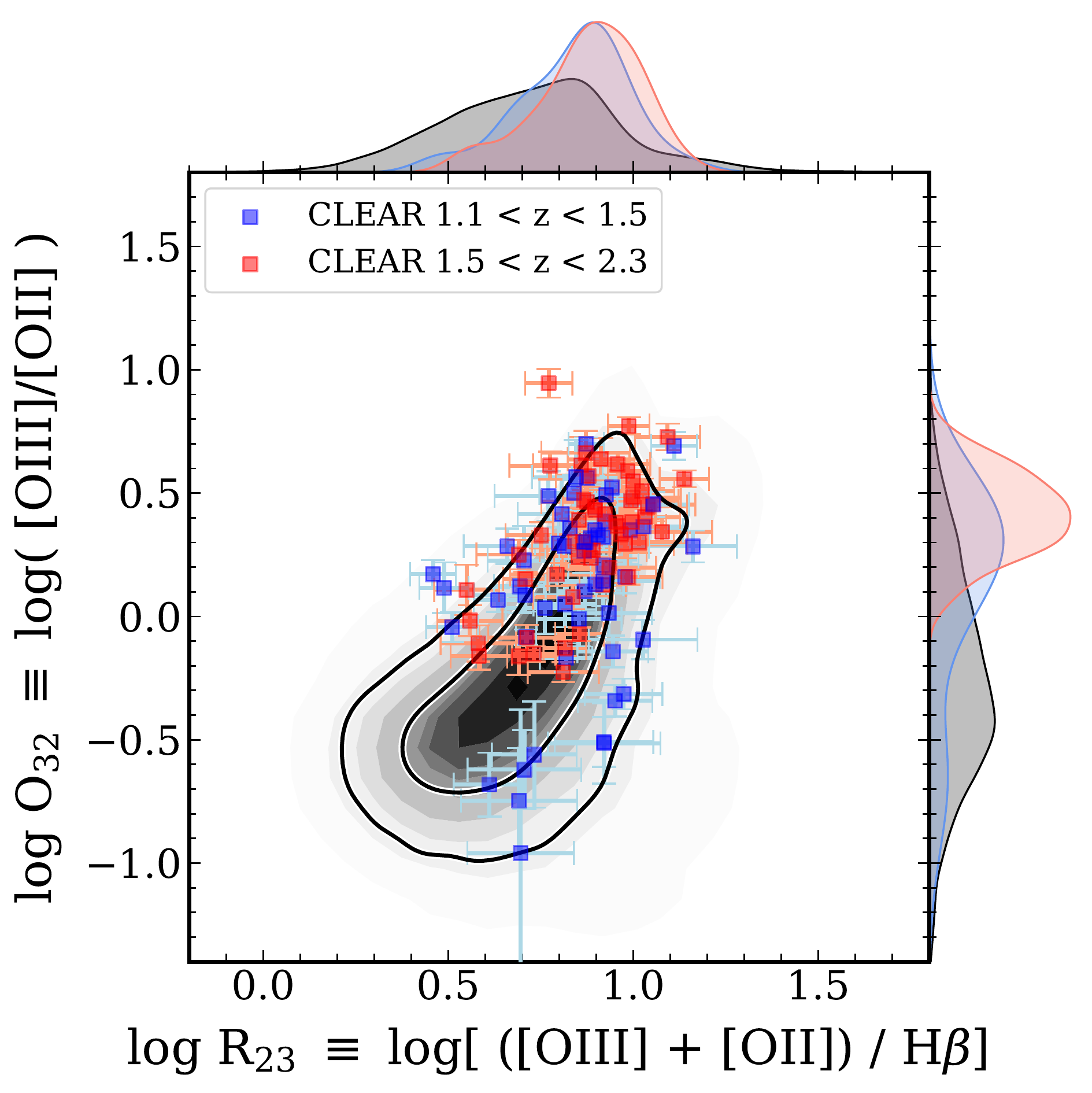}
\includegraphics[width=1\textwidth]{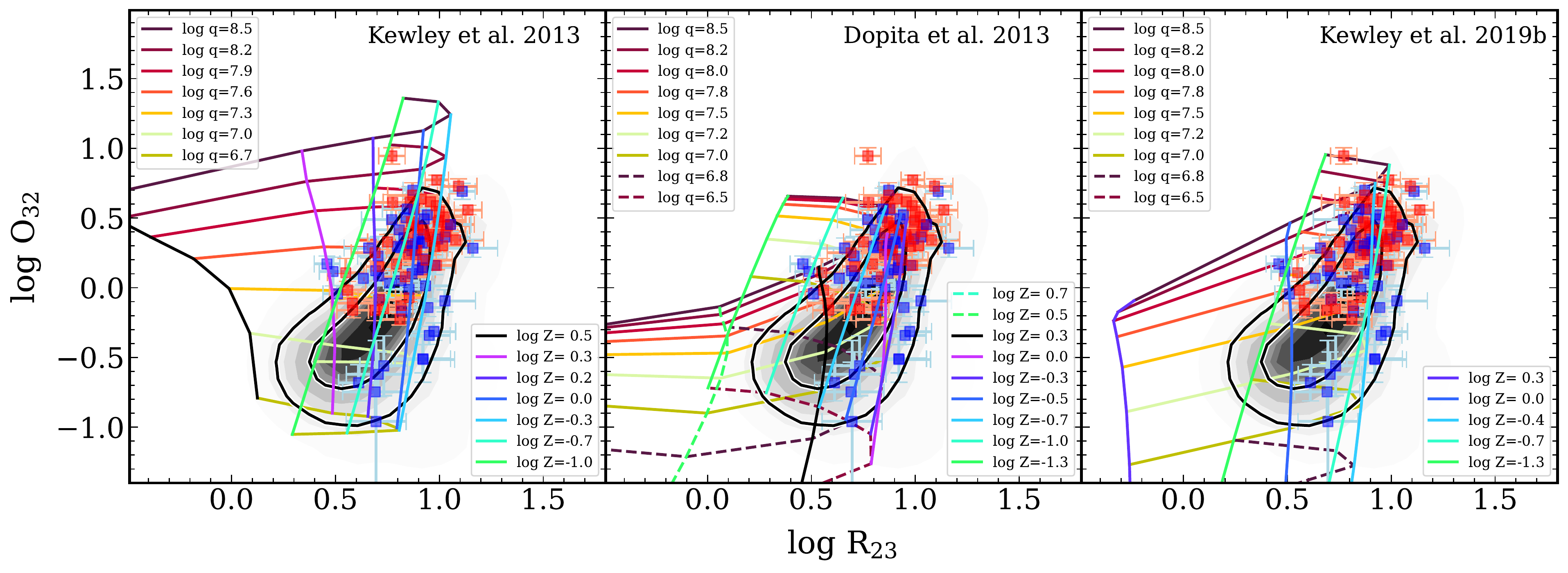}
\caption{Top panel: Distribution of O$_{32}$ as a function of $R_{23}$
for galaxies in the CLEAR sample (colored squares) compared to that
from SDSS (using a kernel density estimator [KDE], gray shaded region;
the contours contain 50\% and 80\% of the SDSS sample).  For clarity,
the plot only shows CLEAR galaxies with SNR $>$3 in \oii, \oiii, and
\hb, color-coded by redshift. The data have been corrected for dust
attenuation (see Section~\ref{section:gasproperties}).  The CLEAR
galaxies reside in the upper end of the SDSS distribution.  The panels
to the top and right show the one-dimensional distributions from a
KDE.  Bottom panels: the distribution of the CLEAR and SDSS R$_{23}$
versus O$_{32}$ distributions compared to predictions from
photoionization models considered here.  In this work we adopt the
MAPPINGS~V models \citep{kewley19} as these are the most current at
the time of this writing, and they best cover the line ratios spanned
by the observed galaxies.    \label{fig:o32_r23}}
\end{figure*}

\subsection{Comparison to Photoionization
Models}\label{section:o32_r23}

Figure~\ref{fig:o32_r23} compares the distribution of R$_{23}$ and
O$_{32}$ for galaxies in the CLEAR sample to those from SDSS DR14
\citep{thomas13}.   For both SDSS and CLEAR galaxies, the emission
lines have been corrected for dust extinction.  For SDSS, we used the
values provided by \citeauthor{thomas13}.  For our CLEAR
galaxies, we corrected the line ratios using dust attenuation
estimates from our SED fitting to the grism spectra and photometry
(see Section~\ref{section:sedfit}) as most galaxies in our sample do
not have measurements of multiple Balmer emission lines.  We
also assume the \citet{calz00} attenuation law and that the
attenuation in the nebular gas is the same as for the stellar
continuum \citep[c.f.][]{reddy15}.  

The CLEAR galaxies at $1.1 < z < 2.3$  lie at the upper end of the
R$_{23}$--O$_{32}$ distribution defined by SDSS (the latter is shown
using a kernel density estimator [KDE]).  This is similar to the
findings of other studies of high redshift galaxies
\citep[e.g.,][]{sand16a,strom18,runco21}, which interpret the data as
an increase in ionization parameter, harder ionizing spectrum,
decrease in metallicity, and possibly elevated nitrogen abundances and
higher $\alpha$/Fe abundance ratios.  The CLEAR galaxies support many
of these assertions and we discuss these further below (see,
Section~\ref{section:ZQR}).\footnote{\citet{garg22} recently argued
that high-redshift surveys may be missing lower-ionization galaxies
that fall in the lower-left portion of the R$_{23}$--O$_{32}$
parameter space.  We argue this is not the case for the majority of
the galaxy population (as our data detect galaxies over the majority
of the distribution of the SFR--mass relation,  see
Fig~\ref{fig:sfms}), unless there is a significant population of
galaxies on this relation that are undetected in emission lines. This
will be testable in future studies, e.g., from \textit{JWST}.}

Figure~\ref{fig:o32_r23} also compares the line ratios to
photoionization models.   The models include the \citet{dopita13}
models, which used the MAPPINGS IV code (bottom row, middle panel of
the Figure).   The \citeauthor{dopita13} model includes updated atomic
data, elemental abundance measurements, and modeling prescriptions.
The input ionizing spectrum uses  the STARBURST99 population synthesis
model \citep{leit14} for a stellar population with a constant SFR,
observed at an age of 4~Myr, with a Salpeter IMF with an upper-mass
cutoff of 120~\msol.  The nebular region also assumes spherical
geometry, isobaric photoionization, and that the distribution of
electron velocities allows for an extended tail to higher energies (a
so-called ``$\kappa$'' distribution).    These models span the range
of R$_{23}$--O$_{32}$ observed in the SDSS and CLEAR data, although
the models are unable to produce the highest ratios (e.g., the models
are limited to R$_{23} \lesssim 1$ while galaxies with R$_{23} > 1$
are evident in the SDSS and CLEAR data).

The \citet{kewley13} models in Figure~\ref{fig:o32_r23} (bottom row,
left panel) show the effects of using the PEGASE~2 \citep{fioc99}
population synthetic models that include harder ionizing spectra
(e.g., \citeauthor{kewley13} add the spectra of planetary nebular
nuclei for stars with high effective temperatures, $T_e > 50,000$~K to
estimate for the effects of the stellar photospheres of massive
stars).  These models increase the ratios of O$_{32}$ in response to
the increased ionization parameter.  Nevertheless, these models also
have difficulty achieving the highest R$_{23}$ values seen in the
data.\footnote{See, e.g., \citet{dargos19}, who consider a large range
of stellar-population parameters.  They show that only very young,
$\lesssim$2~Myr, stellar populations formed in bursts are capable of
producing the highest line ratios.}

Figure~\ref{fig:o32_r23} also shows line ratios for the MAPPINGS~V
models \citep{kewley19,kewley19b} (bottom row, right panel).  The
MAPPINGS~V models include updates with the latest atomic data and
relative abundances \citep[see,][]{nich17}.  The input spectrum is
based on the STARBURST99 stellar population synthesis models (as
above) using models for massive stars \citep{hill98,paul01} that are
able to produce better the ratios of blue/red supergiants in
low-metallicity regions, such as the Magellanic Clouds.

Importantly, the MAPPINGS~V models are isobaric, and consider the
effects of different values for the ISM pressure, here defined as $P/k
= n_e T$ (in units of K cm$^{-3}$, where $n_e$ and $T$ are the nebular
electron density and temperature).  In the present work, we adopt
$\log P/k = 6.5$ as this represents well the expected conditions in
high-redshift galaxies \citep[see, e.g.,][]{acha19}.  For example,
\citet{sand16a} and \citet{kaas17} find evidence for higher median
electron density, $n_e \simeq 250-300$~cm$^{-3}$ (where $n_e \approx
n_\mathrm{H}$ for ionized gas) for $\sim 1-3$ galaxies \citep[see
also,][]{runco21}.  Combined with the expected nebular temperature of
$\sim$10,000--20,000~K \citep[see,][and references
therein]{andrews13,sand20}, this implies a gas pressure of $\log P/k
\sim 6.5-7.0$.  Figure~\ref{fig:o32_r23} shows the MAPPINGS V models
assuming $\log P/k = 6.5$, but we observe similar results for $6 <
\log P/k < 7.5$.  Models with $\log P/k$=6.5 and 7.0 reproduce the
span of the data as illustrated in the Figure.  Models with $\log P/k
= 6.0$ do not reproduce the data with the highest O$_{32}$ ratios,
while models with higher pressure ($\log P/k$=7.5)  produce lower
R$_{23}$.  We therefore adopt $\log P/k = 6.5$ for our analysis while
noting that  changing this from 6.0--7.5 does not alter our
conclusions.

The MAPPINGS~V models still have difficulty producing the highest
R$_{23}$ values seen in the data (i.e., those with $\log R_{23}
\gtrsim 1$ in both SDSS and CLEAR, see Fig.~\ref{fig:o32_r23}).  This
effect has been seen in other studies.  To explain this offset could
require \editone{stellar populations with enhanced $\alpha$/Fe ratios \cite[e.g.,][]{sand16a,stei16a,strom22}}, or a change in nebular
geometry (e.g., ``density'' bounded nebula
[\citealt{brin08,nakajima14,kashino19}], or  clumpy geometries
[\citealt{jin22}]).  This highlights the need for improvements in
photoionization models to fully account the range of line emission
observed in high redshift galaxies.  We plan to investigate this in a
future study. 

\begin{figure*}
  \gridline{
    \fig{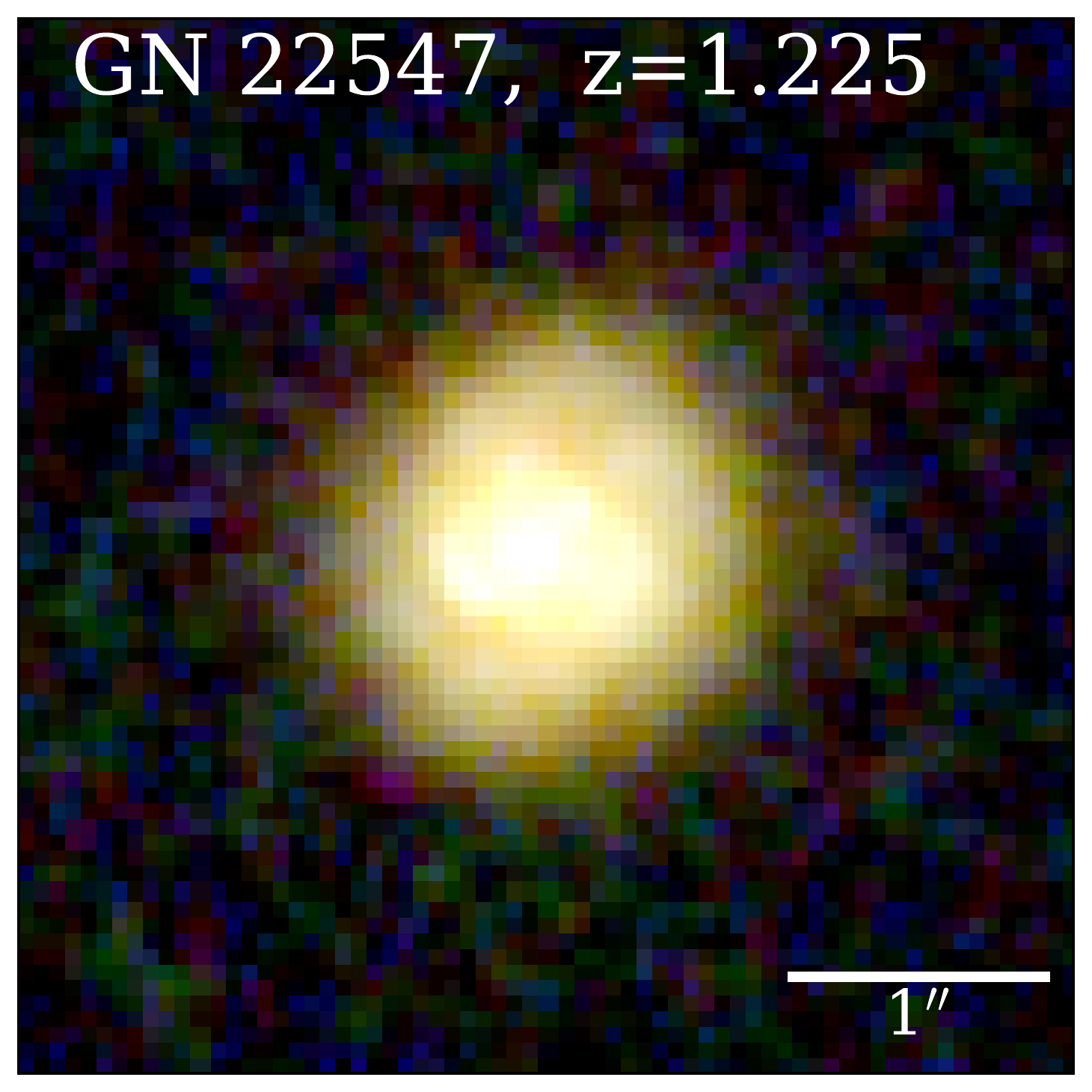}{0.20\textwidth}{}
    \fig{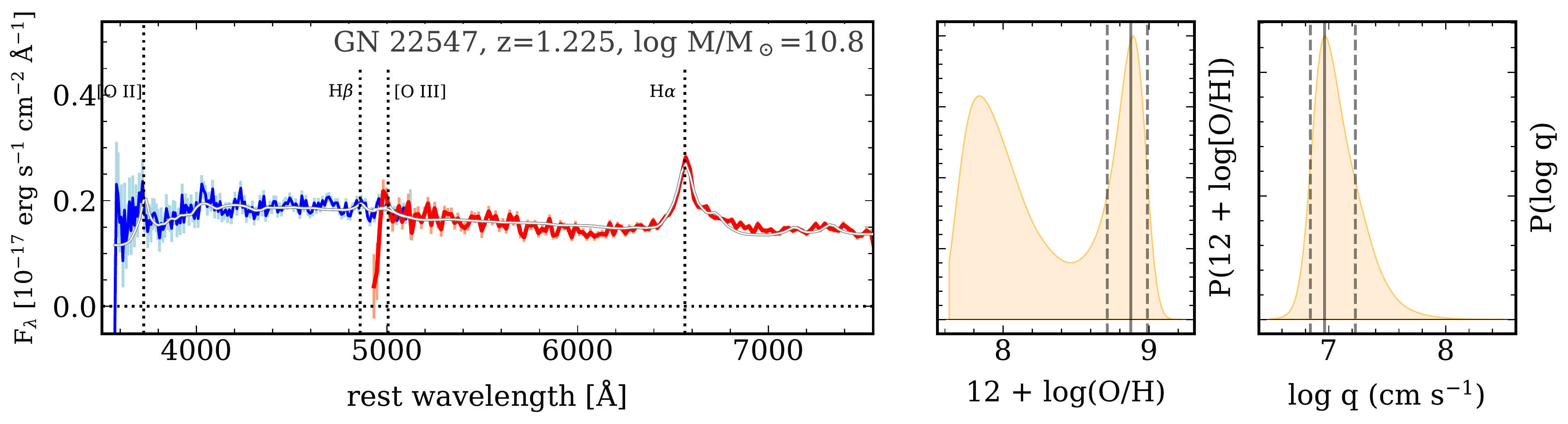}{0.75\textwidth}{}
  }
 \vspace{-18pt}
  \gridline{
    \fig{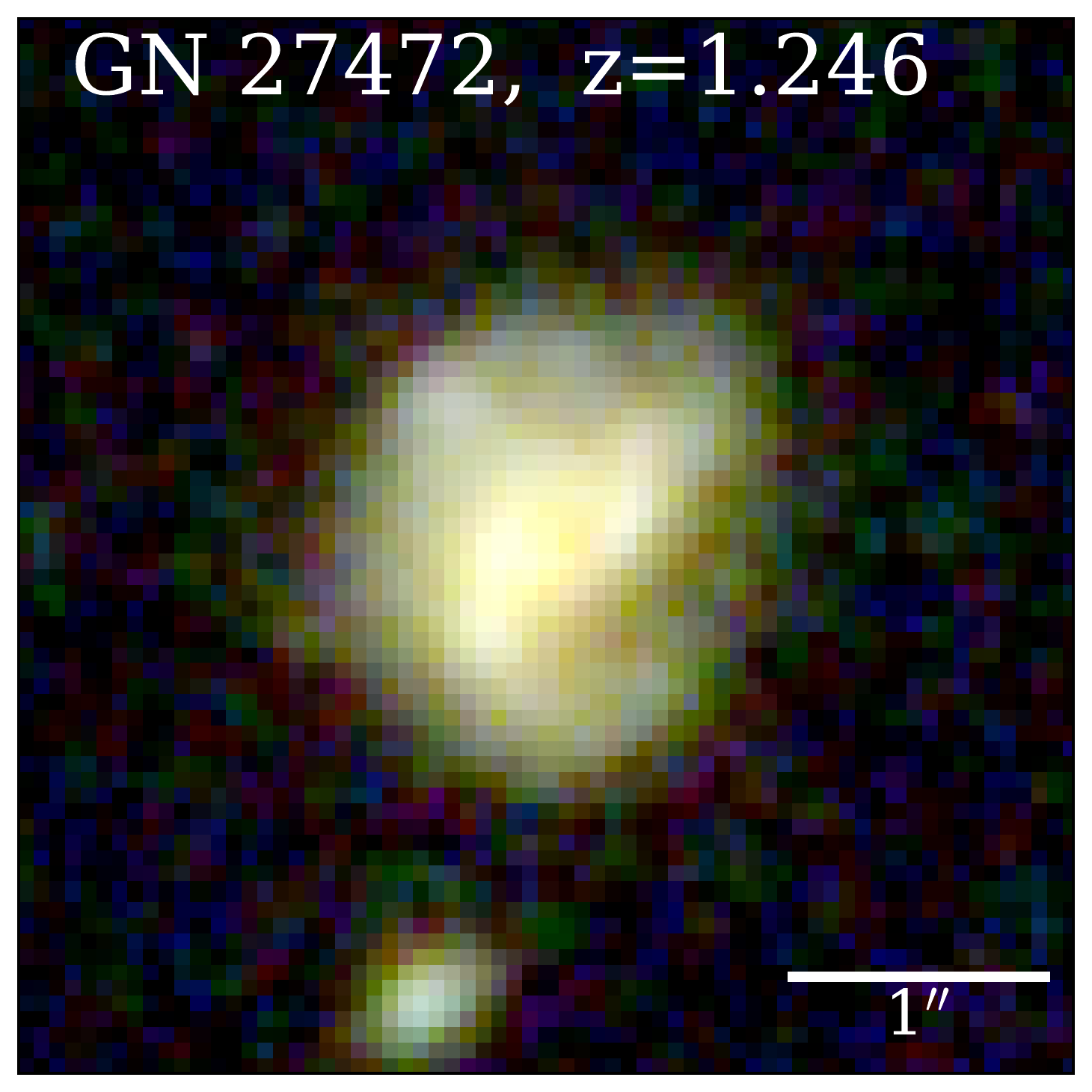}{0.20\textwidth}{}
    \fig{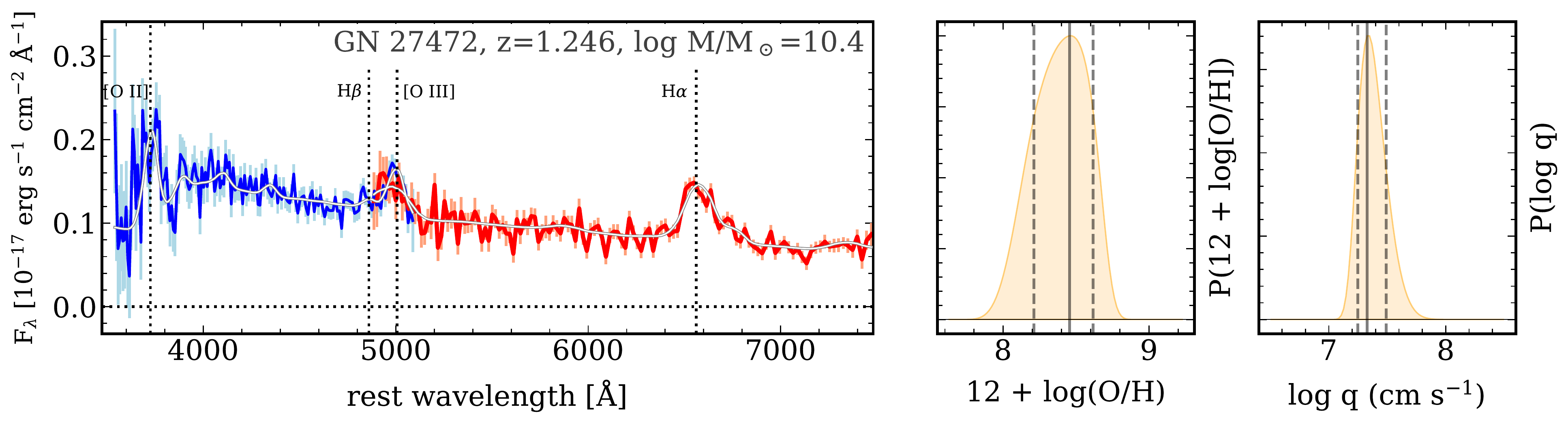}{0.75\textwidth}{}
  }
 \vspace{-18pt}
  \gridline{
    \fig{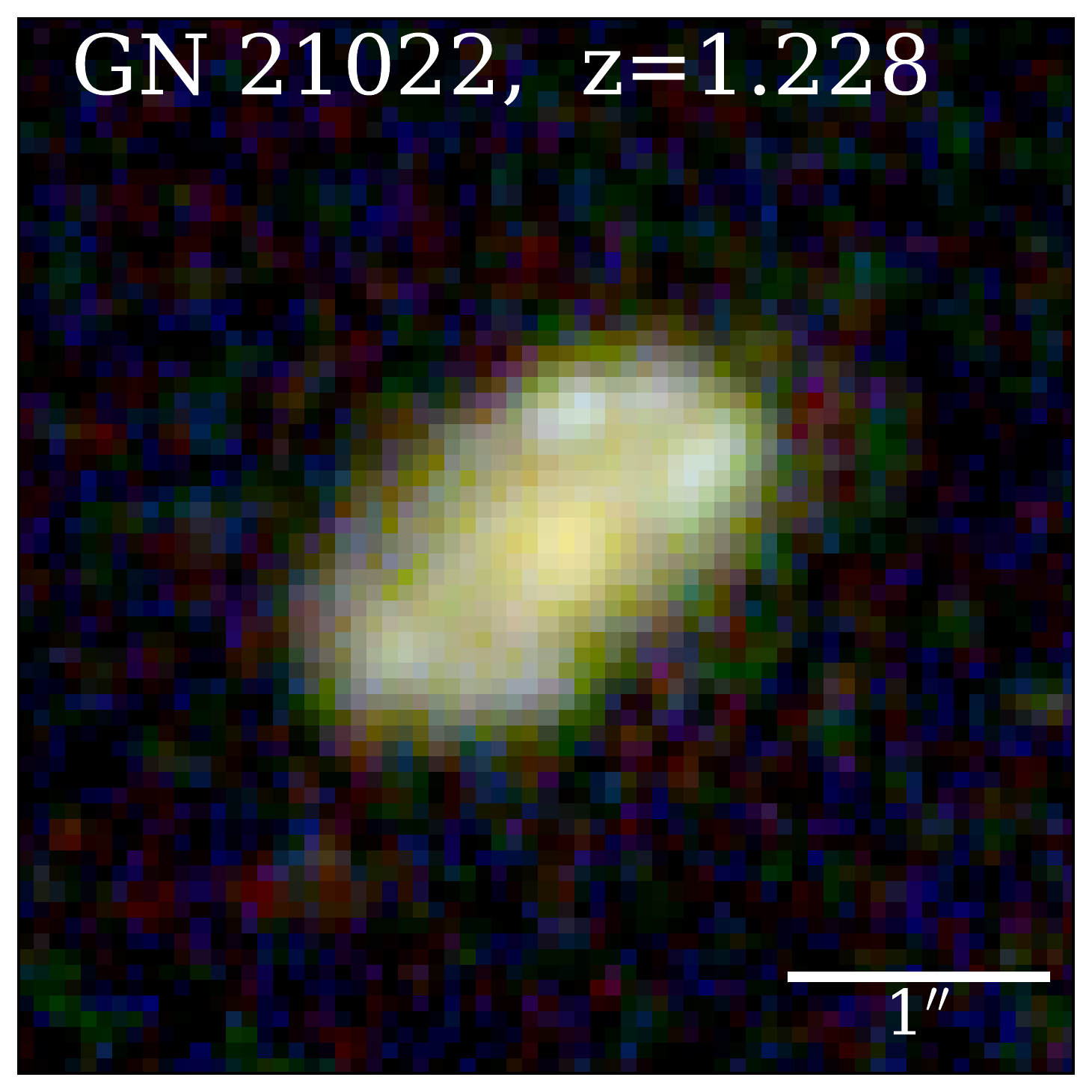}{0.20\textwidth}{}
    \fig{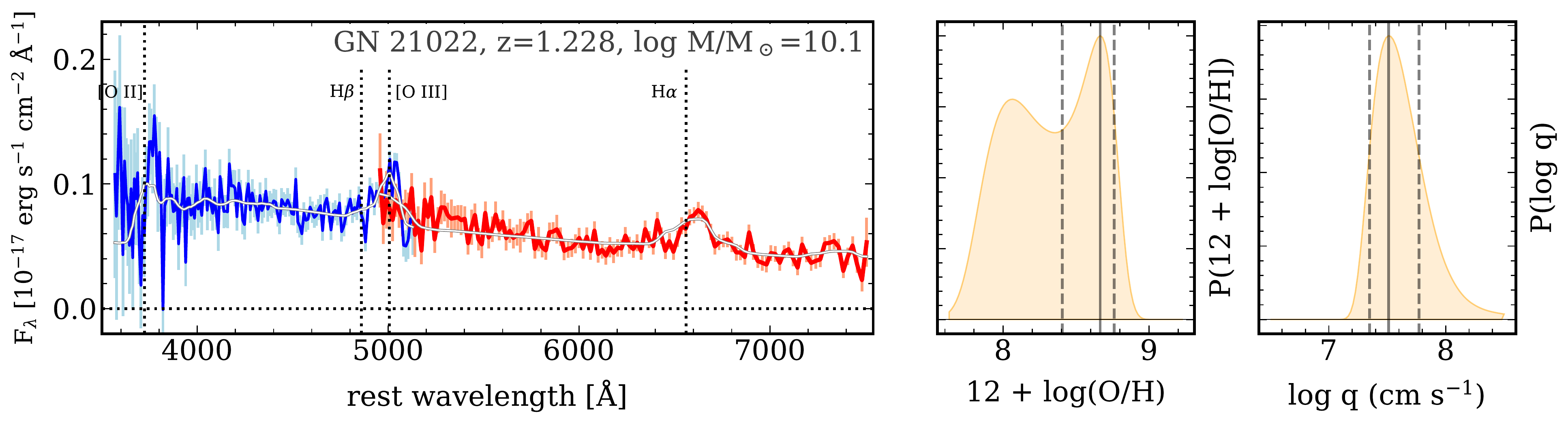}{0.75\textwidth}{}
  }
 \vspace{-18pt}
  \gridline{
    \fig{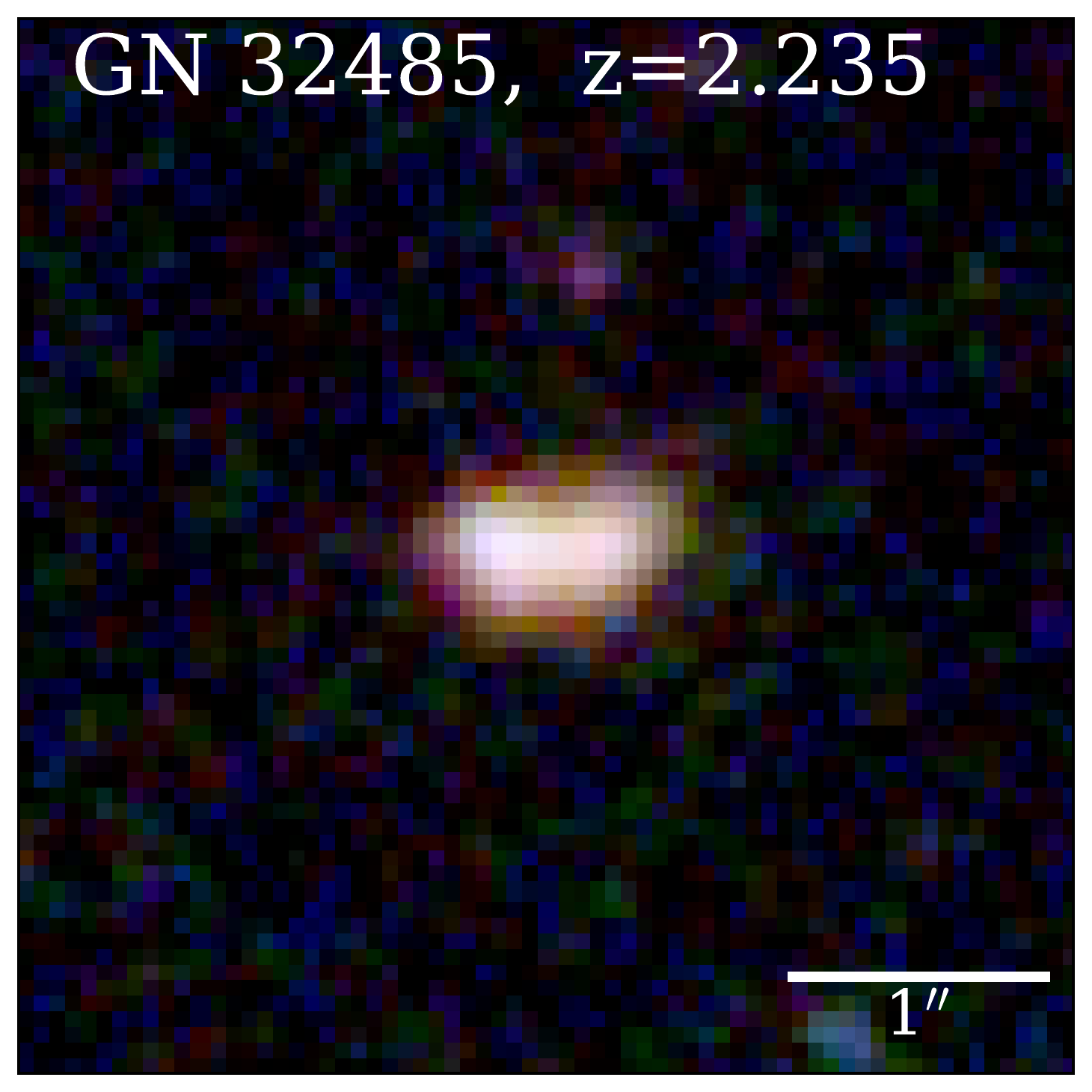}{0.20\textwidth}{}
    \fig{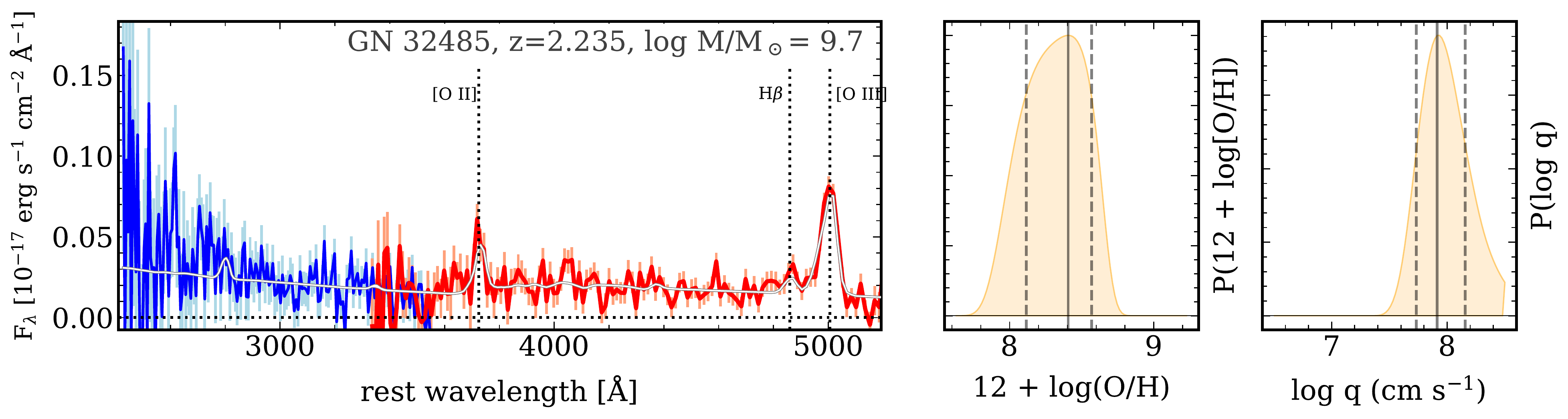}{0.75\textwidth}{}
  }
 \vspace{-18pt}
  \gridline{
    \fig{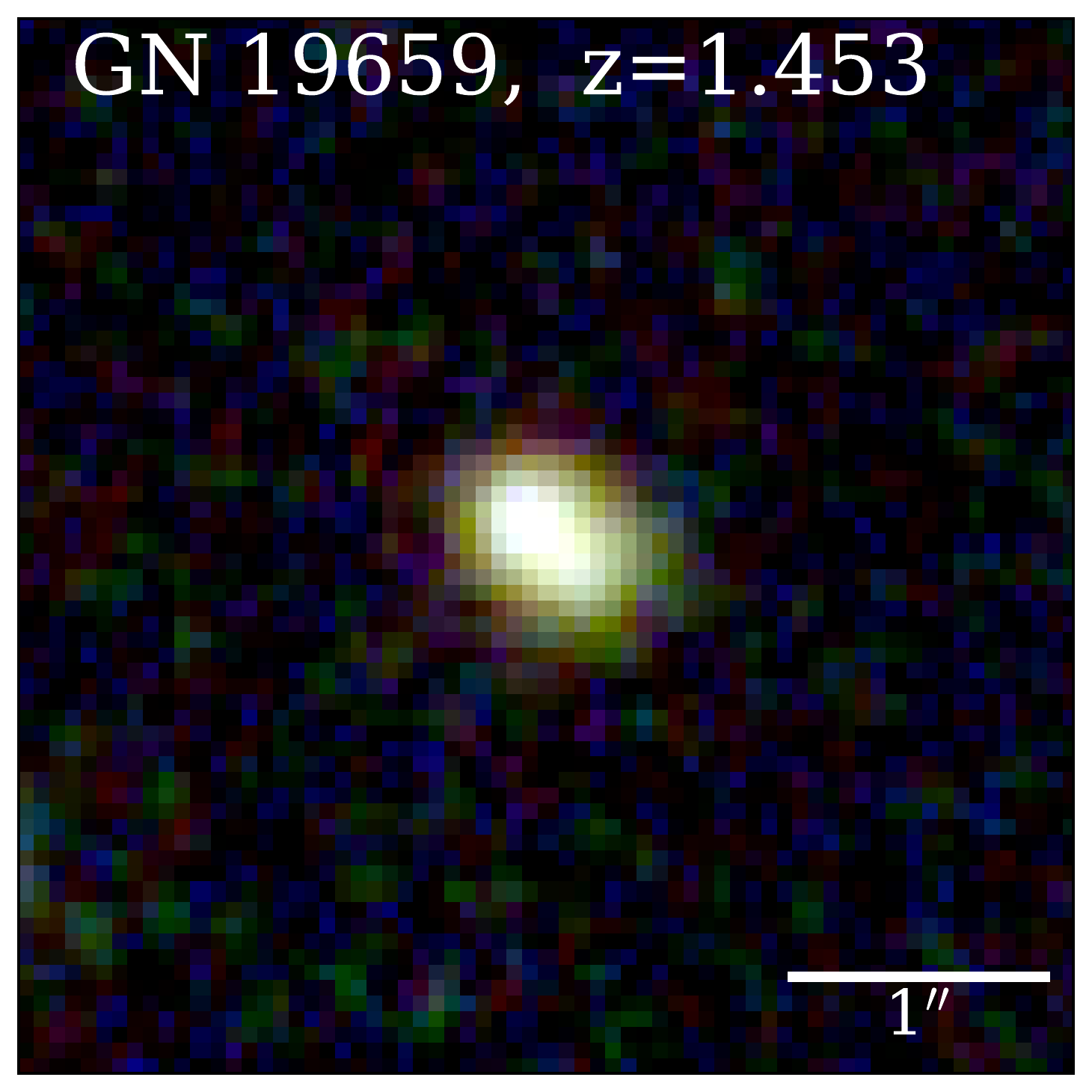}{0.20\textwidth}{}
    \fig{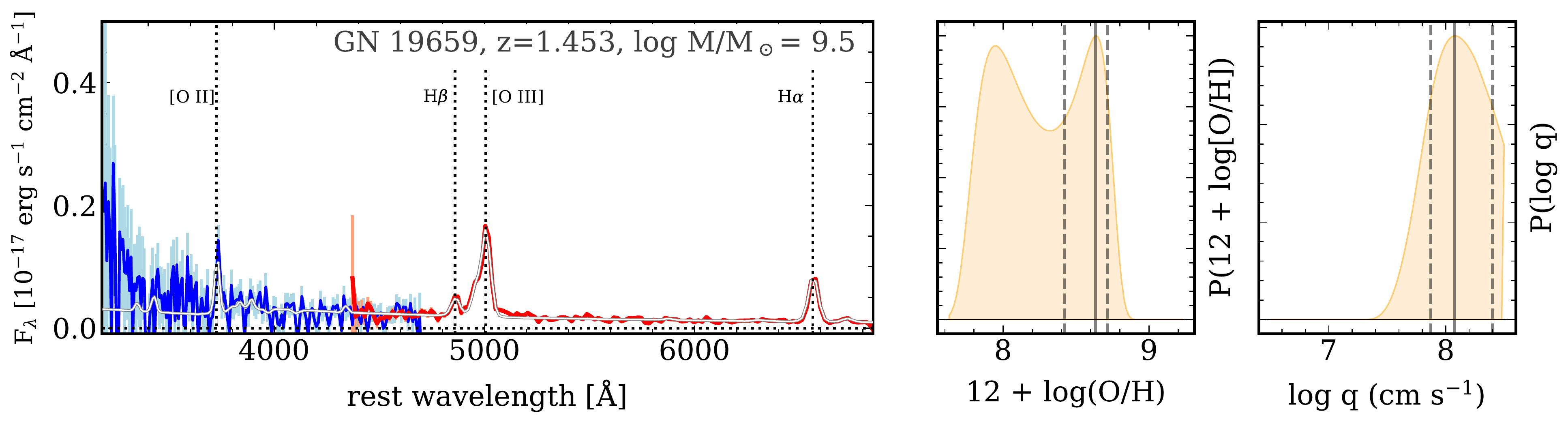}{0.75\textwidth}{}
  }
  \caption{Gallery of \hst\ RGB images (using ACS F775W, WFC3 F105W
and F160W imaging), grism spectra, and derived constraints on the
gas-phase metallicity ($12 + \log$ O/H) and ionization ($\log q$).
The figure includes galaxies in the CLEAR GOODS-N pointings with
redshift $1.2 < z < 1.5$ such that the spectra contain \oii, \hb,
\oiii, and \ha+\nii.   Each row shows one galaxy.  The left-most panel
shows the RGB image.  The scale bar corresponds to one arcsecond.  The
middle panel shows the 1D, G102 (blue) and G141 (red), extracted
spectra, prominent emission features are indicated. The right-most two
panels adjacent to each spectrum show the posterior on the gas-phase
metallicity (\OH) and ionization parameter ($\log q$) derived by
comparing the \oii, \hb, and \oiii\ line fluxes against the MAPPINGS V
photoionization models.  The solid and dashed vertical lines show the
mode and 68\% range derived for the highest density interval
\citep[HDI][]{bailer18}.  The galaxies are sorted as a function of decreasing
stellar mass (top to bottom). \label{fig:spectra_izi_results}}
  \end{figure*}

 \begin{figure*}
  \gridline{
    \fig{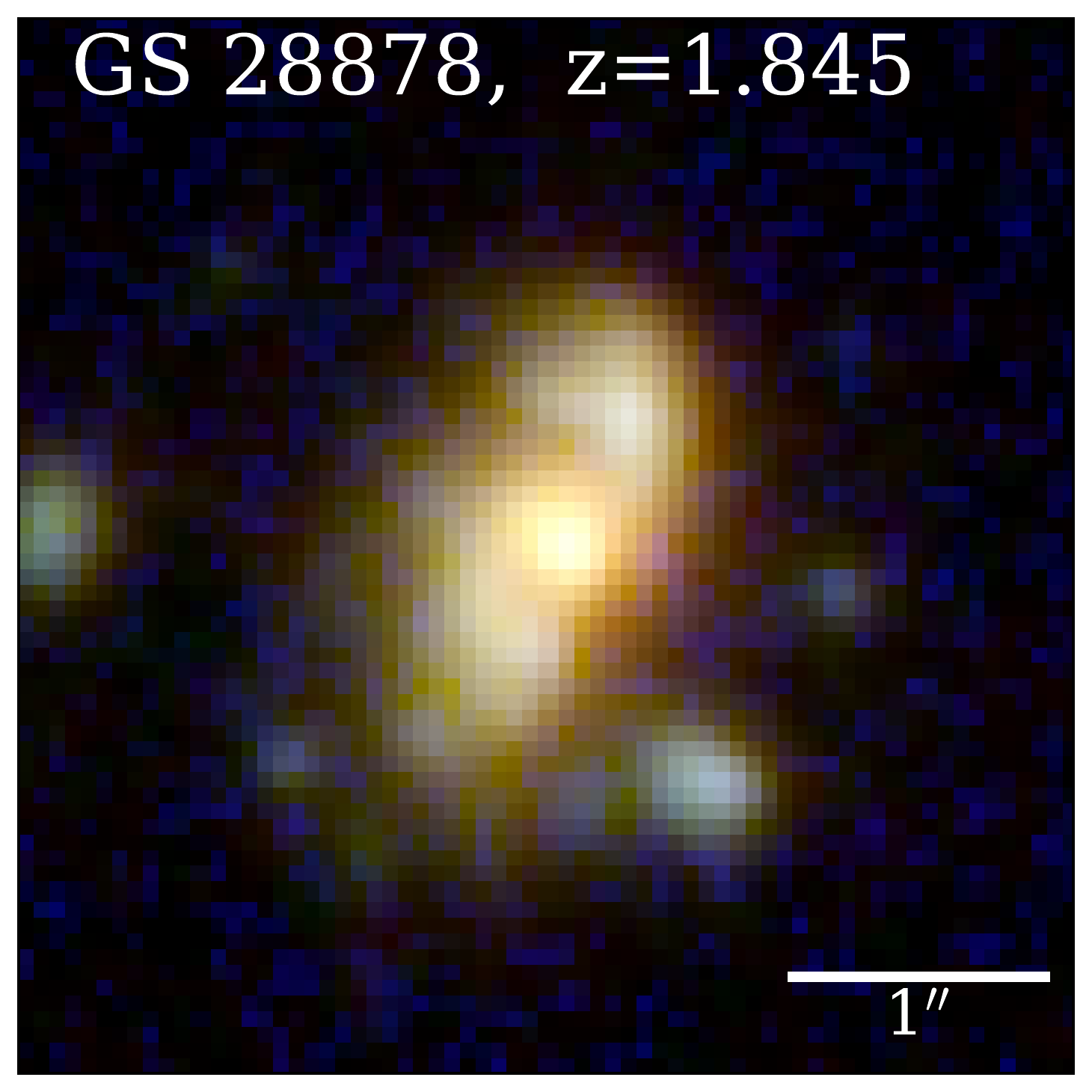}{0.20\textwidth}{}
    \fig{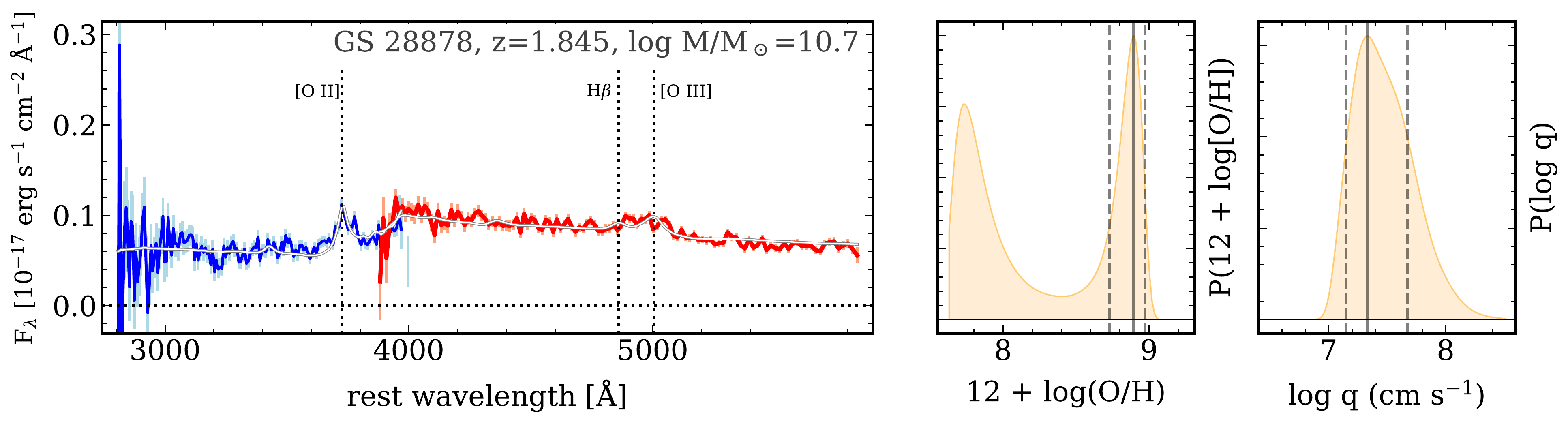}{0.75\textwidth}{}
  }
 \vspace{-18pt}
  \gridline{
    \fig{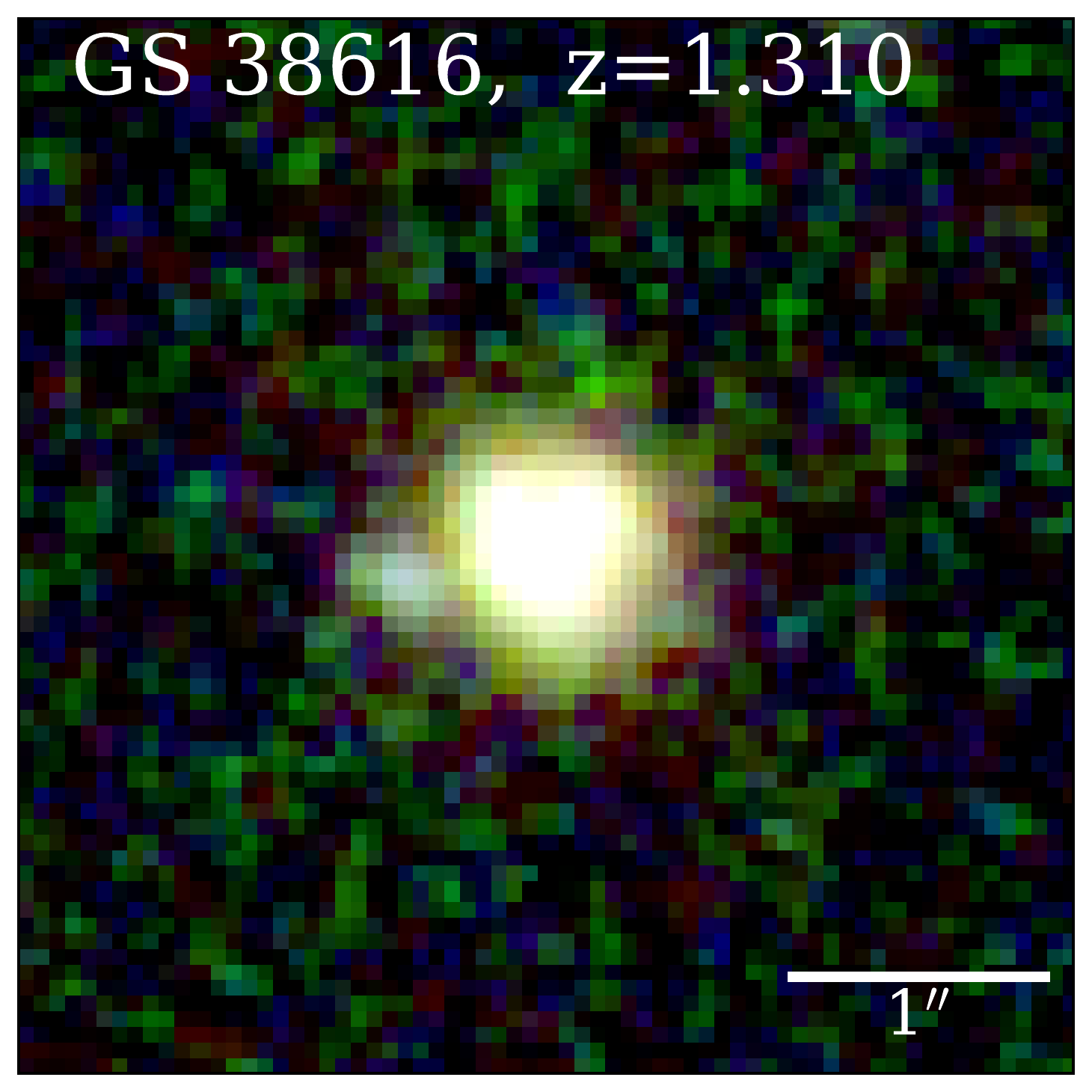}{0.20\textwidth}{}
    \fig{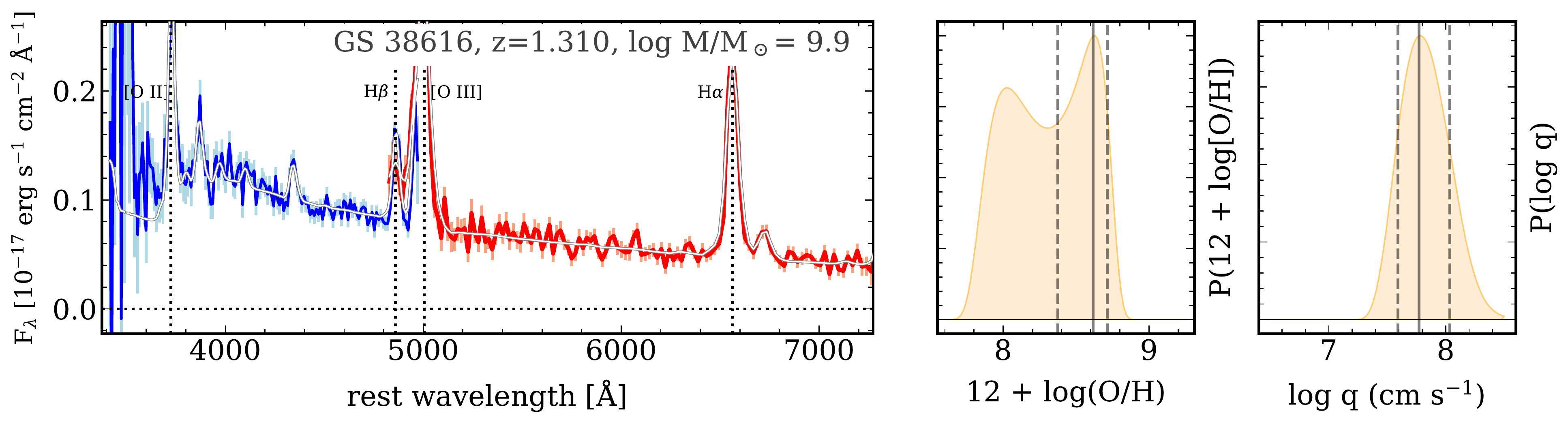}{0.75\textwidth}{}
    }
 \vspace{-18pt}
  \gridline{
    \fig{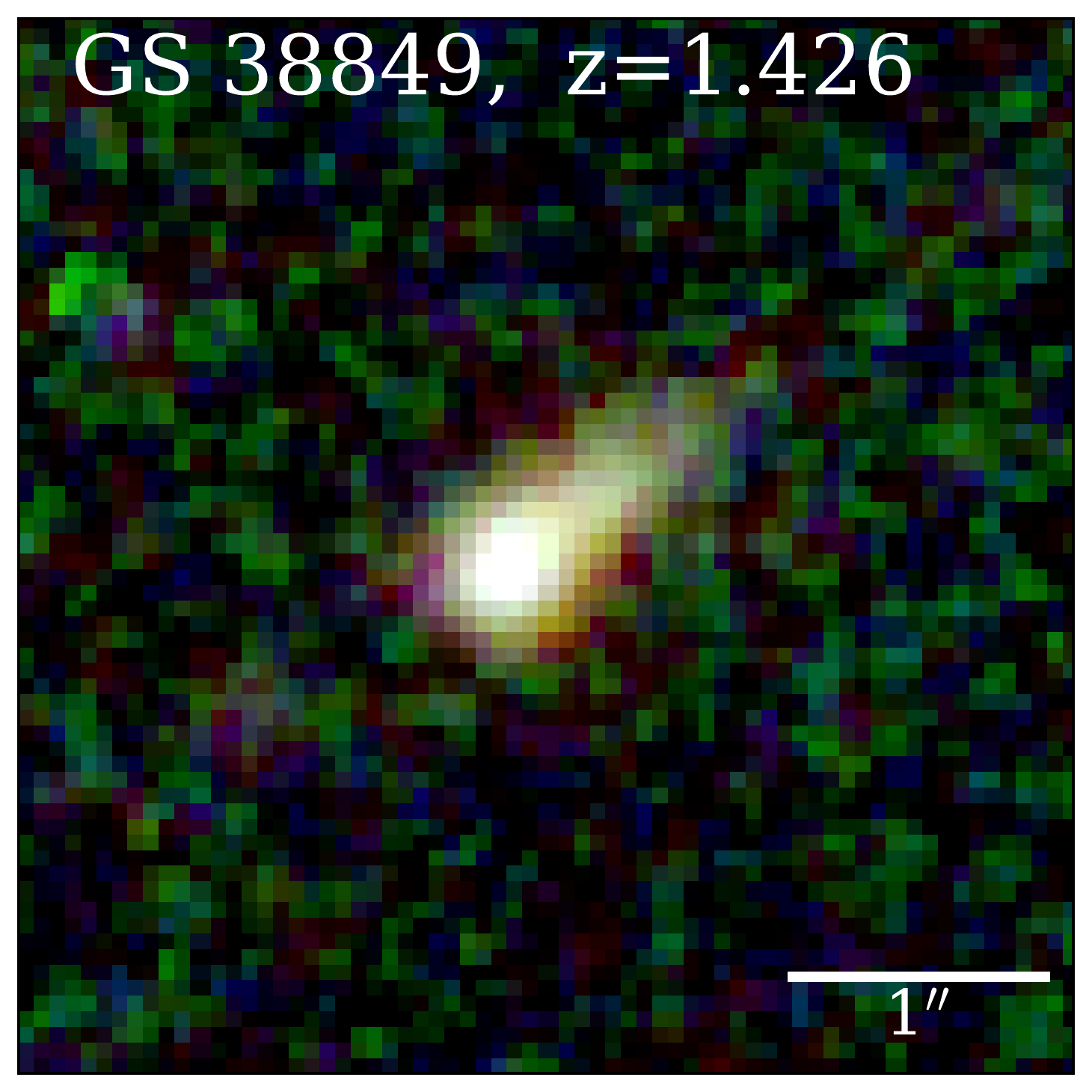}{0.20\textwidth}{}
    \fig{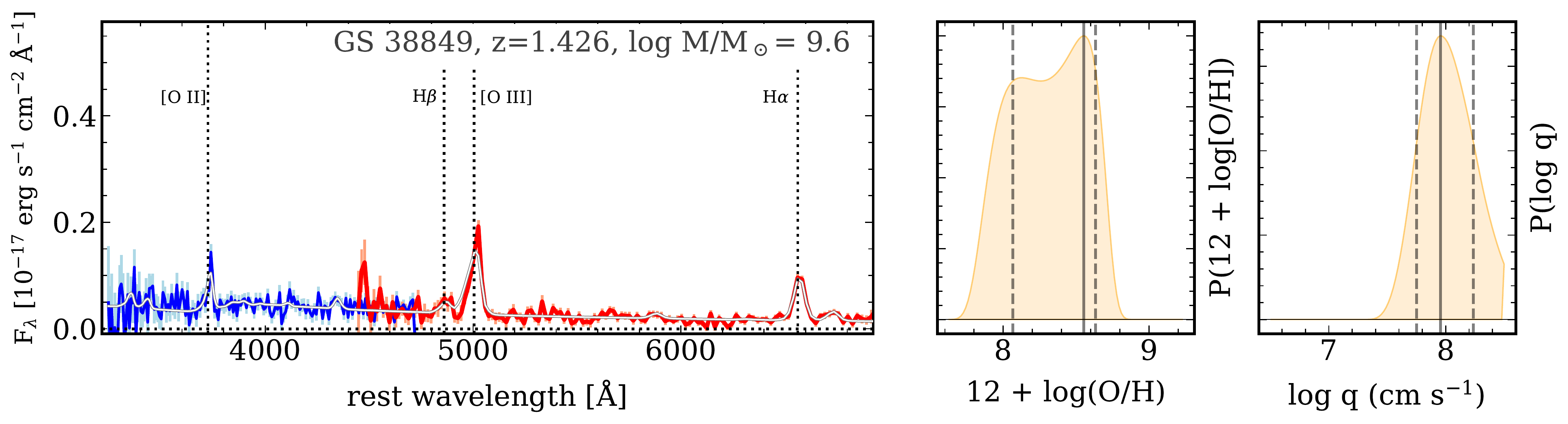}{0.75\textwidth}{}
    }
 \vspace{-18pt}
  \gridline{
    \fig{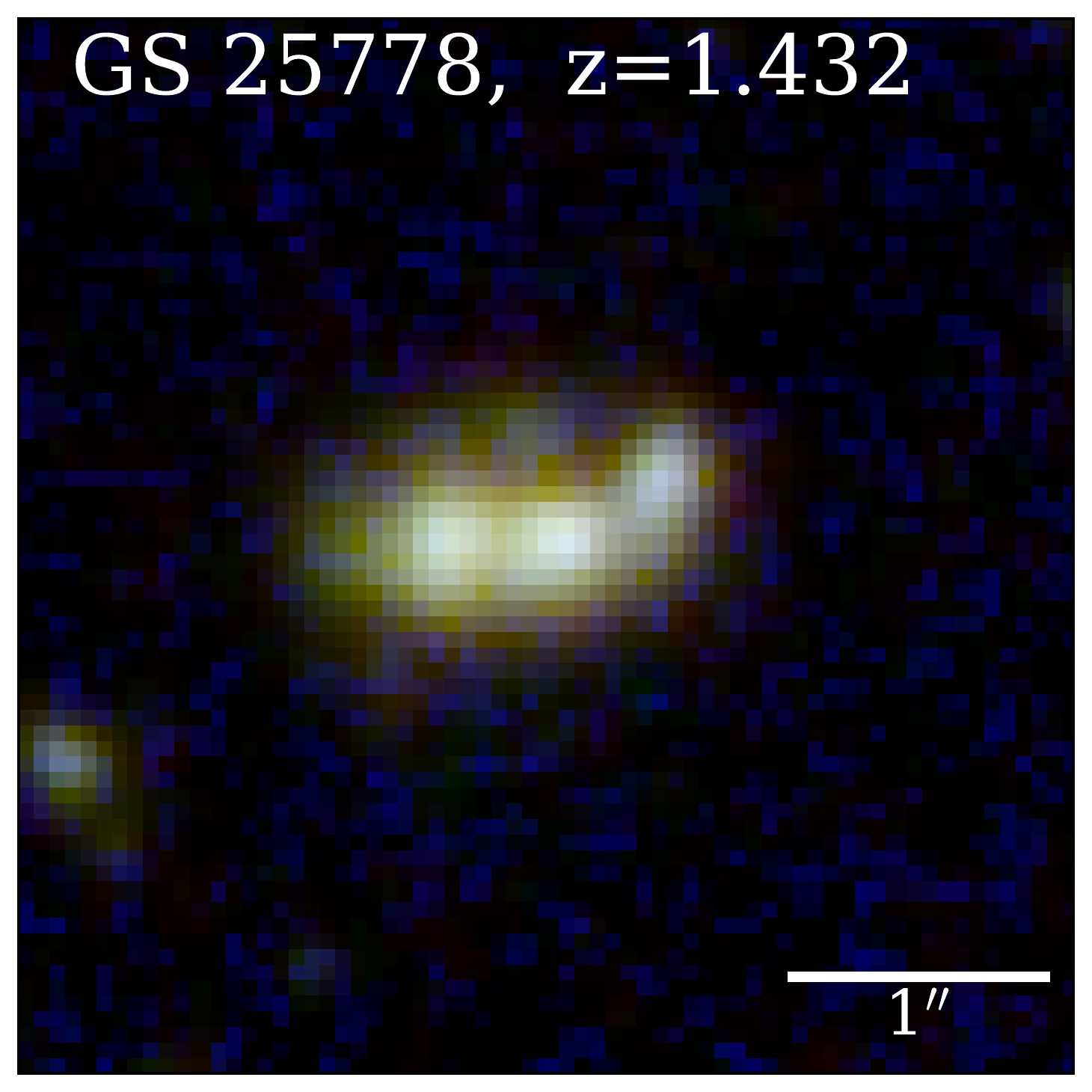}{0.20\textwidth}{}
    \fig{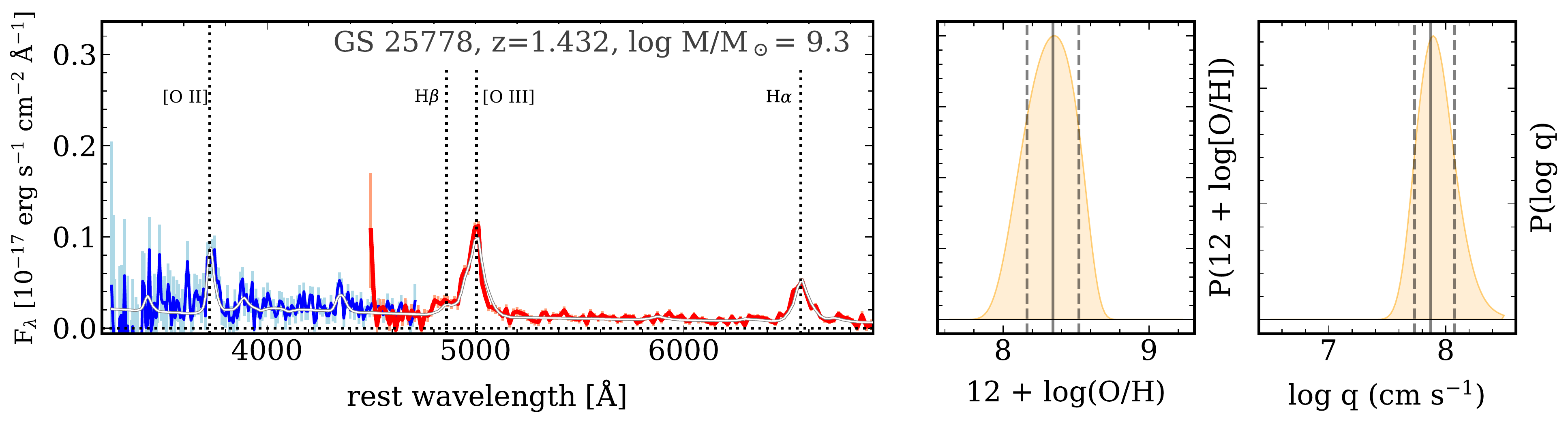}{0.75\textwidth}{}
  }
 \vspace{-18pt}
  \gridline{
    \fig{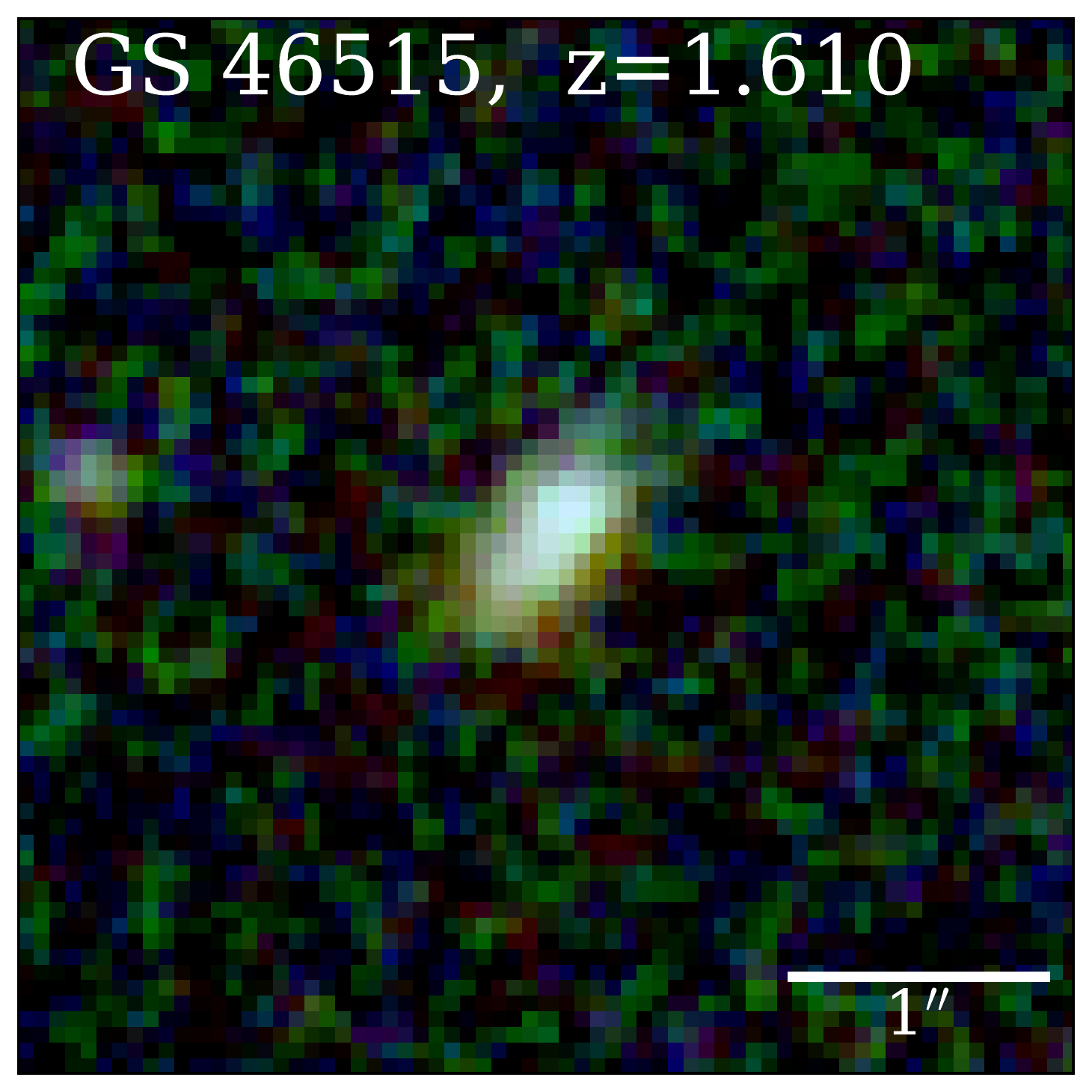}{0.20\textwidth}{}
    \fig{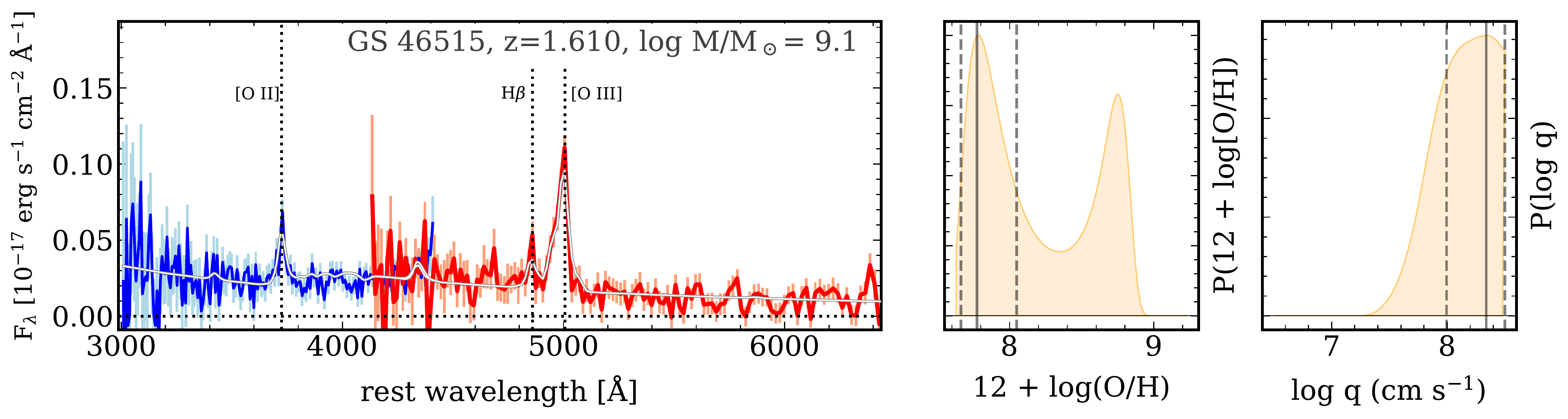}{0.75\textwidth}{}
    }
\caption{Same as Figure~\ref{fig:spectra_izi_results}, but showing
example galaxies with spectroscopic redshifts $1.3 < z < 1.9$ in the
CLEAR GOODS-S pointings.  The galaxies are sorted as a function of
decreasing stellar mass (top to
bottom). \label{fig:spectra_izi_results_gs}}
\end{figure*}

\begin{figure*}[t]
\centering
\includegraphics[height=0.45\textwidth]{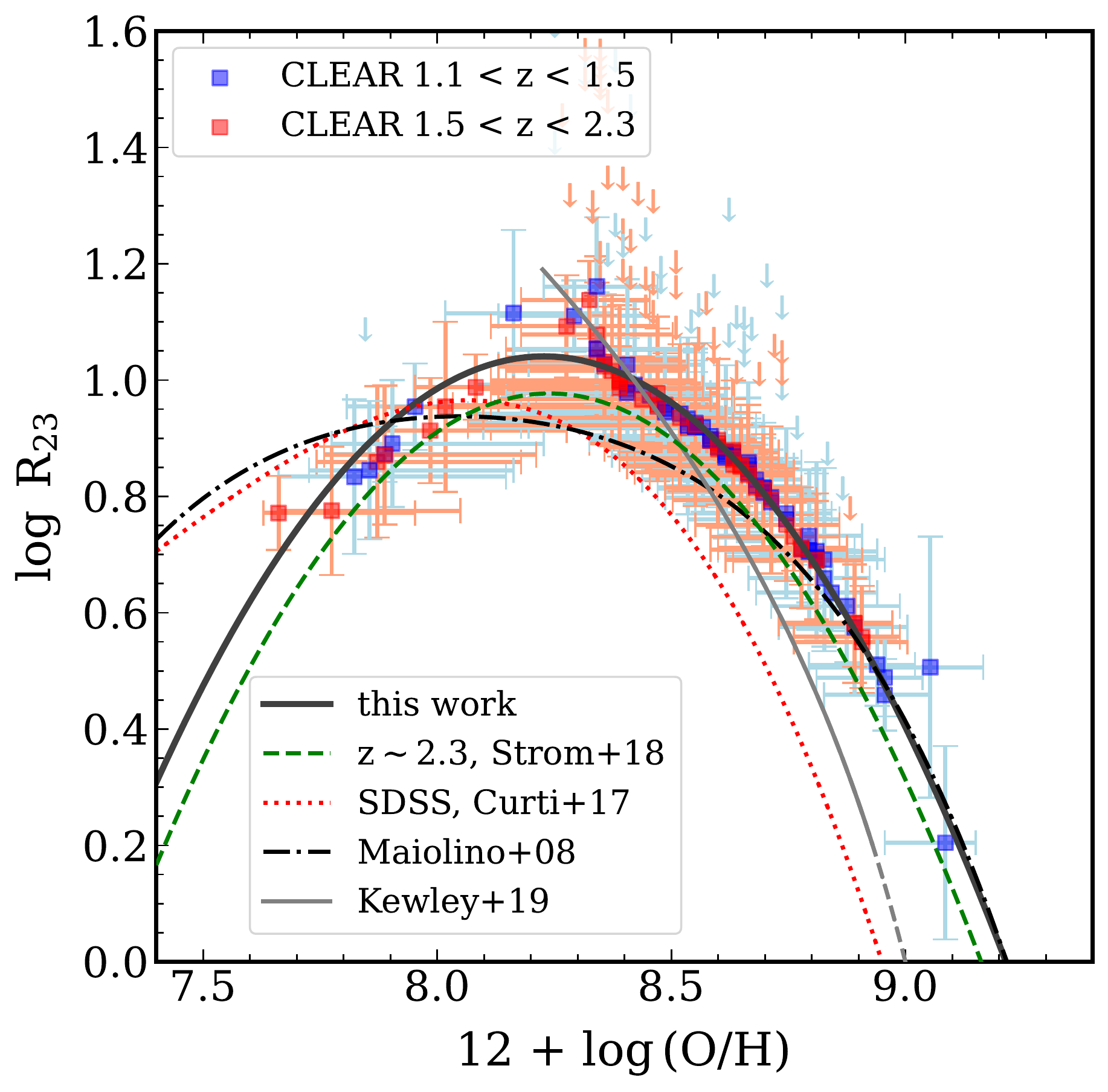}
\includegraphics[height=0.45\textwidth]{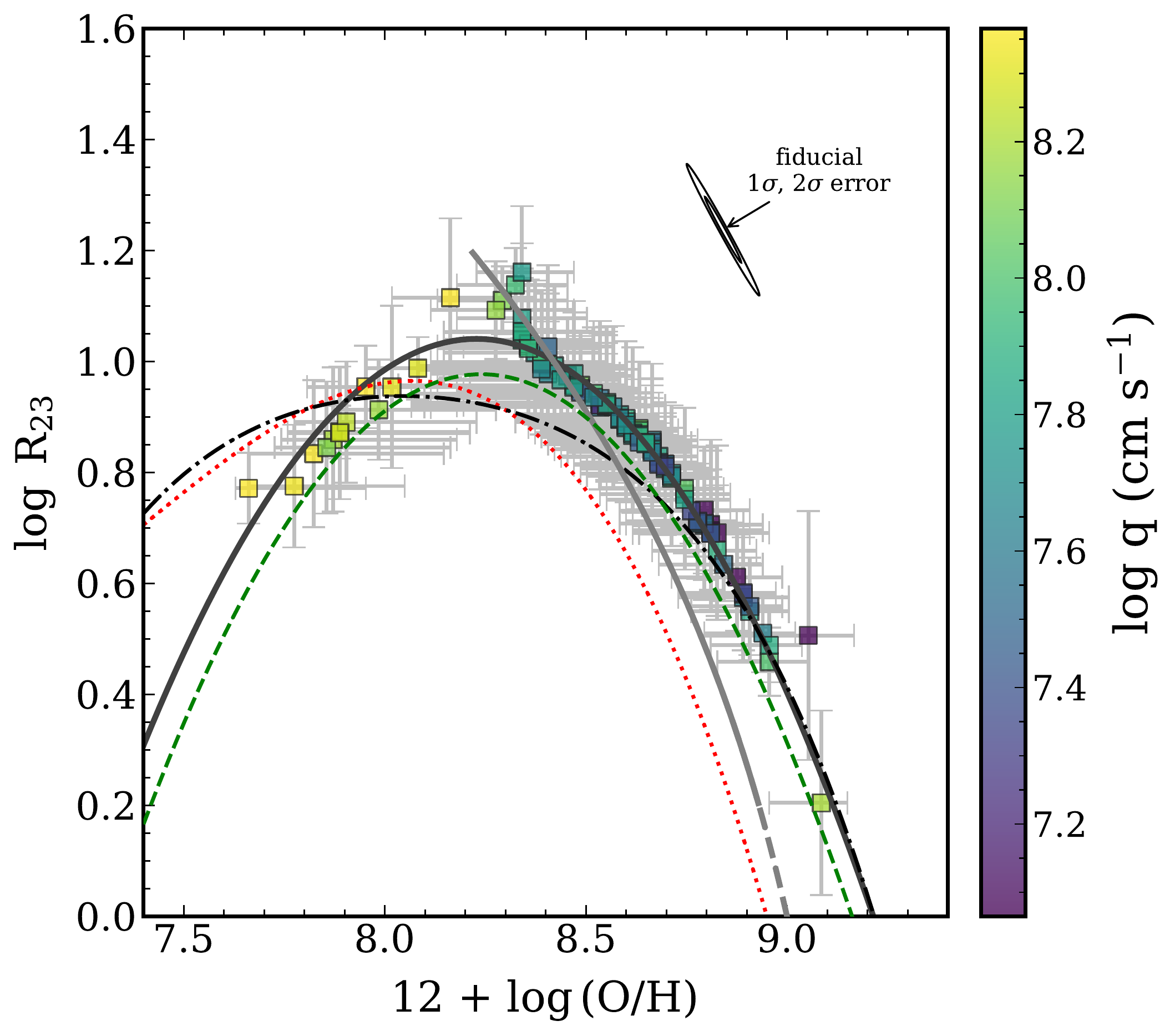}
\caption{Metallicity versus $R_{23}$ for the CLEAR sample derived by
comparing the observed  \oii, \hb, and \oiii\ line fluxes to
MAPPINGS V photoionization models.   The data points show the mode of
the posterior likelihoods on metallicity.  The error bars
denote the inner-68 percentiles derived from the highest density
intervals; arrows show $1\sigma$  limits.  The curves denote various
calibrations in the literature \citep{maio08, curti17,
strom18, kewley19}; the solid-thick line is our fit to the CLEAR
results (see Eq.~\ref{eqn:r23_Z}). The left panel shows the CLEAR
galaxies colored by redshift.    The right panel shows the CLEAR
galaxies colored by ionization, $\log q$.  While there is a trend
between $R_{23}$ and metallicity, there is an additional dependence on
ionization $q$.  The \OH\ values are derived using the line fluxes in
R$_{23}$, which causes the apparently ``tight'' scatter data points
relative to their error bars.  To illustrate this, we show example
error ellipses for a fiducial galaxy with 12 + log(O/H) = 8.8.  These
illustrate the covariance between R$_{23}$ and \OH\ lies in the
direction of the relation. 
\label{fig:r23_Z}}
\end{figure*}

\subsection{Estimating the Gas-Phase Metallicity and
Ionization Parameter}\label{section:izi}

We use the code, ``Inferring the gas phase metallicity ($Z$) and
Ionization parameter'' (\izi) developed by \citet{blanc15} to estimate
the metallicity ($Z$, which we take to be the nebular oxygen
abundance, 12 + $\log$(O/H)) and ionization parameter
($q$).\footnote{We use $q$ as the ionization parameter, where $q$ is
related to $U$ (the \textit{dimensionless} ionization parameter)
through $q/c \equiv U$, where $c$ is the speed of light. $U$ is
normally defined as the ratio of the number density of ionizing
photons ($n_\gamma$, those with $E_\gamma > 13.6$~eV) to the number
density of Hydrogen atoms, $n_{H}$, $U\equiv n_\gamma / n_\mathrm{H}$.
As discussed in \citet{kewley19}, $q \equiv \Phi / n_\mathrm{H}$,
where $\Phi$ is the ionizing photon flux in units of photons per
cm$^{2}$ per s.  Physically,  $q$ has units of cm s$^{-1}$ and can be
therefore considered as the speed at which the ionization front moves
into the surrounding neutral medium. }   \izi\ is a Bayesian code
which computes posterior likelihoods for $Z$ and $q$ by comparing the
measured emission line fluxes for our galaxies to predictions from the
photoionization models.  \izi\ is flexible in the sense that it can
use any combinations of lines, including the summed fluxes of multiple
lines (which is useful in our case where emission lines  are blended
at the resolution of the WFC3 grisms).

We use MAPPINGS V models \citep{kewley19} assuming a isobaric pressure
in the nebula of $\log P/k = 6.5$ (K cm$^{-3}$ for the reasons in
Section~\ref{section:o32_r23}. We adapted the output of the MAPPINGS V
models into the format required for \izi.  We assume a ``flat'' prior
for both $\log Z$ and $\log q$, spanning the ranges $-1.3 \le \log
Z/Z_\odot \le 0.3$ and $6.5 < \log q / (\mathrm{cm~s^{-1}}) < 8.5$.
We tested alternative pressure values from $\log P/k = 6 - 7.5$ and
find no substantial differences in our conclusions.   We also tested
the use of different priors (including a prior with the shape of the
local- and high-redshift mass--metallicity relation
[\citealt{maio08,andrews13,henry21,sand21}]).  These priors only
change slightly the shape of the posteriors of lower-mass galaxies
(shifting the inter-68 percentile range to higher values of
metallicity by $<$0.1~dex) while leaving the median values nearly
unchanged (at fixed stellar mass).   Moreover, adopting a prior
affects the MZR, changing the average
metallicity of galaxies at $\log M_\ast/M_\odot$=9.5 by $<$0.01 dex
(Section~\ref{section:MZR}).   We therefore adopt the results from the
``flat'' prior to avoid any bias inflicted by the prior information,
but this has a minimal impact on our conclusions. 

We focus on results that use the sum of the emission from
\oii$\lambda\lambda3726$, 3729 doublet, the sum of the emission from
\oiii$\lambda\lambda 4959$, 5008, and the emission from \hb\ as these
lines are available over our full redshift range of our sample, and
they constrain both the ionization state of the gas and trace the
majority of nebular oxygen.  In all cases we correct the line emission
for dust attenuation as in Section~\ref{section:o32_r23}.    In
Appendix~\ref{section:appendix}, we compare these results to the case
where we also include the emission from \ha+\nii$\lambda\lambda$6548,
6583 and \sii$\lambda\lambda$6716, 6731 with the emission from \oii,
\oiii, and \hb\ for galaxies at $1.1 < z \lsim 1.6$ where all these
lines are also covered by the WFC3 grisms.  Adding these lines does
not change our interpretation, though it does provide additional confidence in the
results derived from only the \oii, \oiii, and \hb\ lines. 

We show examples of the results in
Figures~\ref{fig:spectra_izi_results} and
\ref{fig:spectra_izi_results_gs} for galaxies in GOODS-N and GOODS-S,
respectively.  For each galaxy in the figure we show the image
(ACS F775W, WFC3 F105W and F160W) along with the 1D spectrum from the
G102 + G141 grisms.   The right-most panels show the posteriors on
gas-phase metallicity (12 + log O/H) and ionization $\log q$ (in units
of cm s$^{-1}$) derived from \izi\ for each galaxy.   Because the images show the same
observed bands, color differences indicate the presence of the strong
emission lines, spectral breaks, or spatially variant dust effects as
they redshift through the filters.  For example, for $z \gsim 1.6$ the
4000~\AA/Balmer break shifts redward of the F105W band.   This
accounts for the redder appearance of some galaxies (such as GS
28878).  In other cases, the \oiii\ emission line shifts into the
F160W filter for $z \gsim 1.8$, which accounts for the redder
appearance of other galaxies (such as GN 32485).

In what follows we report the mode for the metallicity and ionization
using the $P(\log Z)$ (where we use $\log Z = 12 + \log \mathrm{O/H}$
here) and $P(\log q)$ distributions, along with the 16th percentile and
84th percentile to indicate the uncertainties.    We derive the latter
using the HDI, which is the smallest region that
contains 68\% of the probability density
\citep[see][]{bailer18,estr20}.  These are indicated by the vertical
lines in the $P(\log Z)$ and $P(\log q)$ panels for each galaxy in
Figures~\ref{fig:spectra_izi_results} and
\ref{fig:spectra_izi_results_gs}.

\subsection{The relation between strong-line ratios and metallicity}\label{section:r23_Z}

Figure~\ref{fig:r23_Z} shows our results for the metallicity ($12 +
\log \mathrm{O/H}$) derived from the R$_{23}$ line ratio.  The data
points show the results for individual CLEAR galaxies with
metallicities and ionization derived using \izi\ from the strong
emission lines.  The curves in the Figure show calibrations from the
literature, derived using different methods and galaxy samples.   It
should be noted that the errors on the data points in the Figure are
highly covariant (as the metallicity values are \textit{derived from}
the $R_{23}$ ratios).  To illustrate this, the right panel of
Figure~\ref{fig:r23_Z} shows the equivalent $1\sigma$ and $2\sigma$
error ellipses derived from the covariance between these parameters
for a fiducial galaxy in the sample with $\OH = 8.8$.   This accounts
for the lack of perceived scatter in the results (the scatter is
covariant as indicated by the error ellipse, and is directed along the
observed sequence between R$_{23}$ and metallicity).  

The R$_{23}$--metallicity calibrations include those based on nearby
star-forming galaxies (e.g., SDSS galaxies at $z\sim 0.1$,
\citealt{maio08,curti17}).  \citeauthor{maio08} used a combination of
direct measurements (for low-metallicity galaxies) with metallicities
inferred from photoionization models for higher metallicity galaxies
in SDSS.  These results show that R$_{23}$ is famously
``double-valued'' with a maximum and inflection point around $\log$
R$_{23} \simeq 1$ at $12 + \log \mathrm{O/H} \simeq 8.2$.   On the
high-metallicity branch of R$_{23}$, other calibrations find lower
metallicity at fixed R$_{23}$ using direct $T_e$ metallicity methods
\citep{curti17}, but those authors caution that the \oiii\
$\lambda4363$ emission exhibits contamination from [\ion{Fe}{2}]
emission in higher metallicity regions.  In addition, other studies
have argued that some of the offset may result from a contribution to
the emission from diffuse interstellar gas (DIG, \citealt{sand17}),
but at these redshifts this effect is expected to be small
\citep{sand21}).  Yet other studies find that the assumption about the
ionization of the gas leads to biases that cause the metallicity
derived from R$_{23}$ to be \textit{undervalued} \citep{berg21}.  

The effect of the (isobaric) pressure of the nebular gas is
also important.  \citet{kewley19} use the suite of predictions from
MAPPINGS V with a gas pressure of $\log P/k = 5$, valid for $8.53 <
\OH < 9.23$.  These calibrations have a dependence on ionization
parameter.   These models produce the calibration illustrated in
Figure~\ref{fig:r23_Z} (thick gray line).  These are consistent with
the \citet{curti17} calibration.    Increasing the pressure to $\log
P/k = 7$ has a strong impact on $\log R_{23}$ for metallicities $\OH
\gtrsim 8.5$ (see \citealt{kewley19}, their Figure 9), increasing
$\log R_{23}$ by nearly $\sim$1~dex at $\OH \sim 9$.  Because we
use an isobaric pressure of $\log P/k=6.5$, this explains the
offset in our calibration and that of \citeauthor{kewley19} and
\citeauthor{curti17},  illustrated in the figure.  Our calibration is
more consistent with that derived independently by \citet{strom18}.

\begin{figure*}[t]
\centering
\includegraphics[height=0.39\textwidth]{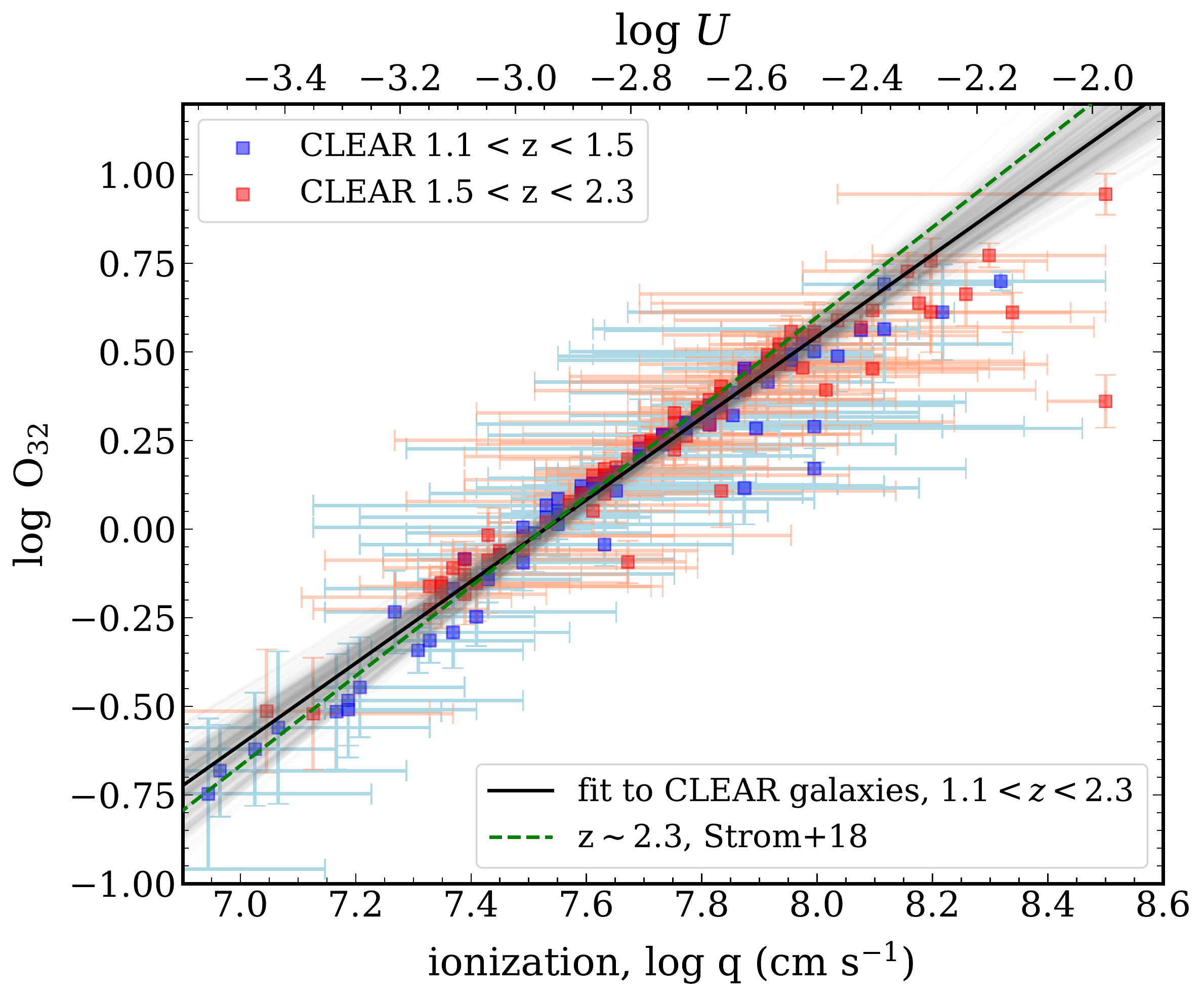}
\includegraphics[height=0.39\textwidth]{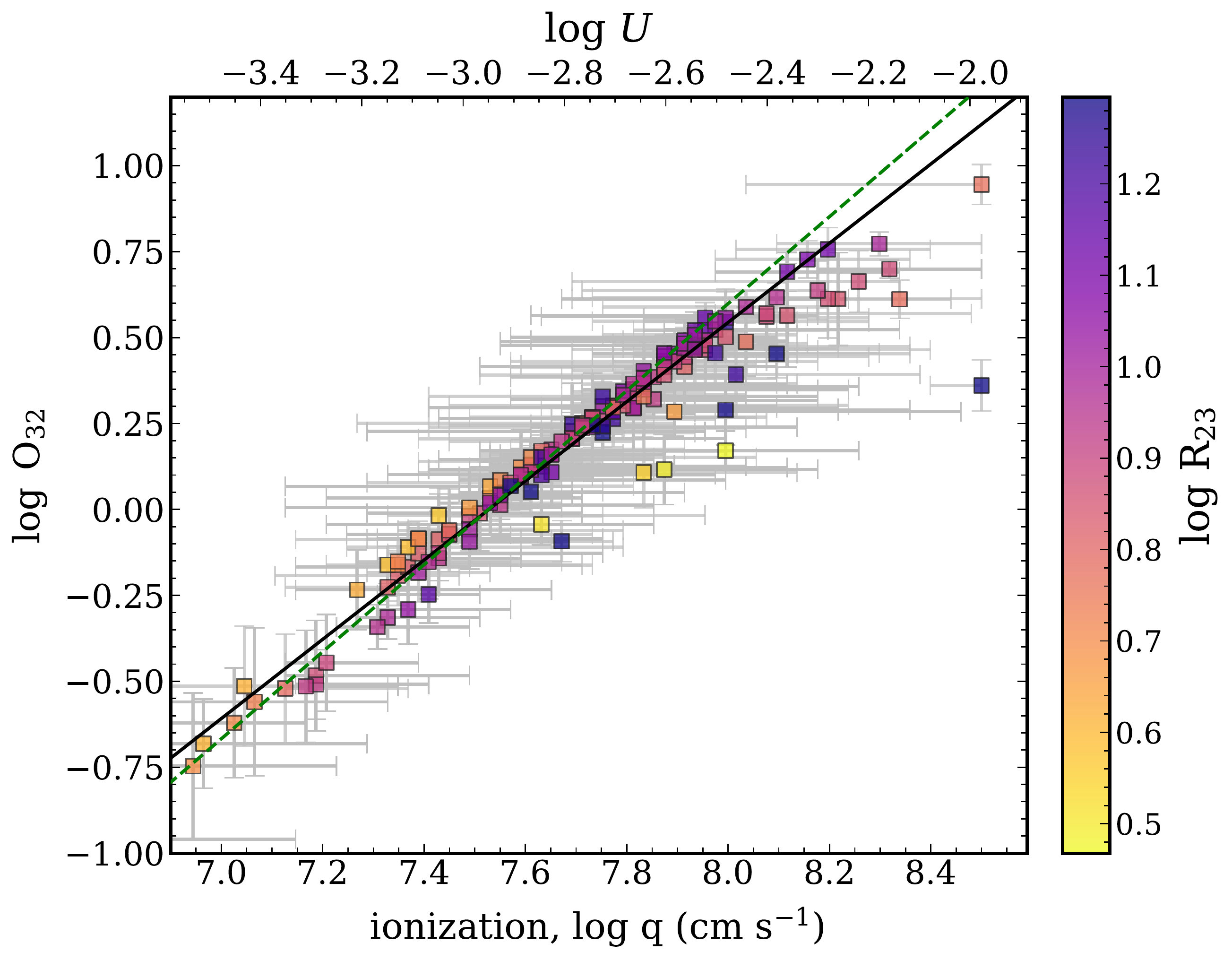}
\caption{O$_{32}$ versus ionization  for the CLEAR sample, derived by
comparing the observed \oii, \hb, and \oiii\ line fluxes to the
MAPPINGS V photoionization models.   The left panel shows CLEAR galaxies
color coded by redshift.   The right panel shows the CLEAR
galaxies color coded by R$_{23}$ value.   
In each panel, the top axis shows the corresponding
dimensionless ionization parameter ($U = q / c$).  The dashed line
shows the linear relation calibrated by \citet{strom18} using
independent photoionization modeling  for $z\sim 2.3$ galaxies.  The
solid line shows a linear fit to the CLEAR galaxies derived from a
Bayesian method.  The shaded gray region shows 400 random draws from
the posterior of the linear fit.  The O$_{32}$ ratio
correlates with ionization parameter.    There is a secondary
dependence on R$_{23}$ (which translates to a dependence on gas-phase
metallicity).   \label{fig:o32_q}}
\end{figure*}

We fit a quadratic function to the R$_{23}$--$Z$
relation derived for the CLEAR galaxies of the form,
\begin{equation}\label{eqn:r23_Z}
  \log \mathrm{R}_{23} = A \times  (\log Z-\log Z_0)^2 + \log \mathrm{R}_0,
\end{equation} 
where $\log Z \equiv 12 + \log(\mathrm{O/H})$ and $A = -1.07 \pm 0.03$,
$\log \mathrm{R}_0=1.041 \pm 0.004$, and $\log Z_0=8.228 \pm 0.006$.
\added{We include the statistical uncertainties when performing the fit.
However, we have not included the effects of the covariance between
R$_{23}$ and $Z$ (as discussed above), and therefore the uncertainties
on these parameters are overestimated. }

This relation we observe for CLEAR is consistent with other studies of
high-redshift galaxies, that also show larger R$_{23}$ at fixed
metallicity compared to calibrations derived for nearby
galaxies.\footnote{Note that the relation we derive between R$_{23}$
and $Z$ is not, strictly, an independent calibration.  Rather, it is a
relation appropriate for $1.1 < z < 2.3$ galaxies based on their
observed emission lines (\oii, \hbeta, \oiii) and the MAPPINGS V
models given our assumption of ISM pressure.  The same note applies to
the relation between O$_{32}$ and $\log q$.\label{footnote:modeling}}
For example,  \citet{strom18} fit strong emission lines measured in
$z\sim 2.3$ galaxies using predictions from photoionization models
that allow for a range of stellar and nebular metallicity, ionization
parameter and N/O abundance.  They then parameterize R$_{23}$--$Z$ as
a quadratic expression, which is also shown in Figure~\ref{fig:r23_Z}.
The \citeauthor{strom18} relation is similar to the one for CLEAR,
with an offset of 0.05--0.1 dex.  The results from
\citeauthor{strom18} are similar to those from \citeauthor{maio08}
(although \citeauthor{strom18} argue that the \citeauthor{maio08} and
other calibrations based on direct $T_e$ measurements need to be
revised upwards by 0.24 dex).

Figure~\ref{fig:r23_Z} also shows that for the CLEAR sample the
ionization of the gas increases as the metallicity decreases. The
ionization constraints (primarily from $\oiii/\oii$) within the
\texttt{IZI} fitting allow us to determine a likelihood that  galaxies
fall on the upper or lower branch of the R$_{23}$--$Z$ relation.
Galaxies on the lower-metallicity branch of R$_{23}$ have
significantly higher ionization than galaxies on the upper-metallicity
branch of R$_{23}$.   Physically, the change in ionization causes an
increase in the fraction of doubly ionized oxygen (O$^{++}$ ) at the
expense of singly ionized oxygen (O$^+$), which therefore leaves the
numerator in the definition of R$_{23}$ mostly unchanged.  However,
the change in ionization should then be apparent in the ratio of
O$_{32}$, which we discuss in the next subsection. 

\subsection{The relation between strong-line ratios and ionization}\label{section:o32_q}

Figure~\ref{fig:o32_q} shows there is a tight correlation between the
O$_{32}$ ratio and the ionization parameter, $q$, derived by modeling
the emission lines of the CLEAR galaxies with the MAPPINGS V
photoionization models (Section~\ref{section:izi}).  The correlation is
expected as an increase in ionization parameter corresponds to an
increase in the ratio of O$^{++}$ to O$^{+}$.    Low-redshift
star-forming galaxies typically have ionization parameters in the
range, $7.3 < \log q / (\mathrm{cm~s^{-1}}) < 7.6$
\citep{mous10,poet18}, while \ion{H}{2} regions and super-star
clusters in starburst galaxies (e.g., M82) have ionization parameters
as high as $\log q / (\mathrm{cm~s^{-1}}) \sim 8.2-8.7$
\citep{lsmith06,perez14}, which is observed in other low redshift,
extreme star-forming galaxies \citep[e.g.,][]{berg21,oliv21}.  For the
CLEAR galaxies at $1.1 < z < 2.3$, the range of ionization parameter
extends over $7.3 \lsim \log q / (\mathrm{cm~s^{-1}}) \lsim 8.5$,
spanning the full range seen in star-forming galaxies in the local
universe.  This is consistent with other studies of high redshift
galaxies that show evidence for increased ionization
\citep[e.g.,][]{sand16a,strom18}.

We fit a linear relation between $\log$ O$_{32}$ and $\log q$ using
\begin{equation}\label{eqn:o32_q}  \log q = A\times \log \mathrm{O}_{32} + \log q_0.
\end{equation}
We fit the relation using a Gaussian Mixture Model (\texttt{linmix},
\citealt{kelly07}), which yields $A = 0.86 \pm 0.07$ and $\log q_0 =
7.53 \pm 0.02$.   This line is shown in Figure~\ref{fig:o32_q} (along
with 400 random draws from the posterior).   This linear fit is very
similar to that from \citet{strom18}, who derived their relation from
a sample of $z\sim 2.3$ star-forming galaxies with independent
photoionization modeling (see also Footnote~\ref{footnote:modeling}).

Closer inspection of the CLEAR galaxies in Figure~\ref{fig:o32_q}
shows that while \oiii/\oii\ correlates with ionization parameter,
there is a secondary effect.   The effect is subtle (with galaxies
with lower R$_{23}$ shifting  above the relation at lower ionization
and below the relation at higher ionization).  This effect is likely a
result of the fact that many of the galaxies in our sample lie at the
peak of the R$_{23}$--$Z$ relation ( near R$_{23} \sim 1$ in
Figure~\ref{fig:r23_Z}).  R$_{23}$ correlates with gas-phase
metallicity, this translates to a secondary effect in \OH.
Empirically, this secondary dependence on R$_{23}$ comes from the
relative strength of \hb\ compared to \oii\ + \oiii.  Galaxies with
stronger \hb\ push toward lower R$_{23}$.   This is predicted by the
photoionization modeling (see Figure~7 of \citealt{kewley19}), and
physically is a result of the fact that gas cooling is more efficient
in higher metallicity environments.  While the CLEAR galaxies in
Figure~\ref{fig:o32_q} show an apparent ceiling, with $\log
\mathrm{O}_{32} \lsim 0.75$, this seems only a property of our sample.
For example, some extreme galaxies at $z\sim 0.01$ with low
metallicity ($\OH \simeq 7.4-7.6$)  have even higher line ratios
($\log \mathrm{O}_{32} \simeq 1.1-1.3$) with high ionization
parameters ($\log U \simeq -2.4 $ to $-1.8$, \citealt{berg21}).  This
is consistent with our relation (Equation~\ref{eqn:o32_q}), which
would predict $\log U \simeq -2.0$ to $-1.8$ for these galaxies.
Taken together, this is evidence that while to first order the $\log$
O$_{32}$ ratio traces gas ionization strongly there is a secondary
dependence on metallicity that contributes to the scatter in this
relation, and this persists at high redshift ($1 \lsim z \lsim 2$).
%

\section{Results}\label{section:results}

\subsection{On the Mass-Metallicity Relation}\label{section:MZR}

\begin{figure*}[th]
\centering
\includegraphics[width=0.87\textwidth]{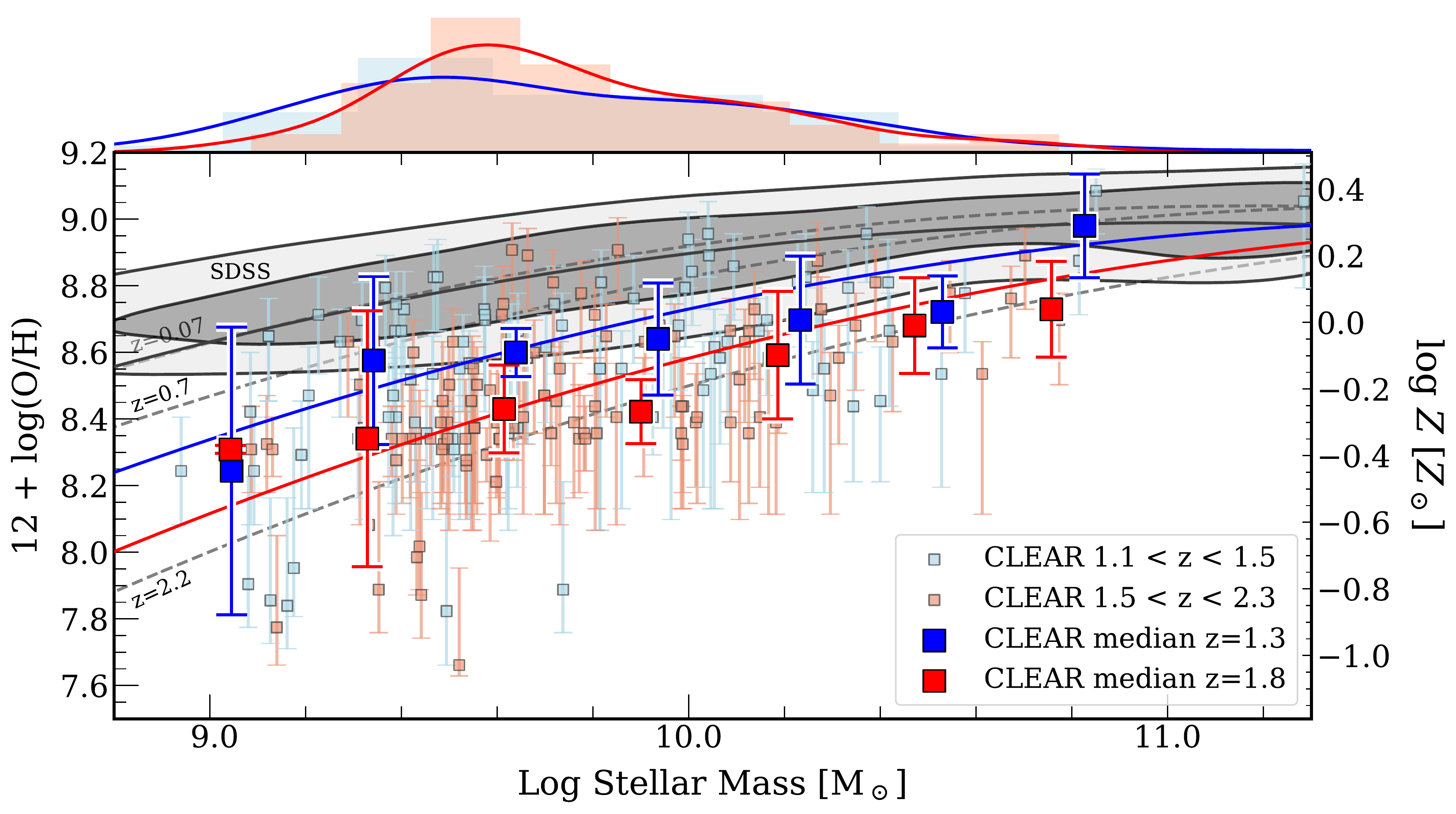}
\includegraphics[height=0.35\textwidth]{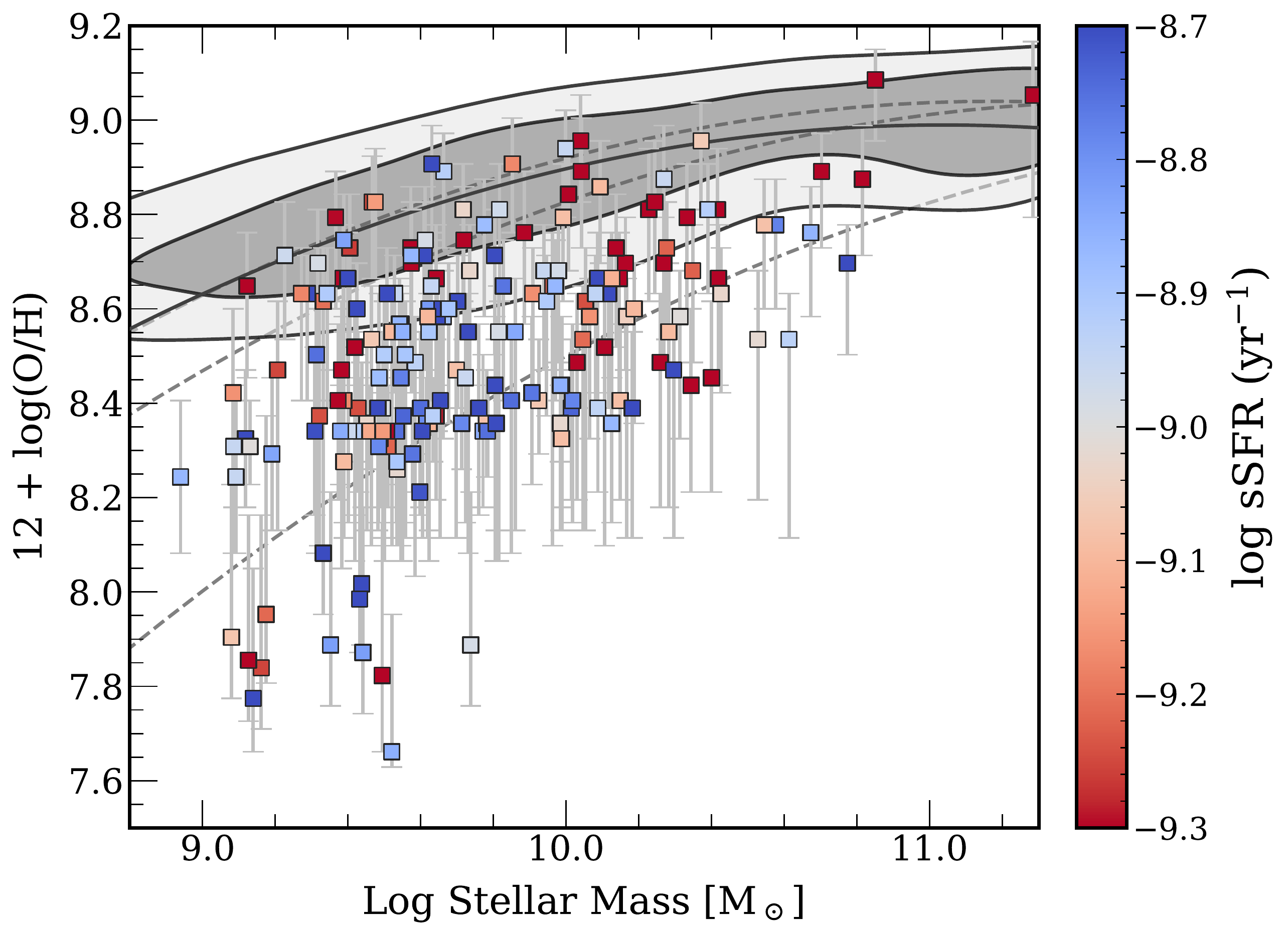}
\hspace{0.25in}
\includegraphics[height=0.35\textwidth]{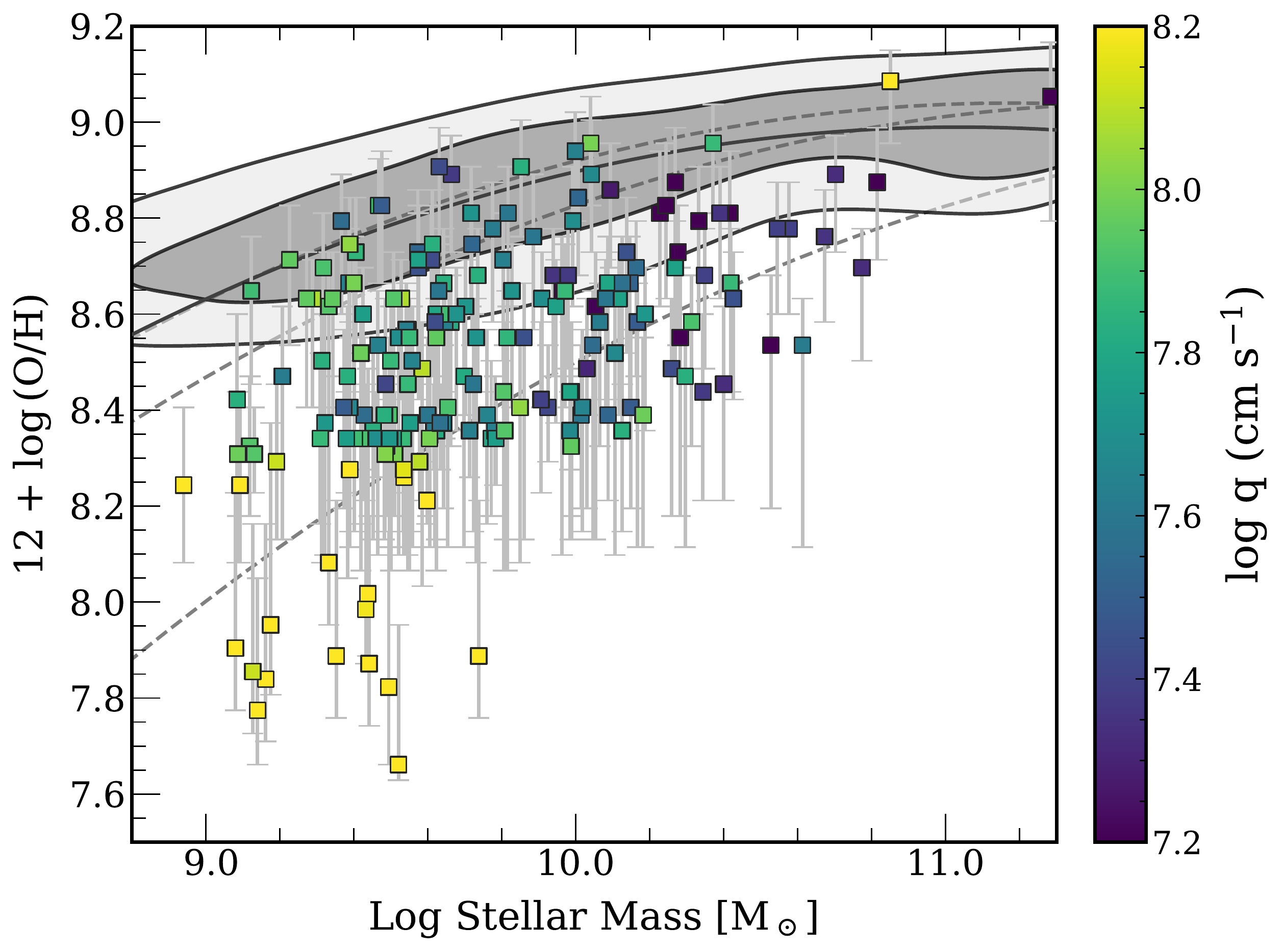}
\caption{The stellar-mass, gas-phase-metallicity relation (MZR) for
CLEAR galaxies at $1.1 < z < 2.3$.  The top plot shows the MZR
with the CLEAR galaxies color coded by redshift. Large squares
show medians in bins of stellar mass.  Error bars show the
scatter in each bin.    The red and blue solid lines show
fits to the CLEAR galaxies.  The gray-shaded region shows the
distribution for SDSS galaxies derived using our analysis (identical to
that applied to CLEAR).   The black solid curve shows our quadratic fit to
the SDSS galaxies (Equation~\ref{eqn:MZR} and
Table~\ref{table:MZR_fit}).  The dotted lines show the MZR relation at
$z=0.07$, 0.7, and 2.2 derived by \citet{maio08}.  The histogram and
curve along the top of the panel shows the distribution of stellar
mass in the $z\sim 1.3$ (blue) and $z\sim 1.8$
CLEAR samples.   The bottom panels show the MZR with galaxies
color-coded by specific SFR (bottom left; $\log$
sSFR ) and nebular ionization (bottom right; $\log
q$).   \label{fig:MZR}}
\end{figure*}

Figure~\ref{fig:MZR} shows the relation between stellar--mass, and
gas-phase metallicity (the ``MZR'') derived for the galaxies in CLEAR
at $1.1 < z < 2.3$ compared to some relations in the literature at low
and high redshift.  We also show results for SDSS DR14 using results
derived from the same set of photoionization models and emission line
fluxes as for CLEAR (see Section~\ref{section:izi}).
Figure~\ref{fig:MZR} shows that the gas-phase metallicities for the
SDSS galaxies derived for these models mostly follows those derived by
other methods (notably, \citealt{trem04} and \citealt{maio08}, the
latter is illustrated in the figure).  Comparing these results, there
is evolution from the relation from SDSS ($z\sim 0.2$) compared to
CLEAR.  Galaxies at higher redshift have lower metallicities, and this
has been observed by multiple studies (using a myriad of methods to
derive gas-phase abundances, e.g.,
\citealt{maio08,maio19,henry21,sand21}, and references therein).
%

To measure the evolution in the MZR, we parameterize the relation
using the prescription of \citet{maio08}, 
\begin{equation}\label{eqn:MZR} 12+ \log \mathrm{O/H} = -0.0864(\log
M_\ast - \log M_0)^2 + K_0.
\end{equation}
Here, $\log M_0$ is a scale stellar mass (in units of Solar masses)
when the relation achieves metallicity $K_0$.  Figure~\ref{fig:MZR}
shows the relations derived by \citeauthor{maio08} for SDSS and the
AMAZE samples at $z=0.07$, 0.7, and 2.2 (as labeled), using an
independent strong-line calibration \citep{kewley02b} calibrated
against SDSS DR4 observations.

We fit Equation~\ref{eqn:MZR} to the results for our SDSS and CLEAR
samples.  For both SDSS and CLEAR we have modeled the same set of
emission lines (\oiii $\lambda\lambda$4959+5007, \oii\ $\lambda\lambda
3726$+3728, \hb) with the same photoionization models (see
Section~\ref{section:izi}). These are independent from the calibration
used by others (e.g., \added{\citealt{maio08}}).    For SDSS, because we
have sufficient dynamical range in stellar mass we fit for both $M_0$
and $K_0$.   However, because the stellar-mass distribution of CLEAR
is strongly focused on $\log M_\ast/M_\odot \sim 9.3-9.9$ we fix $K_0
= 9.00$ at the value derived we derive from SDSS and equal to that
obtained by \citet{maio08} at $z=2.2$.   For
CLEAR, we also fit for $M_0$ for subsamples of galaxies split in redshift
for $1.1 < z < 1.5$ and $1.5 < z < 2.3$.  
%
%

It is worth noting that the MZR relation we derive for SDSS
(Figure~\ref{fig:MZR}, solid gray line) agrees well with the relation
derived by \citet{maio08} (dashed gray line).   This comparison is
important because we have used a different photoionization model, and
different choices of strong emission lines.   \added{\citet{maio08}}
calibrate their metallicities using photoionization models that assume
lower $n_e$, which result in lower pressure.  Nevertheless, the
calibrations agree well for R$_{23}$ versus \OH\ for high
metallicities ($\OH > 8.8$).  At lower metallicities, our calibration
shifts to higher \OH\ at fixed R$_{23}$ (see Figure~\ref{fig:r23_Z}).
This accounts for the slight increase in the median of the SDSS
distribution we observed around masses $\log M_\ast / M_\odot \sim 9$
compared to the fitted relation from \added{\citet{maio08}}.  The
differences emphasize the importance of calibrating the relation
between strong emission lines and metallicity in order to study the
absolute evolution in the MZR.  The comparison we measure here is
differential (in that we use the same set of photoionization models
for the galaxies at all redshifts), but this ignores possible
evolution in the physical conditions in the galaxies (in which case
one should use photoionization models whose physical properties
[especially the galaxy density/pressure] also evolve with time).
We plan to explore these effects in a future study).  

Table~\ref{table:MZR_fit} shows the derived values of $M_0$ (and
$K_0$) for our fits to the SDSS and CLEAR samples.   By fixing $K_0$,
we see a steady increase in $\log M_0$ with increasing redshift.  This
corresponds to a decrease in the typical gas-phase metallicity with
increasing redshift (at fixed mass).    Using the results of the
analytic fits, at a stellar mass of $\log M_\ast / M_\odot = 9.0$ we
observe a decrease in $12+\log\mathrm{O/H}$ of 0.3 dex from $z\sim
0.2$ to $z\sim 1.35$ and an additional decrease
of 0.2 dex $z\sim 1.9$.   This implies that galaxies
like a progenitor of the Milky Way \citep{papo15} had metallicities of
$\OH \simeq 8.3$ at $z=1.9$ and $8.5$ at $z=1.4$ (or restated as
saying the Milky Way progenitor had roughly $Z_\mathrm{gas} \approx
0.4-0.6$~\zsol\ at 9--10 Gyr in the past).  This is consistent with
direct measurements of abundances in stars of this age within the
Milky Way \citep[e.g.,][]{berg14} and implies we are building a
coherent picture of the chemical enrichment of galaxies like our own. 
%

\begin{deluxetable}{lccc}
\tabletypesize{\footnotesize}
\tablecolumns{4}
\tablewidth{0pt}
\tablecaption{Fitted parameters for the analytic form of the mass-metallicity relation
  using Equation~\ref{eqn:MZR}. \label{table:MZR_fit}}
\tablehead{
\colhead{sample}  & \colhead{Redshift Range} &
\colhead{$\log M_0$} & \colhead{$K_0$} }
\startdata
SDSS  & $z\sim 0.2$ & 11.04 $\pm$ 0.01 & 9.000 $\pm$ 0.002 \\ 
CLEAR  & $1.1 < z < 1.5$ & 11.77 $\pm$ 0.06 & 9.0\tablenotemark{$\ast$} \\ 
CLEAR & $1.5 < z < 2.3$ & 12.20 $\pm$ 0.04  & 9.0\tablenotemark{$\ast$}  \\ 
CLEAR\tablenotemark{$\dag$} & $1.1 < z < 1.5$ & 12.06 $\pm$  0.06 & 9.0\tablenotemark{$\ast$}  \\ 
\enddata
\tablecomments{Except where noted, all fits are derived using the metallicities derived from
  the \oii, \oiii, and \hb\ emission lines compared to the MAPPINGS V
  models. }
\tablenotetext{\ast}{Value fixed for fit (see text).}
\tablenotetext{$\dag$}{Using
the \oii, \oiii, \hb, \ha+\nii, and \sii\ emission-line fluxes, see
Appendix.}
\vspace{0pt}
\end{deluxetable}

In Figure~\ref{fig:MZR}, the lower two panels show the MZR with the
CLEAR galaxies color-coded by sSFR and ionization parameter ($\log
q$).   There is an apparent dependence on sSFR, in that galaxies with
higher sSFR have lower gas-phase metallicity (and because we show this
as a function of \textit{specific} SFR this relation is at fixed
\textit{mass} by construction).  The dependence of the MZR on SFR has
been observed previously both at high and low redshift, and several
studies have argued that the``fundamental'' MZR--SFR relation is
independent of redshift
\citep[e.g.,][]{elli08,mann10,andrews13,sand18,sand21,cresci19,curti20,henry21}.

Similarly, there is an apparent trend between the galaxy MZR and
ionization.  Figure~\ref{fig:MZR} shows that  galaxies in CLEAR with
higher stellar mass are generally only found to have lower ionization
($\log q$): indeed, galaxies with the lowest ionization ($\log q /
\mathrm{(cm~s^{-1})} \lesssim 7.3$) are only found with higher stellar
masses ($\log M_\ast/M_\odot > 10.2$).  Galaxies with the highest
ionization ($\log q / (\mathrm{cm~s^{-1}}) \gtrsim 8.1$) are generally
only found at lower stellar masses, $\log M_\ast / M_\odot < 9.9$.
There is also qualitative evidence that at \textit{fixed stellar mass}
galaxies with higher metallicity have lower ionization parameters.
This is similar to the relation between SFR and the MZR discussed
above, and implies that ionization, metallicity and sSFR are
intertwined.  

\subsection{On the Mass--Ionization Relation} \label{section:MQR}

\begin{figure*}[th]
\centering
\includegraphics[width=0.87\textwidth]{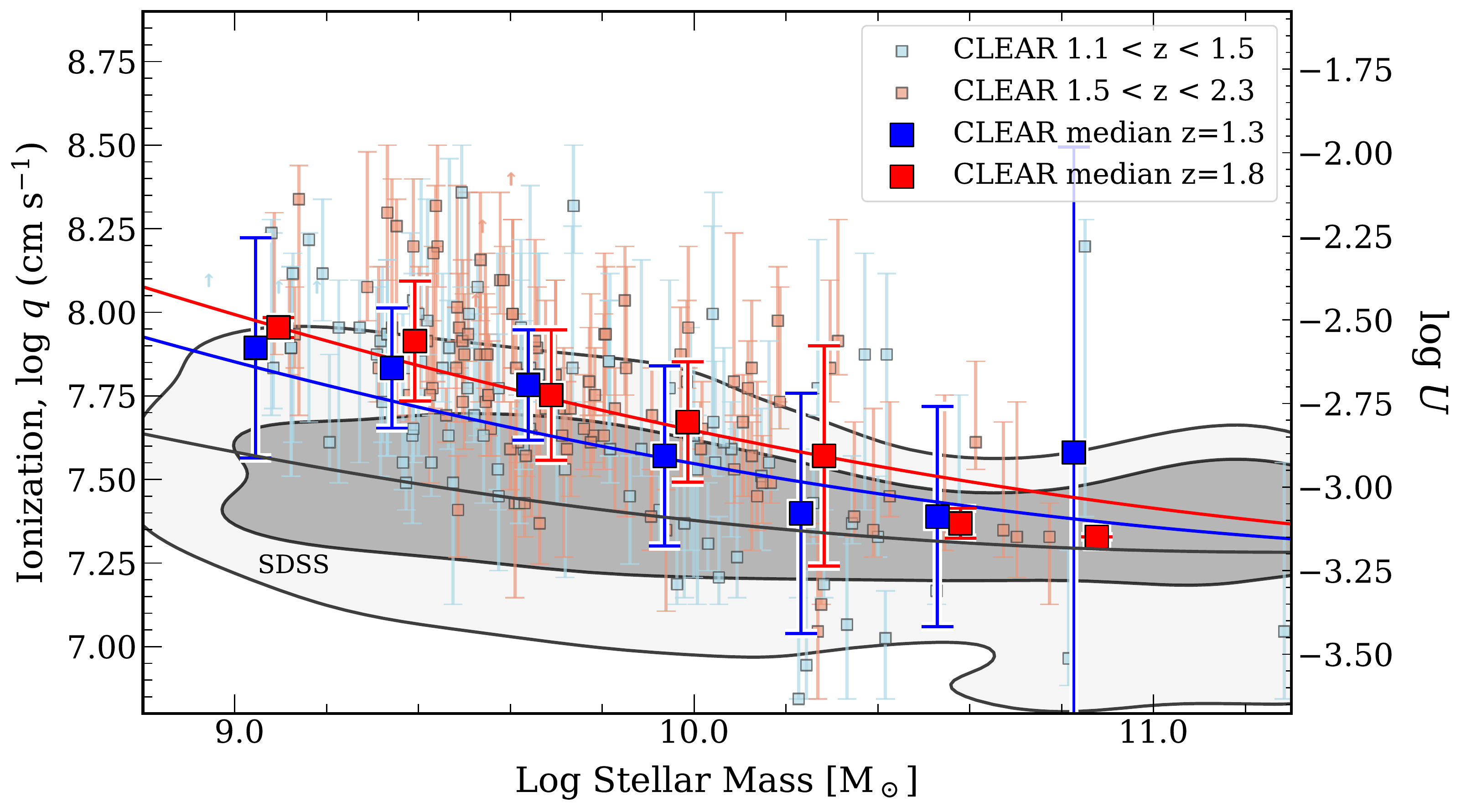}

\vspace{12pt}

\includegraphics[height=0.33\textwidth]{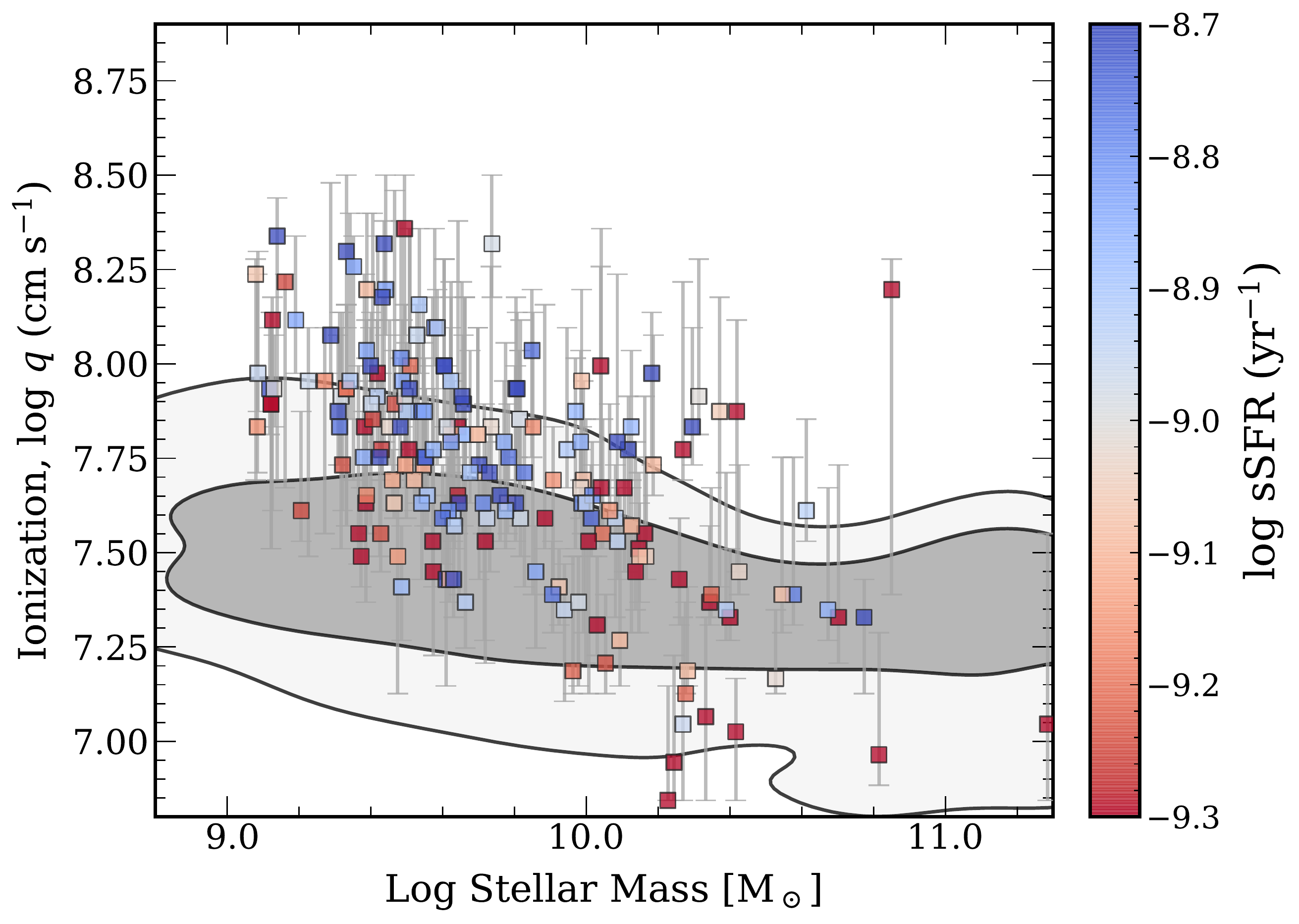}
\hspace{0.125in}
\includegraphics[height=0.33\textwidth]{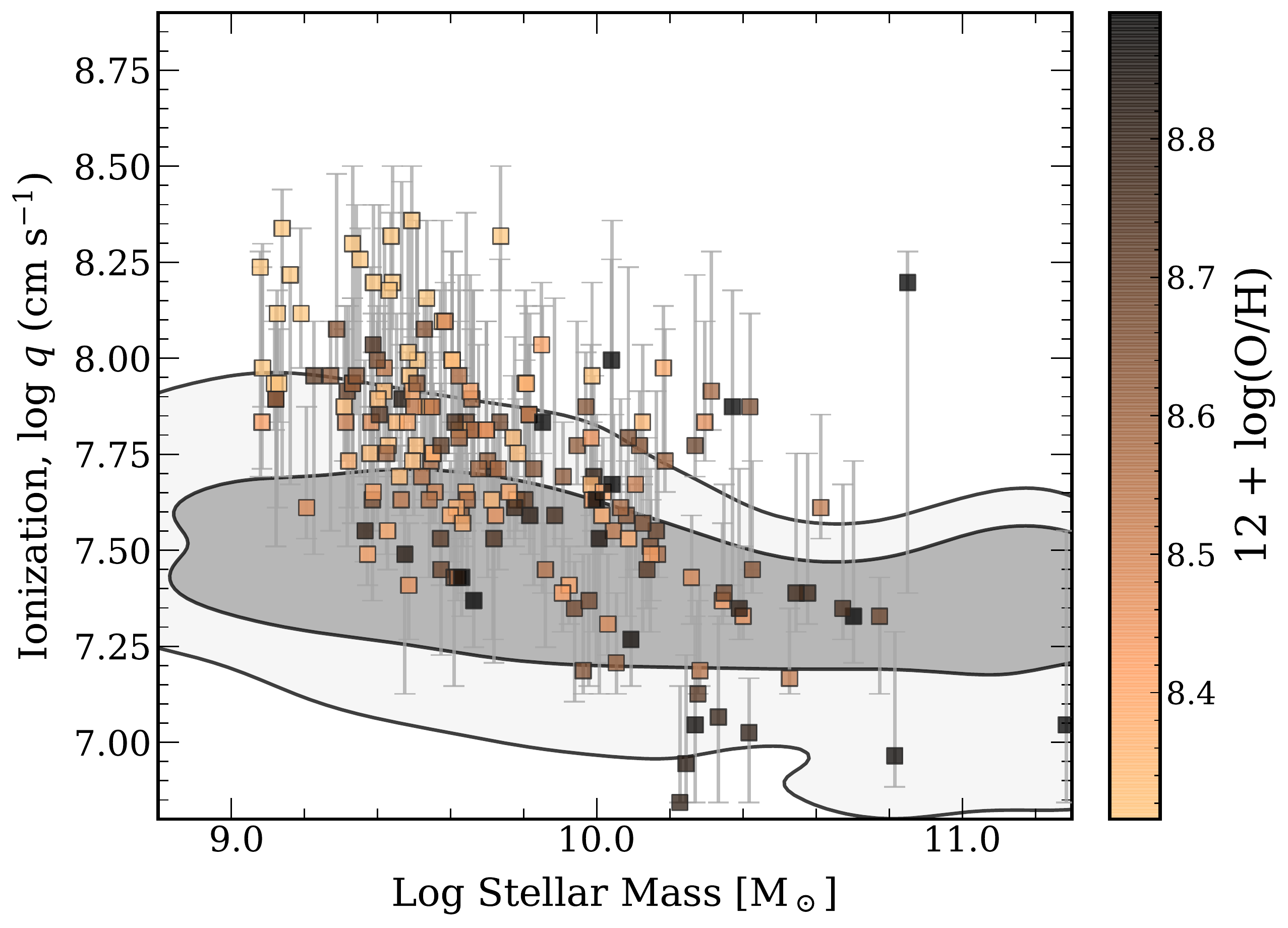}
\caption{The stellar-mass, ionization relation (MQR) for CLEAR
galaxies at $1.1 < z < 2.3$.  The top plot shows the MQR with the
CLEAR galaxies color-coded by redshift.  Large squares show medians in
bins of stellar mass.  The error bars show the scatter in each bin.
The shaded region denotes the SDSS galaxy distribution of ionization
parameters derived using our analysis (identical to that applied to the
CLEAR galaxies).   The solid lines show analytic fits to the data for
the SDSS (black), CLEAR $z=1.3$ (blue), and CLEAR $z=1.8$ samples.
The bottom panels show the MQR with galaxies color-coded by specific
SFR (bottom left; $\log$ sSFR ) and gas-phase metallicity, $12 +
\log(\mathrm{O/H})$.  \label{fig:MQR}}
\end{figure*}

Figure~\ref{fig:MQR} shows the relation between stellar--mass and
nebular ionization parameter (i.e.,  the
``mass-ionization-relation'', or MQR) derived for the galaxies in
CLEAR at $1.1 < z < 2.3$.  In comparison, we also show the MQR for
galaxies in SDSS analyzed using the same set of emission lines and
photoionization models used for the analysis of the CLEAR galaxies
(see Section~\ref{section:izi}).  The MQR clearly evolves from $z\sim
0.2$ (from SDSS) to $z\sim 1.3$ and to $z\sim 1.8$ (from CLEAR).   At
fixed stellar mass, galaxies have higher ionization parameter at
higher redshift, where the effect is stronger for galaxies of lower
stellar mass.   This extends trends seen both at lower redshift and
higher stellar masses  (see also  \citealt{kewley15,kaas18,strom22}).  

We parameterize the MQR using a simple quadratic relation inspired by
the MZR \citep{maio08},
\begin{equation}\label{eqn:MQR}
  \log q = 0.0571 \times (\log M_\ast - \log M_0)^2 + \log q_0.
\end{equation}
Here, $\log M_0$ is a scale stellar mass (in units of $M_\odot$) when
the relation achieves ionization $\log q_0$, and $q$ is measured in
units of cm s$^{-1}$.  For our SDSS sample, we fit for $M_0$ and $\log
q_0$.  For CLEAR, we fix $\log q_0$ to the value derived for the SDSS
sample ($\log q_0 = 7.282$) because the stellar-mass distribution of
CLEAR peaks at $\log M_\ast/M_\odot \sim 9.3-9.9$.  For CLEAR we also
fit $M_0$ for galaxies in subsamples of redshift, $1.1 < z < 1.5$ and
$1.5 < z < 2.3$. 
%

\begin{deluxetable}{lccc}
\tabletypesize{\footnotesize}
\tablecolumns{4}
\tablewidth{0pt}
\tablecaption{Fitted parameters for the analytic form of the mass-ionization relation
  using Equation~\ref{eqn:MQR}. \label{table:MQR_fit}}
\tablehead{
\colhead{sample}  & \colhead{Redshift Range} &
\colhead{$\log M_0$} & \colhead{$\log q_0$} }
\startdata
SDSS  & $z\sim 0.2$ & 11.29 $\pm$ 0.11 & 7.282 $\pm$ 0.007 \\ 
CLEAR  & $1.1 < z < 1.5$ & 12.22 $\pm$ 0.10 & 7.282\tablenotemark{$\ast$} \\ 
CLEAR & $1.5 < z < 2.3$ & 12.56 $\pm$ 0.06  & 7.282\tablenotemark{$\ast$}  \\ 
\enddata
\tablecomments{All fits derived using the nebular ionization parameters derived from
  the \oii, \oiii, and \hb\ emission lines compared to the MAPPINGS V
  models. }
\tablenotetext{\ast}{Value fixed for fit (see text).}
\end{deluxetable}

Table~\ref{table:MQR_fit} shows the best-fit values and their
uncertainties for $M_0$ (and $Q_0$) for our fits to the SDSS and CLEAR
samples.   There is a steady increase in $\log M_0$ with increasing
redshift.  This corresponds to an increase in the typical nebular
ionization with increasing redshift (at fixed mass).    Using these
fits, at a stellar mass of $\log M_\ast / M_\odot = 9.0$ we observe an
increase in $\log q$ of 0.29 $\pm$ 0.05 dex from $z\sim 0.2$ to $z\sim
1.35$ and an additional increase of 0.14 $\pm$ 0.04 dex from $z\sim
1.3$ to $z\sim 1.8$.   

At a stellar mass of $\log M_\ast / M_\odot = 10.0$ the evolution is
weaker, as we observe a decrease in ionization parameter of 0.19 $\pm$
0.03 dex from $z\sim 0.2$ to $z\sim 1.3$.  At higher stellar masses
$\log M_\ast/M_\odot \gsim 10.5$ there is very little evidence  that
the MQR evolves from $z\sim 0.2$ to $z\sim 1.1-2.3$, but sample is
limited by smaller numbers of massive galaxies.  For example, 
\citet{kaas18} find an increase in the ionization parameter of
$\simeq$ 0.3 dex for galaxies with $\log M_\ast/M_\odot \simeq 10.4 -
10.9$ from $z\sim 0.2$ to 1.5, only slightly stronger than the trend
we see here.  Our results are also similar to the independent
measurements derived by \citet{strom22} at $z\sim 2.3$.  

\added{ One potential source of concern in the interpretation of the
MQR (in Figure~\ref{fig:MQR}) is if our sample is biased by sources
with strong emission lines.  Our samples are selected with
$m(\mathrm{F105W}) < 25$ mag, so sources with strong emission lines
could be overrepresented (particularly near our magnitude limit).  To
test this scenario, we used the measurements of the EW for strong
lines (\oii,  \hb, and \oiii) for sources in our sample (measured from
their G102 and G141 spectra) that have redshifts that place these
lines in the F105W passband.  We then correct the F105W magnitude
(using e.g., Eqn.~2 of \citealt{papo01}), and exclude any object with
$m(\mathrm{F105W}) > 25$~mag.  This removes only 9 sources (a loss of
$<$5\% of the sample).  Excluding these 9 sources, we refit the MQR
using Equation~\ref{eqn:MQR}, and find that $\log M_0$ is reduced by
0.05 and 0.04~dex for CLEAR at $z\sim 1.3$ and 1.8, respectively
(this is less than the statistical uncertainty in
Table~\ref{table:MQR_fit}).  Therefore, the evolution in the MQR is
not seriously impacted by the presence of strong emission lines in the
sample.   }

The bottom panels of Figure~\ref{fig:MQR} show the dependence
of the MQR on sSFR and on the gas-phase metallicity.
There is a an overall trend of increasing  metallicity with increasing
stellar mass, which is a consequence of the MZR (see
Section~\ref{section:MZR}).  There is no identifiable trend between
metallicity with ionization at fixed mass: for example, galaxies with
$\log M_\ast/M_\odot = 9.4-9.8$ span a wide range of \OH\ and $\log
q$. Qualitatively, in this stellar-mass range, galaxies with the
highest ionization-parameters have lower metallicity (and vice versa)
but this statement is limited by the size of our sample.  We return to
these points in Section~\ref{section:discussion}.

\subsection{Ionization Parameter Dependence on Specific SFR}\label{section:logq_ssfr}

Another interesting question is to what extent the change in
ionization parameter is driven by changes in the SFR (or the changes
in SFR at fixed stellar mass, which is the sSFR).  \citet{kaas18}
considered the correlation between ionization parameter and sSFR for
SDSS galaxies and galaxies at $z\sim 1.5$.  They concluded that higher
$\log q$ is directly linked to an increase in SFR (see also
\citealt{brin08}).   

\begin{figure}[th]
\centering
\includegraphics[width=0.48\textwidth]{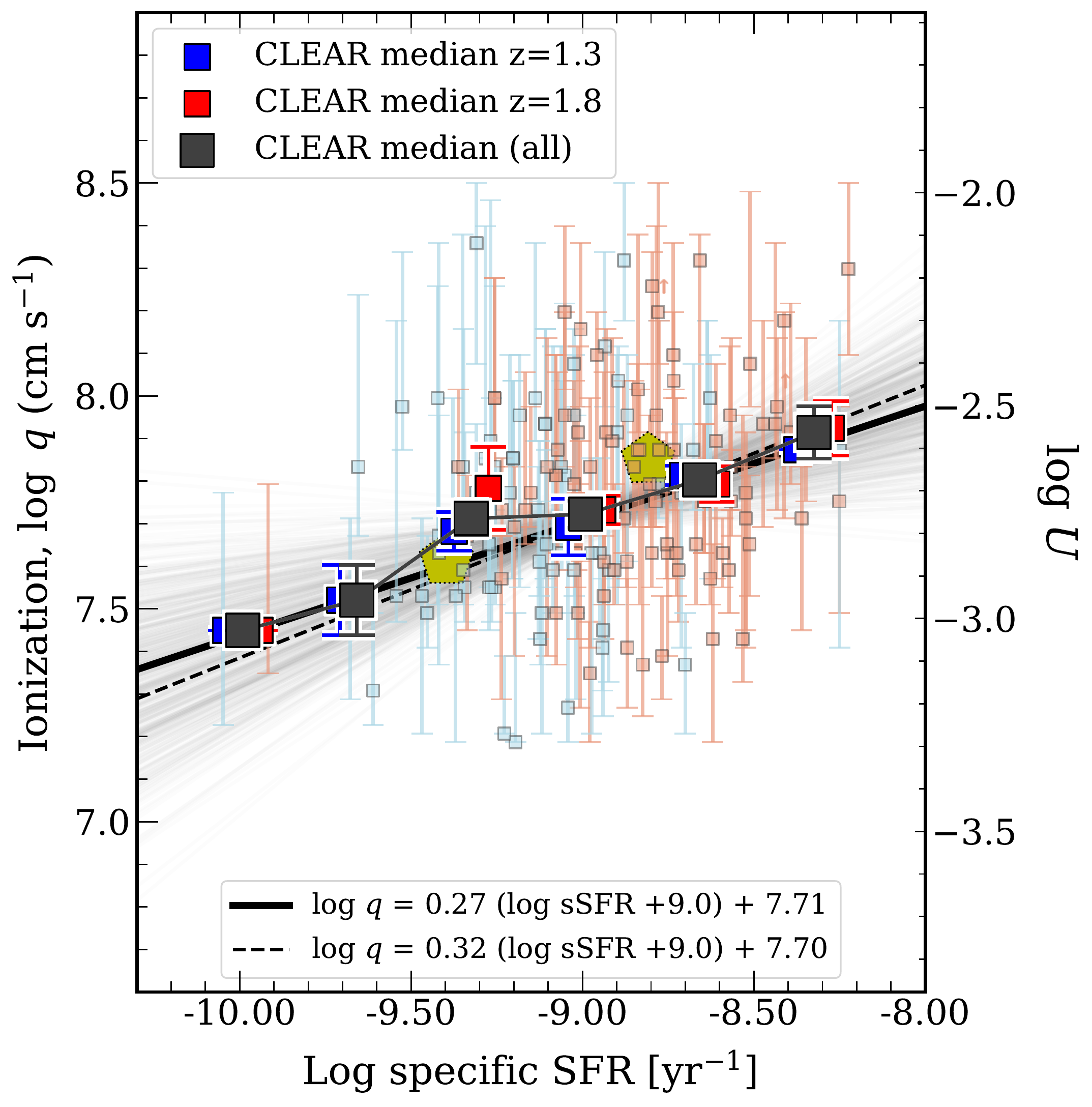}
\caption{ Relation between the specific SFR \textit{derived from SED
fitting} (see Section~\ref{section:sedfit}) and the gas phase
ionization for CLEAR galaxies (see Section~\ref{section:izi}).  The
right axis shows the equivalent dimensionless ionization parameter, $U
\equiv q/ c$.   The CLEAR galaxies are color coded by redshift; large
squares show medians binned by sSFR (error bars show errors on the
median).  The CLEAR sample includes all galaxies with stellar masses,
$\log M/M_\odot = [9.2, 10.2]$.  The dashed line is a linear fit to
the CLEAR galaxies.  The thick, solid line shows the fit to a
subsample with an additional, log sSFR / yr$^{-1}$
$>$ $-9.5$ (the light solid lines show 500 random draws from the
posterior).  The large yellow pentagons show measurements for
higher--mass galaxies at $z\sim 1.5$ ($\log
M_\ast/M_\odot = 10.4-11$, \citealt{kaas18}).   There is evidence of a
correlation between sSFR and gas ionization.  The trend becomes
stronger when considering sSFR as traced by the \hb\ emission
equivalent width (Section~\ref{section:ZQR}).  \label{fig:sSFR_q}}
\end{figure}

To investigate the relation between nebular ionization and sSFR, we
focus on CLEAR galaxies in the stellar mass range, $\log
M_\ast/M_\odot = 9.2-10.2$ (where the majority of our sample resides).
Figure~\ref{fig:sSFR_q} shows the trend between ionization and
sSFR for these galaxies in CLEAR.  We fit a linear relation, 
\begin{equation}\label{eqn:sSFR_q}
\log q = A \times (\log [\mathrm{sSFR}] + 9.0)~ +~ \log q_0,
\end{equation}
where $q$ is the ionization parameter (in units of cm
s$^{-1}$) and the sSFR  is estimated from the stellar
population fits to the galaxy SEDs (see Section~\ref{section:sedfit})
in units of yr$^{-1}$.  Table~\ref{table:sSFR_q} provides the
fitted values and their uncertainties for $A$ and $\log q_0$.

In all cases there is a significant correlation between nebular
ionization and specific SFR.   For fits to the full galaxy sample (not
shown in the figure) and for galaxy subsamples, there is a significant
correlation (see Table~\ref{table:sSFR_q}).  The full galaxy sample
gives a slope of $A=0.50 \pm 0.10$.   We also estimate the
significance using Pearson's correlation coefficient, which gives
$r=0.29$ (with a $p$-value of $8.2\times 10^{-5}$ [$3.9\sigma$ for a
Gaussian distribution]).  Limiting the fit to the subsample of
galaxies with stellar mass in the range $\log M_\ast/M_\odot$ = [9.2,
10.2] yields a weaker correlation coefficient ($r=0.27$), yet still
significant ($p=9.2\times 10^{-4}$).   Restricting the fit to the
subsample with $\log M_\ast/M_\odot$ = [9.2, 10.2] and $\log$ sSFR /
yr$^{-1}$ $< -9.5$) yields a correlation coefficient of only $r=0.21$
(with a $p$-value of $0.013$ [about $2.5\sigma$]).  While still
significant, it is slightly weaker than the relation when considering
the full sample, indicating that the range of galaxy stellar masses
and/or SFR differences  likely account at least partly for the
correlation.  \citet{kaas18} also observe a correlation between
ionization parameter and sSFR with a very similar slope for
measurements from the stacked spectra for $z\sim 1.5$ galaxies at
higher stellar masses.    Motivated by these results, we explore this
relation more below (in Section~\ref{section:ZQR}).

\section{Discussion}\label{section:discussion}

In the previous sections we described correlations between strong-line
emission-line ratios, gas-phase metallicity (\OH), gas ionization
parameter ($\log q$), stellar mass and SFR.  These trends lead to
interesting conclusions about the nature of star-forming galaxies at
high redshift, but these depend on the application of the
photo-ionization models to the data.  In the sections that follow we
discuss these factors, and we discuss the implications this makes for
the evolution of metallicity and ionization in galaxies.

\subsection{Evolution in Mass-Metallicity Relation and Caveats}

The evolution in the MZR (Figure~\ref{fig:MZR}) has been observed for
star-forming galaxies previously,  and has been used to constrain the
evolution of metals and feedback effects in galaxies as a function of
redshift and stellar mass \citep[see, e.g.,][]{maio19}.   The most
recent measurements find that over the mass range $\log M_\ast /
M_\odot = 9.5-10$ the evolution of the gas-phase metallicity evolves
by $\Delta \log \mathrm{O/H} $ = 0.2--0.3~dex from $z\sim 2.3$ to $z
\sim 0.1$ \citep[e.g.,][]{henry21,sand21}.  This is generally
consistent with our findings in CLEAR, where we see that galaxies at
$z=1.35$ (1.90) have metallicity lower by $\Delta\log \mathrm{O/H}$ =
0.25 (0.35) dex compared to SDSS galaxies at $z\sim 0.2$ at stellar
masses, $\log M_\ast/M_\odot = 9.4-9.8$.  

Some comparisons to other studies of the MZR are useful as
they highlight systematics in the analyses.
\citet{sand21} studied the evolution of the MZR to $z > 3$ using
Keck/MOSFIRE observations of $\simeq$450 galaxies.  They find that the
low-mass slope of the MZR is consistent with no evolution, following
$\Delta (\log \mathrm{O/H})/\Delta \log M_\ast \sim 0.30$ from $z\sim
0$ to 3.3. However, the  normalization of the MZR evolves strongly,
with $\Delta (\log \mathrm{O/H})/\Delta z \simeq -0.11$ (at fixed
$M_\ast$) with a small uncertainty ($\simeq$0.02 dex).  They argue
this is consistent with the idea that at fixed stellar mass the galaxy
gas fractions and metal removal efficiencies increase at higher
redshift.  

We find stronger evolution with redshift in the normalization in our
CLEAR and SDSS samples,  $\Delta(\log \mathrm{O/H})/\Delta z = -0.21
\pm 0.02$ dex from $z=0.2$ to $z=1.3$ and $\Delta(\log
\mathrm{O/H})/\Delta z = -0.34 \pm 0.04$ dex from $z=1.35$ to
$z=1.90$.  This is higher than that from \citet{sand21} by $\sim
0.2-0.3$~dex and is likely related to the use of strong-line
metallicity calibrators: \citeauthor{sand21} take average
metallicities derived from multiple strong-line indicators, while we
derive metallicities from the same set of lines fit to the MAPPINGS V
photoionization models. \added{\citet{sand21} also use models that allow for increasing $\alpha$/Fe, which could account for some of the offset in the evolution.} We take this difference as an estimate of the
systematics in the strong-line calibrators  \citep[e.g., see][who show
the normalization of different calibrators can vary by as much as
0.7~dex]{kewley08}. 

The study of \citet{henry21}  is more similar to the analysis
presented here.  They used measurements from stacked \hst/grism
spectra of more than 1000 galaxies at $z\sim 1.3-2.3$ (along with
higher quality spectra of $\sim$50 individual galaxies) to measure the
evolution of the MZR and derived gas metallicities using strong lines
(\oii, \hb, \oiii, \ha+\nii) with the calibration from
\citet{curti17}.
%
%
They derive O/H abundances at $z\sim 1.3 - 2.3$ that are consistent
with those from \citet{sand18} at $z\sim 2.3$, yielding $\OH = 8.28
\pm 0.02$ ($8.37^{+0.01}_{-0.02}$) for $\log M_\ast/M_\odot = 9-9.5$
($9.5-10$).   Comparing to the $z\sim 0.1$ results from
\citeauthor{curti17} they find that the normalization of the MZR
evolves by $\Delta \log \mathrm{O/H} \simeq 0.3$ dex at a fixed mass
$\log M_\ast/M_\odot = 9-9.5$.  This yields an evolution of $\Delta
\log (\mathrm{O/H}) / \Delta z \simeq -0.2$~dex from $z\sim 0.1$ to
1.8, consistent with the evolution we derive from our analysis the
CLEAR and SDSS samples.   The fact that \citeauthor{henry21} measure
the same \textit{absolute} gas-phase metallicity as \citet{sand18} at
$z\sim 2$ but different evolution again indicates that systematics are
important, primarily in the absolute normalization of the MZR, both at
high and low redshifts.

\begin{deluxetable}{lcccc}
\tabletypesize{\footnotesize}
\tablecolumns{3}
\tablewidth{0pt}
\tablecaption{Fitted parameters for the linear relation between
specific SFR and nebular ionization for CLEAR galaxies using
Equation~\ref{eqn:sSFR_q}. \label{table:sSFR_q}}
%
\tablehead{
\colhead{sample}  & 
\colhead{slope, $A$} & \colhead{$\log q_0$}  &
\colhead{$r$\tablenotemark{\footnotesize $\dag$}} &
\colhead{$p$-value\tablenotemark{\footnotesize $\ast$}}}
\startdata
\vspace{4pt}
Full sample & 0.50 $\pm$ 0.10  &
7.67 $\pm$ 0.02 & 0.29  & 8.2$\times$10$^{-5}$  \\ [4pt]
$\log M_\ast/M_\odot$=$[9.2,10.2]$ & 0.31 $\pm$ 0.10  &
7.70 $\pm$ 0.02  & 0.27 & 9.2$\times$10$^{-4}$ \\[8pt]
$\log M_\ast/M_\odot$=$[9.2,10.2]$ & \multirow{2}{*}{0.27 $\pm$ 0.12}  &
\multirow{2}{*}{7.71 $\pm$ 0.02} & \multirow{2}{*}{0.21} &
\multirow{2}{*}{0.013} \\ 
\multicolumn{1}{c}{\& $\log \mathrm{sSFR} / \mathrm{yr}^{-1} > -9.5$}  \\
\enddata
\vspace{-4pt}
\tablenotetext{$\dag$}{Pearson's Correlation Coefficient}\vspace{-4pt}
\tablenotetext{\ast}{Two-side probability ($p$-value) of correlation
  occurring by chance (assuming $x$ and $y$ are drawn from independent Gaussian distributions).}
\end{deluxetable}

In summary, our CLEAR results add to the evidence that the MZR evolves
by $\Delta (\log\mathrm{O/H)}/\Delta z \sim 0.2$ dex from $z=0.2$ to
$z=1.3$ and that this rate of evolution in the MZR may increase at
higher $z$ (we observe $\Delta (\log\mathrm{O/H)}/\Delta z  = 0.3$
from $z=1.35$ to $z=1.90$).    The \textit{strength} of our result is
that we have used the same set of photoionization models to convert
the strong emission line ratios to constraints on the gas--phase
metallicities and ionization parameters, both for galaxies in CLEAR
($z \sim 1-2.3$) and SDSS (at $z\sim 0.2$).    However, this strength
is also a \textit{weakness} because using the same set of
photoionization models assumes that the physical conditions in the low
redshift galaxies (SDSS) are the same as in the higher-redshift
galaxies (CLEAR).   For example, we use the MAPPINGS V models with
higher pressure, $P_e/k = n_e T_e = 10^{6.5}$~cm$^{-3}$ K, for the
high-redshift galaxies given the expected temperatures and particle
density of the \ion{H}{2} regions ($T_e \sim 10^4$~K and $n_e \simeq
300$~cm$^{-3}$, see \citealt{sand16a,kaas17,strom17,runco21}), where
the electron density is an order of magnitude larger than observations
at $z\sim 0.1$ \citep[e.g.,][]{sand16a}. As discussed in
\citet{kewley19}, adopting a lower pressure for the SDSS galaxies
would decrease the metallicity of high-$Z$ galaxies at fixed R$_{23}$.
This is  evident in Figure~\ref{fig:r23_Z}, which shows difference
between our calibration and that from \citealt{kewley19} (who use
$\log P_e/k \simeq 5$), and in
Appendix~\ref{section:appendix} which shows the differences between
our MZR and those from \citet[who use the calibration from
\citealt{curti17}, which is similar to \citealt{kewley19}]{henry21}.
Adopting lower pressure for the SDSS galaxies would lower the
metallicity of high-$Z$ galaxies by $\sim 0.1$~dex (see also
\citealt{sand21}).  Understanding these potential sources of
systematic bias are crucial to have an accurate measurement
of the redshift evolution in the MZR.

It is therefore remarkable that even in lieu of the systematic
uncertainties, the MZR evolution in Figure~\ref{fig:MZR} shows
agreement between measurements from SDSS and CLEAR from different
studies
%
%
Therefore, while there is work needed to understand the impacts of the
assumptions about the metallicity calibrations from the strong
emission lines, there appears to be some consensus on the absolute
evolution of the MZR.

\subsection{Evolution in the Mass--Ionization Relation and
Caveats}\label{section:MQRc}

\begin{figure*}[ht]
\centering
\includegraphics[height=0.45\textwidth]{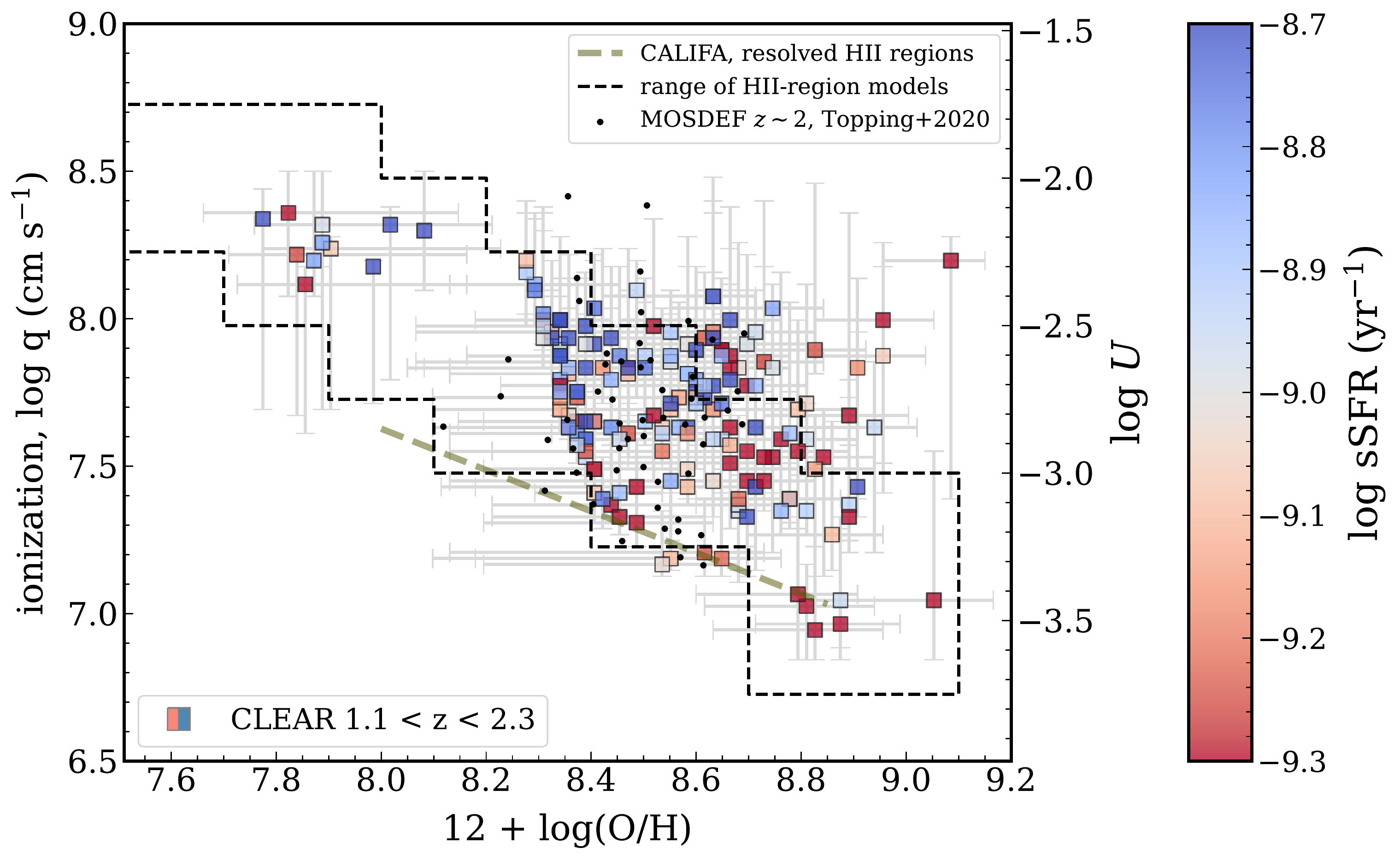}
\vspace{12pt}
\includegraphics[height=0.35\textwidth]{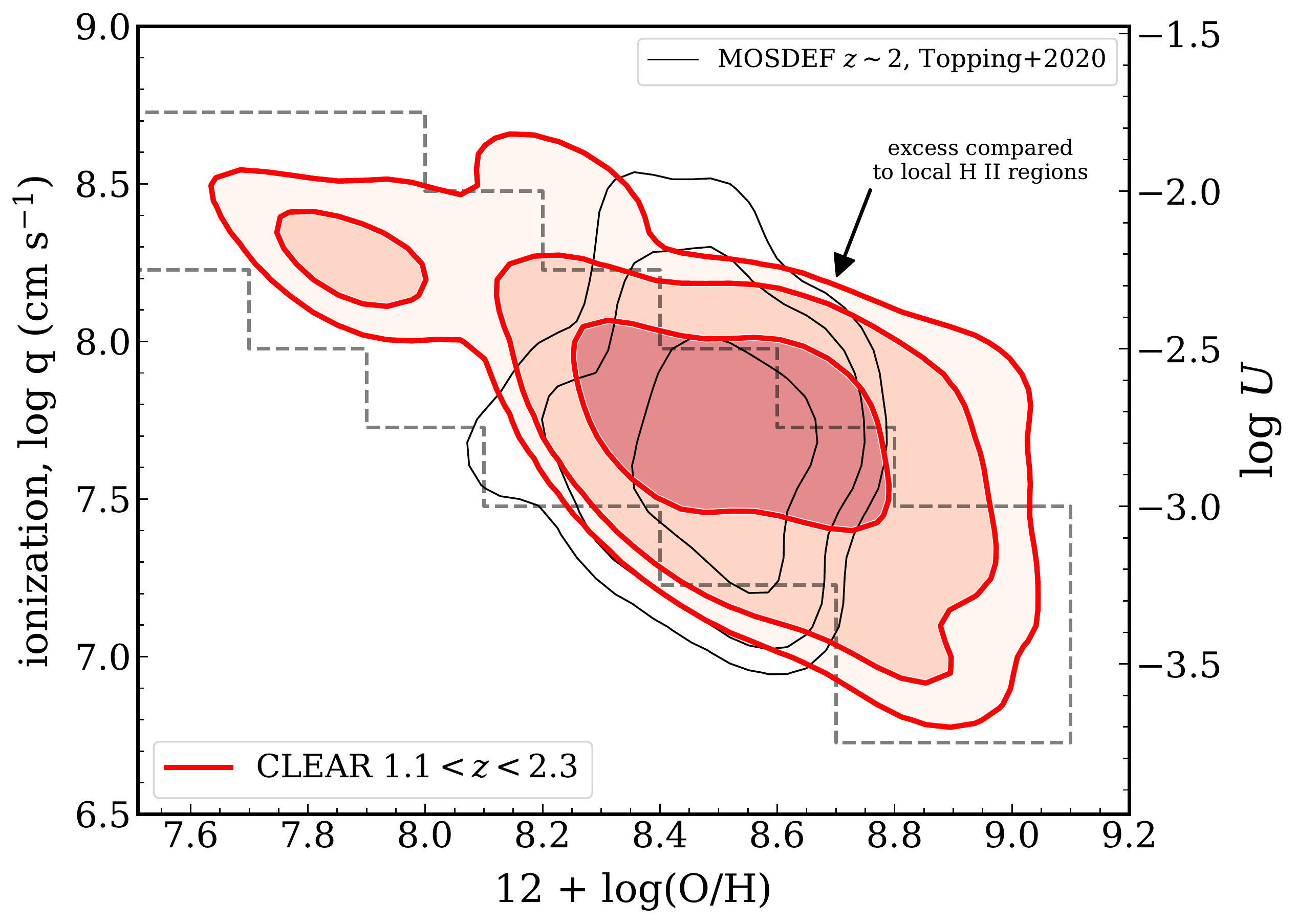}
\caption{Gas-phase metallicities and ionization parameters for CLEAR
  galaxies.    The
top panel shows the data with large squares denoting the CLEAR
galaxies (colored by specific SFR). Smaller points show measurements
from MOSDEF \citep{topp20}.  The dashed lines show the range of parameter
space covered by nearby \ion{H}{2} regions \citet{perez14}.  The thick
solid lines shows measurements for resolved \ion{H}{2} regions in
CALIFA galaxies \citep{espi22}.   The  The right axis shows the
equivalent dimensionless ionization parameter, $U \equiv q / c$. The
bottom panel shows the same information with a KDE derived
distribution (for both the MOSDEF and CLEAR
samples).  \label{fig:logOH_logU}}
\end{figure*}

Figure~\ref{fig:MQR} shows evidence for evolution in the MQR for
star-forming galaxies.    We quantify the evolution in the MQR over
the redshift range $z\sim 0.2$ to $z\sim 2.3$ using our analysis of
the galaxies in CLEAR and SDSS.  At fixed stellar mass, $\log
M_\ast/M_\odot = 9.6$, the differential evolution between the SDSS
galaxies and CLEAR galaxies corresponds to $\Delta \log q / \Delta z
\simeq 0.2$~dex from $z\sim 0.2$ to $z\sim 1.9$. 

Previous studies have seen evidence for this evolution, primarily in
terms of an increase at high redshifts in the strength of
emission-line ratios that are sensitive to the ionization parameter.
\citet{sand18} measured the evolution of O$_{32}$ as a function of
stellar mass and redshift, finding a $0.5$~dex evolution in O$_{32}$
from $z\sim 2.3$ to $\sim 0$ for galaxies at $\log M_\ast/M_\odot
\simeq\,9.5$.  This corresponds to a change in ionization parameter of
$\Delta(\log q) \approx 0.4$~dex (using 
Equation~\ref{eqn:o32_q}).  \citet{kaas18} measured the ionization
parameter from stacked spectra of massive galaxies ($\log
M_\ast/M_\odot \simeq 10.4-11$) at $z\sim 1.5$.  They showed these
exhibit an increase in the ionization parameter of $\simeq$0.4~dex at
fixed stellar mass from $z\sim 0.2$ to 1.5.    This is consistent with
the evolution we measure in CLEAR to $z\simeq 1.9$ and to $z\sim 0.2$
from SDSS. 

Figure~\ref{fig:MQR} (bottom-left panel) also shows that at fixed
stellar mass galaxies in CLEAR span a range of specific SFR from log sSFR
$\sim -9.3$ to $-$8.7, which corresponds to about a factor of $\sim
4$.    Qualitatively, the figure shows that galaxies with higher
sSFRs favor higher ionization parameters.  This is reminiscent of the
well-studied trend between the MZR and SFR in that galaxies with lower
metallicity favor higher SFRs at fixed stellar mass reported in other
studies \citep[see][]{maio19,sand21,henry21}.

Is the MQR a consequence of the evolution in the MZR, or of the
evolution of the SFR-MZR relation?  To investigate this,
Figure~\ref{fig:MQR} (bottom-right panel) shows that at fixed stellar
mass, higher metallicity galaxies in CLEAR generally have lower
ionization parameters, but the scatter is large.  The same Figure
shows that higher ionization parameters correlate with
lower-metallicities.  However, we argue the situation is more nuanced:
in Section~\ref{section:ZQR} we show that galaxies in CLEAR, when
matched in stellar mass and metallicity, show a wide range of
ionization parameter.   Therefore, the metallicity is not solely the
cause of the evolution in the MQR.

\begin{figure*}[th]
\centering
\includegraphics[width=0.6\textwidth]{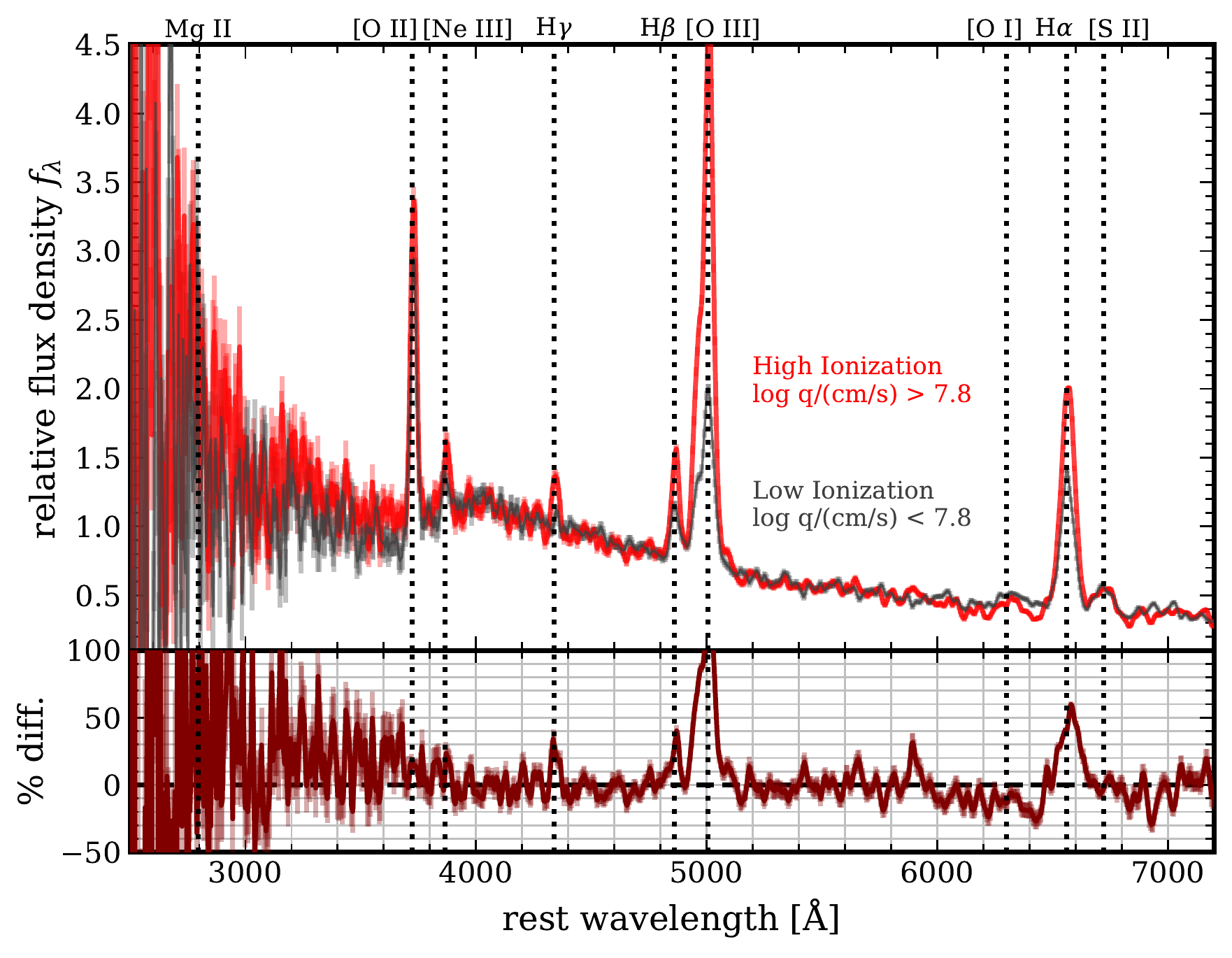}
\vspace{12pt}
\includegraphics[width=0.85\textwidth]{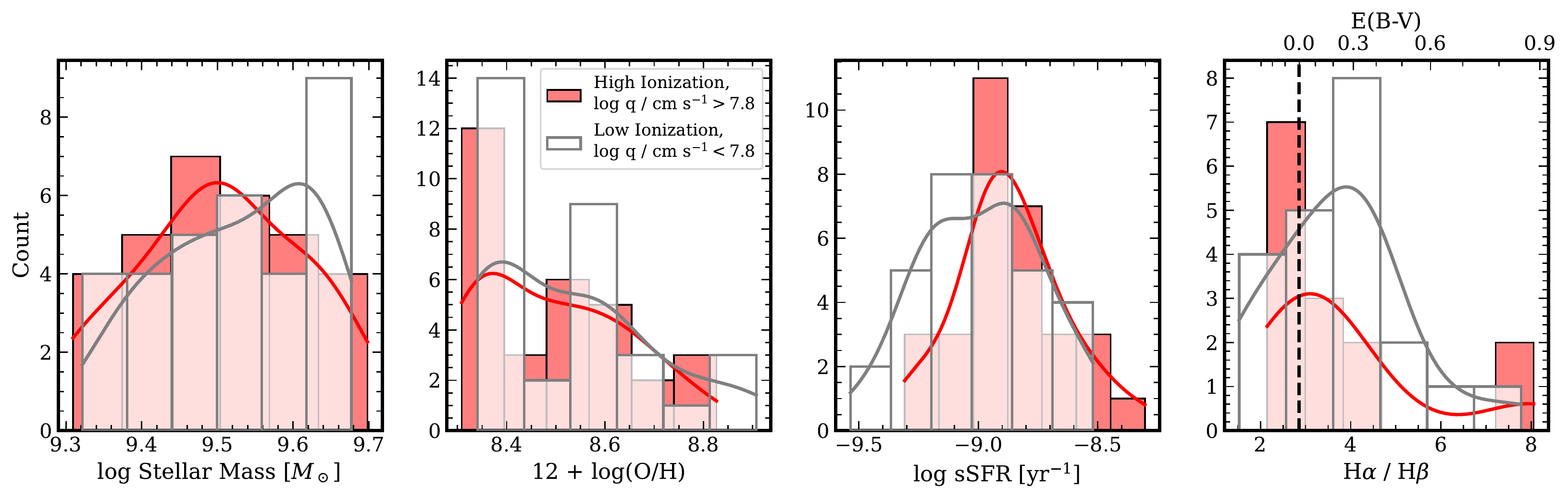}
\caption{Comparison of the properties of galaxies in the
high-ionization subsample ($\log q / [\mathrm{cm~s^{-1}}] > 7.8)$
compared to the low-ionization subsample ($\log q /
[\mathrm{cm~s^{-1}}] < 7.8)$.  The top panel shows the stacked G102
and G141 spectra for each sample (as labeled).  Prominent emission
features are indicated.   All galaxies were selected to have stellar
mass $9.3 < \log M/M_\odot < 9.7$ and $12 + \log(\mathrm{O/H}) >
8.3$.  The bottom panels show distributions of the galaxy properties
in each subsample from this selection, including stellar masses and
specific SFRs (derived from SED fitting), the gas-phase metallicities
(derived from the emission line analysis), and extinction (derived
from the observed \ha/\hb\ ratios).   The stellar masses and
metallicities are similar between the samples, while the sSFR shows differences
(see discussion in the text).  \label{fig:stack_hi_lo}}
\end{figure*}

\subsection{Evolution of the Gas Metallicity and Ionization}\label{section:Zlogq}

Figure~\ref{fig:logOH_logU} shows the relation between the ionization
parameter ($\log q$) and gas--phase metallicity (\OH) for the CLEAR
galaxies.   The figure compares these results to those from MOSDEF
\citep{topp20} at $z\sim 2$ and to measurements for individual
\ion{H}{2} regions in galaxies \citep{perez14,espi22}.   The
metallicity and ionization of the CLEAR galaxies follow the physical
parameter space seen in these other samples.  Therefore, the general
relation between ionization and metallicity seems to describe
star-formation in galaxies at both low and high redshifts.


Inspecting Figure~\ref{fig:logOH_logU} more closely, there is also
evidence that the distribution of \OH\ for the $z > 1$ galaxies skews
to higher ionization parameters at fixed metallicity (at $\OH \gtrsim
8.4$).  This is more apparent in the KDE distributions: compared to
the nearby \ion{H}{2} regions (compared to both \citealt{perez14} and
\citealt{espi22}).   The implication is that galaxies at higher
redshift favor higher ionization parameters, \citep[see
also][\added{see also  \citealt{sand20} who discuss how a decrease in the effective temperature of stars can lead to an increase in ionization parameter.}]{kewley15,strom17,kaas18,topp20,runco21}  Here we see this
trend exists for high redshift galaxies even at fixed metallicity and
stellar mass. 

 What causes the increase in ionization parameter?  To understand the
answer, we divided a sample of galaxies into bins of $\log q$.  We
selected galaxies at $1.1 < z < 2.3$ from our CLEAR sample in a
narrower range of stellar mass, $9.3 \leq \log M_\ast/M_\odot \leq
9.7$, and we required that the galaxy gas-phase
metallicities be $\OH > 8.3$ (to remove low-metallicity galaxies from
consideration).  We then divided this sample into a high-ionization
subsample of galaxies with $\log q > 7.8$ (31 galaxies) and a
low-ionization subsample of galaxies with $\log q < 7.8$ (32
galaxies).    The cuts in stellar mass and ionization have the effect
of making both the stellar-mass distribution and metallicity
distribution approximately the same for the high-- and low--ionization
subsamples (see below, and Figure~\ref{fig:stack_hi_lo}).  So, we are
able to correlate trends between ionization and other galaxy
properties.

We then stacked the WFC3 G102 and G141 dust-corrected spectra for
these subsamples (following the methods in
Section~\ref{section:stacking}).  Figure~\ref{fig:stack_hi_lo} shows
these stacked spectra for the ``low-ionization'' ($\log q < 7.8$) and
``high-ionization'' subsamples ($\log q > 7.8$).   The differences in
the spectra are immediately clear (pun intended).  The two samples
have very different emission-line intensities, while the stellar
continua of the two stacks are nearly identical.
Figure~\ref{fig:stack_hi_lo} also shows the relative differences
between the two spectra.  The strongest difference is in the \oiii\
emission, which is nearly twice as strong at the peak of the line for
the high-ionization galaxies.  The Balmer lines are also stronger by
$\approx$30\% stronger at the peak of the lines for the
high-ionization galaxies.  The \oii\ and \neiii\ lines both show an
increase in the higher-ionization subsample, while the \sii\ lines do
not.
%

The bottom row of panels in Figure~\ref{fig:stack_hi_lo} shows the
distributions of stellar mass, gas-phase metallicity, SED-derived
sSFR, and the \ha/\hb\ ratios for the high- and low-ionization galaxy
subsamples.   Both subsamples were selected to have roughly the same
stellar mass and gas-phase metallicities, and the panels show there
are no substantive differences.  Formally, a Kolmogorov-Smirnov (KS)
test and a Mann-Whitney-$u$ (MWU) test applied to the distributions
return $p$ values of 0.60 and 0.34, respectively, for stellar mass,
and 0.45 and 0.71, respectively, for metallicity.    However, the
specific SFR distributions show evidence they are different.  The KS
and MWU tests return $p$ values 0.07 and 0.08, respectively.  This is
apparent in Figure~\ref{fig:stack_hi_lo} as a shift in the sSFR
distribution of the high-ionization subsample toward higher sSFR.  The
final (right) panel of Figure~\ref{fig:stack_hi_lo} shows the
distribution of the Balmer decrement, but this includes only those
galaxies with $z < 1.63$ for which \ha\ is detected (14 and 23
galaxies in the high-- and low-ionization subsamples, respectively).
The mode of the high-ionization subsample is consistent with
$E(B-V)=0$, and the low-ionization subsample shows slightly higher
attenuation with a mode of $E(B-V)=0.2$, but both have low overall
dust attenuation.  The KS and MWU tests yield $p$ values of 0.45 and
0.69, respectively, providing no evidence they are drawn from
different parent distributions (but this is in part because of the
smaller sample sizes). 

\begin{figure}[t]
\centering
\includegraphics[width=0.48\textwidth, clip=true, trim={6pt, 0, 0, 0}]{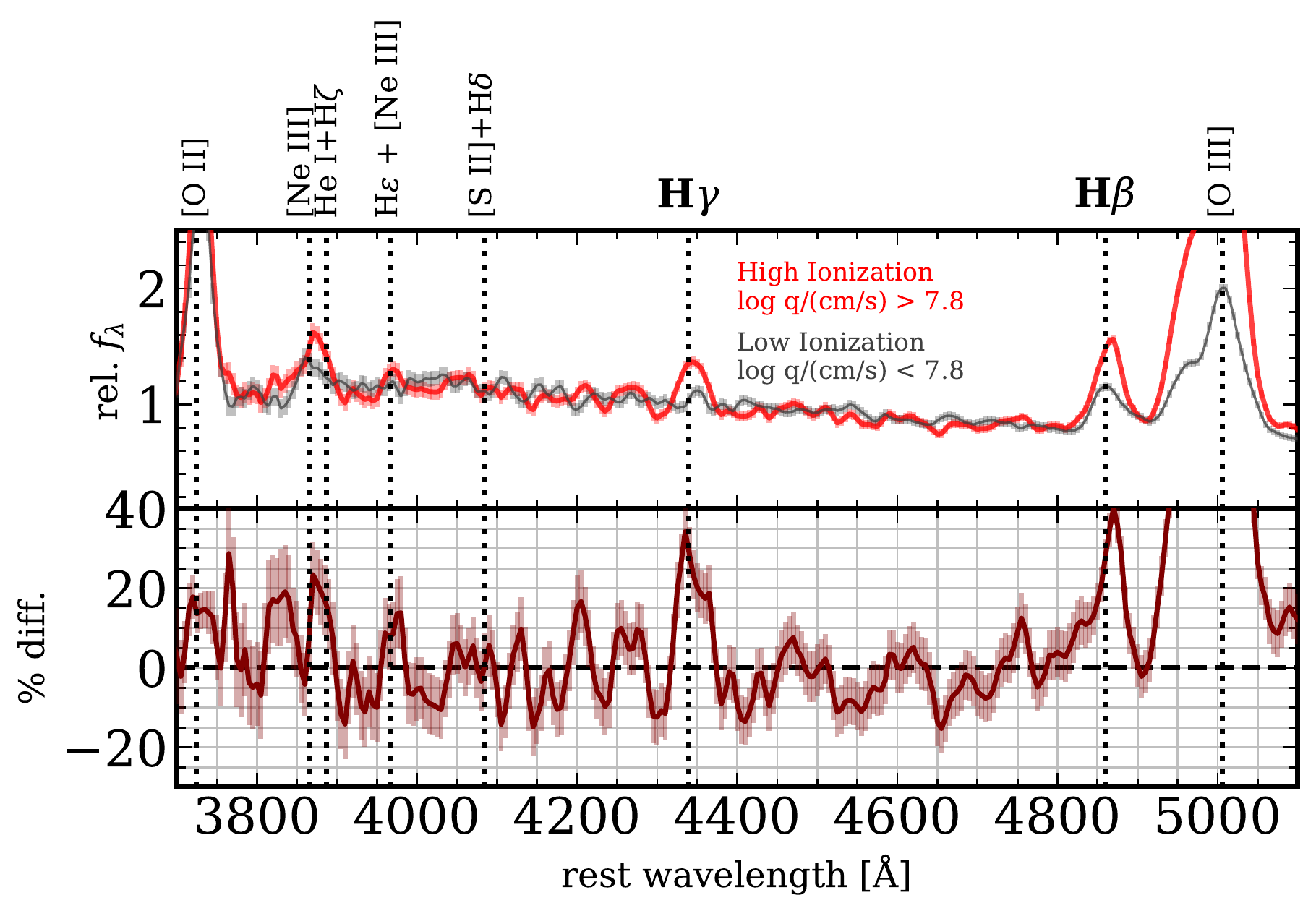}
\caption{Zoom-in on spectroscopic features in the high-ionization
subsample ($\log q / [\mathrm{cm~s^{-1}}] > 7.8)$ compared to those in
the low-ionization subsample ($\log q / [\mathrm{cm~s^{-1}}] < 7.8)$
from Figure~\ref{fig:stack_hi_lo}.  The top panel shows both stacked
spectra.  The bottom panel shows the percent difference.  The \editone{\hb\ and H$\gamma$} features (labeled in bold) are the most
isolated at the resolution of the G102 and G141 grisms (i.e., they are
not blended with other possible features).  These lines all show
excess emission of 20--30\% in the high-ionization subsample compared
to the low-ionization subsample,  This is evidence that
higher-ionization galaxies have higher specific
SFRs.   \label{fig:stack_hi_lo_zoom}}
\end{figure}

\begin{figure*}[t]
\centering
\includegraphics[height=0.4\textwidth, clip=true, trim={0pt, 0, 0,
  0}]{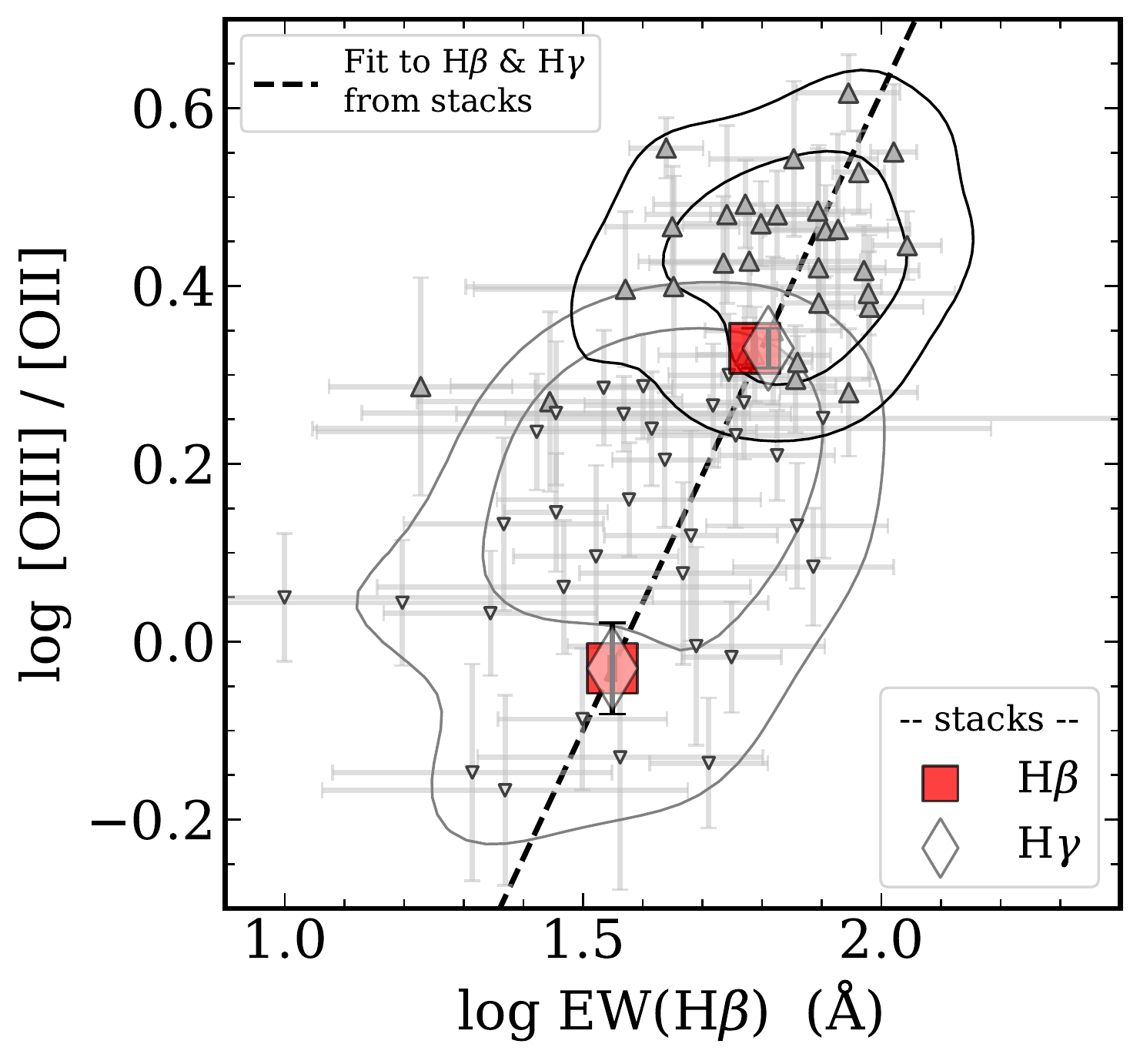}
\includegraphics[height=0.4\textwidth, clip=true, trim={0pt, 0, 0, 0}]{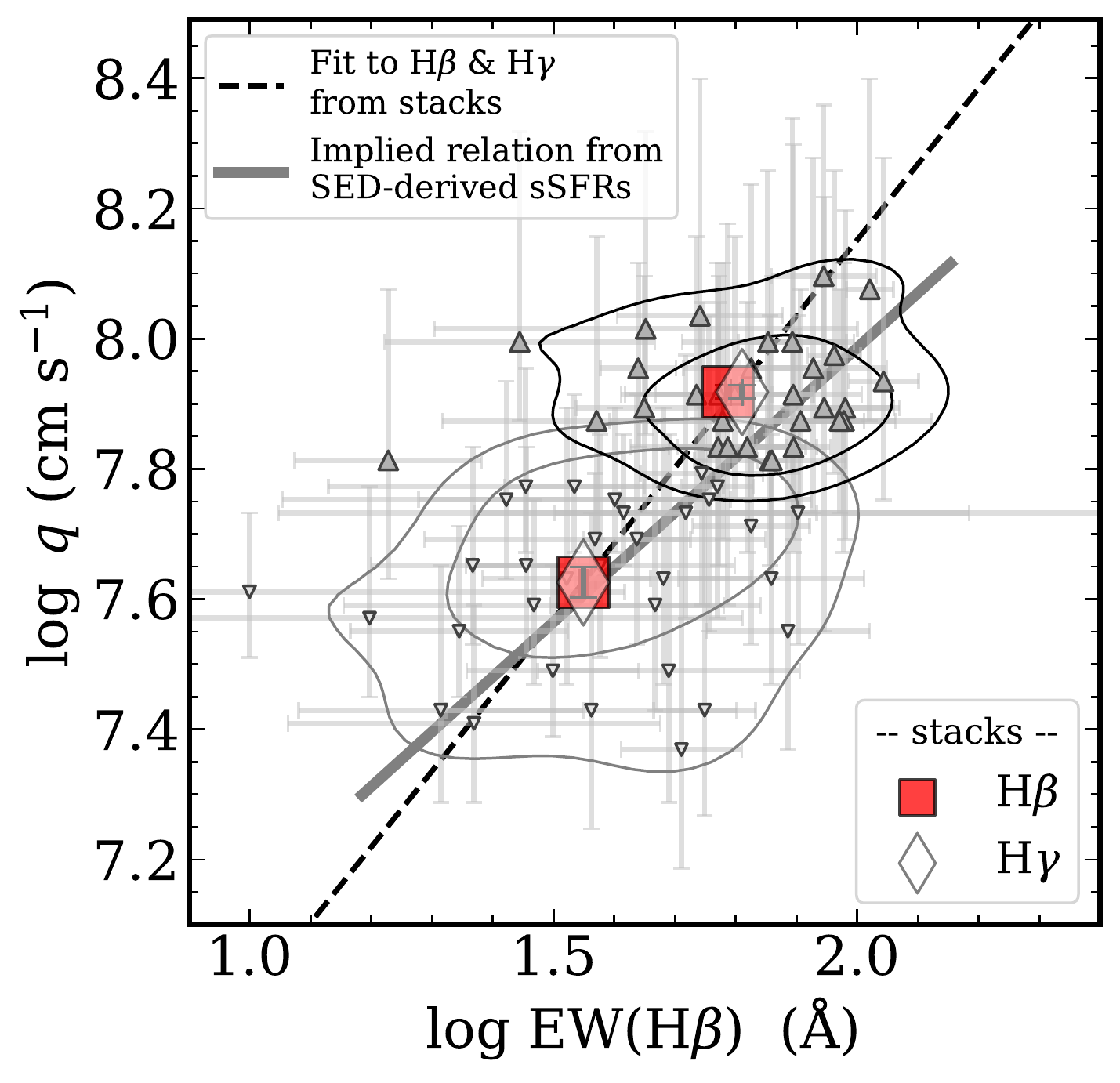}
\caption{Comparison between \hb\ equivalent width, EW, and gas
ionization parameter,$\log q$, for CLEAR galaxies in the
high-ionization ($\log q > 7.8$) and low-ionization subsamples ($\log
q < 7.8$).  The left panel shows the measured \oiii/\oii\
emission-line ratio against the H$\beta$ equivalent width measured
from the stacks (large red squares),  and individual points (gray
downward / upward triangles show galaxies in the low / high ionization
subsamples, respectively).  The contours encompass 50 and 84
percentiles for each sample.  The other large symbols show values
derived from H$\gamma$, scaled by the Case-B ratio
to \hb.  The dashed line is a linear fit to the stacked measurements.
The right panel shows the same distribution against the ionization
parameter $\log q$.  Because the Balmer emission EW scales with
specific SFR, the plots show strong evidence between ionization (and
ionization parameter) and specific SFR galaxies at $z\sim 1-2$.
%
%
\label{fig:balmer_q}}
\end{figure*}

  Why then do the galaxies in the high-ionization subsample have such
stronger emission lines when all other galaxy properties appear to be
similar?    The key may be in the fact that see a correlation between
these (broad-band-derived) specific SFR and ionization parameters (see
Figure~\ref{fig:sSFR_q}).  Figure~\ref{fig:stack_hi_lo_zoom} shows the
region around \editone{\hb\ and \hg} in the stacked spectra of the
high- and low-ionization subsamples.  We focus on these \editone{two} lines as
they are the strongest Balmer emission lines that are unblended at the
WFC3 G102 and G141 spectral resolution (e.g., \ha\ is blended with
\nii; \added{H$\epsilon$ is blended with [\ion{Ne}{3}] 3968}).
\editone{Both} of these Balmer lines (\hb\ \editone{and \hg})
are stronger by $\approx$20-30\% at the peak of the lines in the
high-ionization galaxies. 

We measured the strength of these emission lines using \ppxf\  applied
to the stacked spectra (following Section~\ref{section:ppxf}).
Table~\ref{table:ppxf} reports the emission line equivalent widths.
For all the Balmer lines, the high-ionization subsample has
substantially stronger lines, with the equivalent width of \hb\
\editone{and \hg} higher by a factor of $\simeq$1.7--1.9 than those of
the low-ionization subsample.  Similarly, Figure~\ref{fig:balmer_q}
here shows that correlation between both the Balmer-line equivalent
width (EW) against the ionization parameter ($\log q$) and against the
\oiii/\oii\ ratio.  In the Figure we show the \hb\ EW as it is
measured and we have increased the \hg\ \editone{EW} according to the
theoretical \hb/\hg\ Case-B ratio.

The stronger Balmer emission \added{(as evidenced by higher EW)} in
the high-ionization galaxies indicates that they have a higher
production rate of H-ionizing photons \added{(at fixed galaxy stellar
mass)}.  This translates to higher specific SFRs: for example,
\citet{reddy18} show that $\log$ EW(\hb) $=$ $0.32 \times \log$ sSFR
for galaxies at $z\sim 2$ in MOSDEF.  This translates to a factor of
two higher sSFR for the high-ionization galaxies in our sample
compared to the low-ionization galaxies.  We illustrate this in
Figure~\ref{fig:balmer_q}, which shows the \hb\ EW compared to the
both the O$_{32}$ ratio and the ionization parameter ($\Delta \log q$)
in the high-ionization galaxies relative to the low-ionization
galaxies.  The lines in the Figure show  fits, 
\begin{eqnarray}
  &\log \mathrm{O}_{32} = 1.4 \log \mathrm{EW(H\beta)} - 2.3, \\
  &\log q = 1.2 \log\mathrm{EW(\hb)} + 5.8,
\end{eqnarray}
for the stacked
spectra from the CLEAR subsamples).  Using the relation from
\added{\citet{reddy18}}, we find that $q \sim \mathrm{sSFR}^{0.4}$ for
our CLEAR galaxy samples.  Figure~\ref{fig:balmer_q} also shows how
these relations compare to the $\log q$--sSFR (derived from broad-band
fitting, see Figure~\ref{fig:sSFR_q}, and converted to EW(\hb) using
this relation), which is similar. The conclusion is that the
ionization parameter and specific SFR are correlated for galaxies at
fixed stellar mass and gas-phase metallicity.  

Our result for the CLEAR galaxies is consistent with findings from
some previous studies.  \citet{kewley15} show strong evidence that the
\oiii/\oii\ ratio is correlated with the \hb\ EW in their sample of
$0.2 < z < 0.6$ galaxies (but they did not differentiate by stellar
mass).  This is also seen by \citet{kaas18}, who argue that the
high ionization parameters for galaxies in their $z < 0.3$ sample are
driven by the ratio of the number density of hydrogen--ionizing
photons to the gas density, and not by (lower) metallicities.  In a
study of SDSS galaxies, \citet{kashino19} similarly find that the
ionization parameter is predominantly controlled by the specific SFR,
with $q \sim \mathrm{sSFR}^{0.43}$ (converting to our units),
approximately equal to the relation for the CLEAR galaxies. 

\begin{figure*}[th]
\centering
\includegraphics[width=0.7\textwidth]{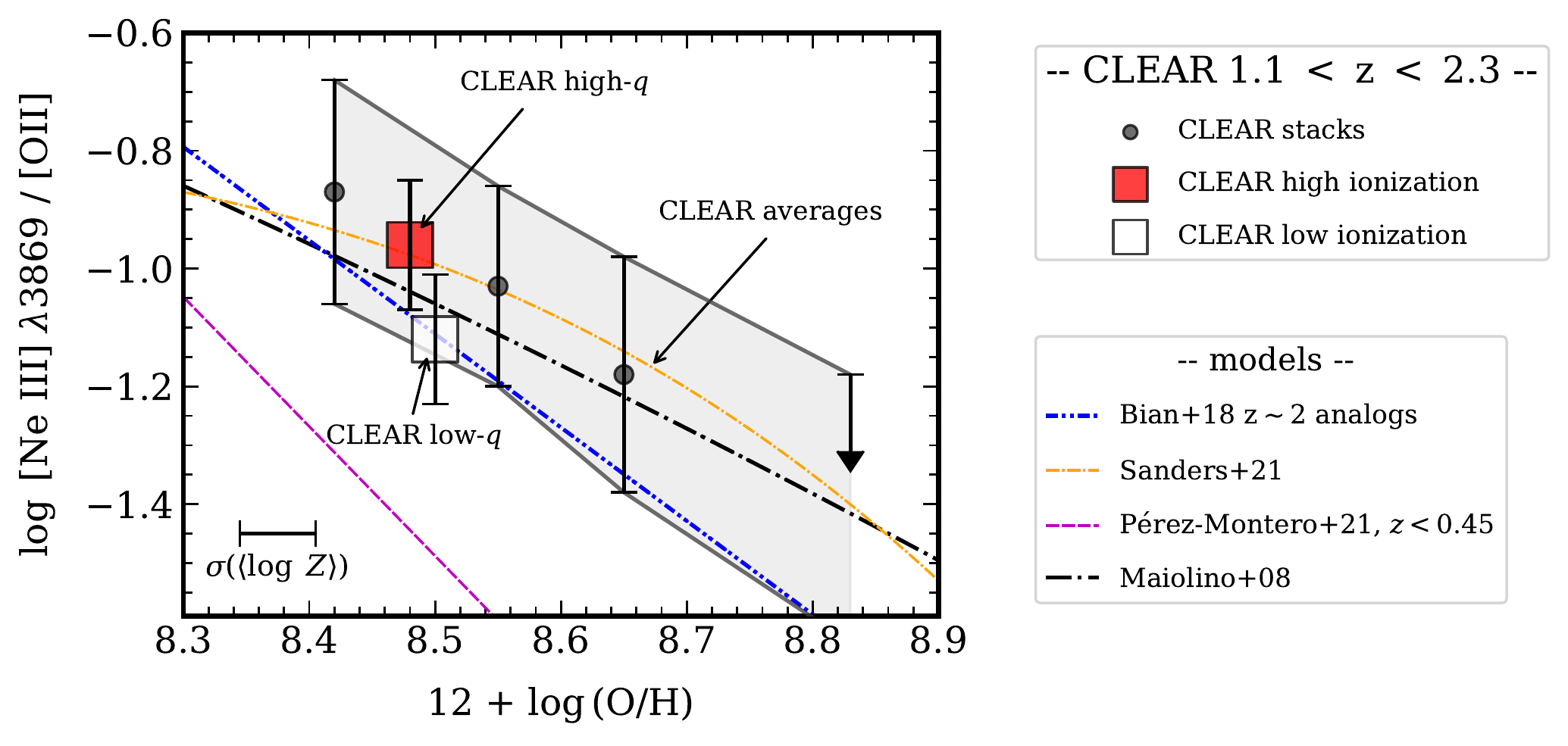}
\caption{Comparison of \neiii/\oii\ emission line ratios and gas-phase
metallicity (12 $+\log$ O/H) derived for CLEAR galaxies and published
relations from the literature.   The data points show mean
metallicities for the CLEAR samples and emission line ratios derived
from the stacked spectra.  The filled circles show values derived for
galaxies in bins of stellar mass (see
Section~\ref{section:spectra_stack}).  The large squares show values
derived for mass-matched samples at high and low ionization ($\log q /
[\mathrm{cm~s^{-1}}] > 7.8$ and $< 7.8$, respectively, see
Section~\ref{section:Zlogq}), where these is no indication the
\neiii/\oii\ ratios are substantially different in these samples (as
would be expected if they had significantly different metallicities,
see e.g., \citealt{sand21}).   The models also include analogs of
$z\sim 2$ galaxies \citep{bian18}, calibrations to SDSS at $z\sim 0$
and higher redshifts
\citep{maio08,curti17,perez21}.  \label{fig:line_ratios}}
\end{figure*}

\subsection{Interpretation of the Specific SFR--Ionization
  Relation at $1.1 < z < 2.3$}\label{section:ZQR}

We are now ready to address the question, what is the origin of the
correlation between gas ionization parameter ($q$) and specific
SFR?   First, we consider several explanations for the correlation,
including selection and physical effects.

\subsubsection{Is a selection effect in stellar mass responsible?}

There is an established correlation between stellar mass and
ionization (see Figure~\ref{fig:MQR}).  This is observed in multiple
studies, including at high redshift \citep[see, e.g., the recent study
of][]{strom22}.   However, in our analysis the stellar masses of the
high-ionization and low-ionization subsamples in
Figure~\ref{fig:stack_hi_lo} are nearly identical (see also the
discussion in Section~\ref{section:Zlogq}).  We constructed both
subsamples to have $9.3 < \log M_\ast/M_\odot < 9.7$, where the median
stellar masses of both the high and low-ionization samples are both $\langle
\log M_\ast/M_\odot \rangle = 9.5$
Figure~\ref{fig:stack_hi_lo}).  If the stellar mass was the sole
driver, Eq.~\ref{eqn:MQR} predicts there should be a difference in
the in ionization parameter of only $\Delta \log q$ = 0.03~dex.  This
is much smaller than the $\Delta \log q \approx 0.3$~dex measured
between the two samples (see Figure~\ref{fig:sSFR_q}). Therefore, we
disfavor stellar mass as the primary driver. 

   \subsubsection{Is a selection effect in gas-phase metallicity
responsible? }

Figure~\ref{fig:logOH_logU} shows there is a correlation between
gas-phase metallicity (\OH) and ionization parameter ($\log q$).   The
distributions of metallicities  of the high- and low-ionization are
again highly similar (Figure~\ref{fig:stack_hi_lo}, and
discussion in Section~\ref{section:Zlogq}).  A
potential selection effect arises because the \OH\ values are derived
from what surmounts to the O$_{32}$ and R$_{23}$ line ratios.
R$_{23}$ is famously double valued, and the fitting method we employed
could skew the metallicities to higher values (as the line ratios are
unable to get ``over the R$_{23}$ hump'' to lower-metallicity
values). Indeed, this accounts for the ``ridgeline'' in the modes of
the metallicity--ionization-parameter distributions seen in
Figure~\ref{fig:logOH_logU}.  (This is in essence asking if our modeling
prefers the ``high-$Z$'' solution.)  

To test if this potential bias affects the correlation, we used the
\neiii\ $\lambda$3868 / \oii\ ratio an as alternative tracer of
gas-phase metallicity.   Multiple studies have demonstrated the
correlation between \neiii/\oii\ and \OH\
\citep[e.g.,][]{maio08,bian18,perez21,sand21}.   \neiii/\oii\ is
particularly useful as the lines are close in wavelength (mitigating
extinction effects) and are accessible at rest-frame optical
wavelengths. While the \neiii\ ratio is often too weak to detect in
individual galaxies ($\log \neiii/\oii \sim -1$, see
Table~\ref{table:ppxf}, although see \citealt{back21}), it is
detectable in the stacked spectra.  

Figure~\ref{fig:stack_hi_lo} shows that the \neiii/\oii\ ratios are
similar in the stacked spectra of the low- and high-ionization
samples. This is supported by the measured \neiii/\oii\ ratios
measured from the stacked spectra in Table~\ref{table:ppxf}.
Figure~\ref{fig:line_ratios} shows the \neiii/\oii\ against \OH\
derived from the stacked spectra (compared to the CLEAR stacks for the
full sample in bins of stellar mass).  Based on the \neiii/\oii\
ratios, there is only a small difference between the values for the
high- and low-subsamples ($\log \neiii/\oii\ \simeq -1.0$ for both),
and this is likely attributed to the difference in the ionization (see
\citealt{kewley19,maio19}).  This is important as the \neiii\ fluxes
are not used in the measurements of \OH.  Indeed, we measured
\neiii/\oii\ in a stacked spectrum of CLEAR galaxies selected to have
low metallicity, \OH $<$ 8.2, which shows they have much higher
ratios, $\log \neiii/\oii\ \simeq -0.6\pm 0.1$.  This is consistent
with other studies of low-$Z$ galaxies at these redshifts
\citep[e.g.,][]{sand21}, and this is substantially higher than that
observed in the high- and low-ionization subsamples.  Therefore, we
disfavor the explanation that metallicity is responsible for the
correlation between $\log q$ and sSFR.

\subsubsection{Physical connections between the specific SFR and
ionization parameter}

There are physical reasons to expect a correlation between the
specific SFR and the ionization parameter.   Such mechanisms need to
relate the sSFR (or specifically, the production of ionizing photons
per unit mass) to the number density of gas particles.  \citet{brin08}
show this can account for the emission-line ratios of low redshift
galaxies (from SDSS, at $z\sim 0.1$).  They
further argue the elevated ionization parameters result from higher
gas densities possibly combined with higher escape fractions of
Hydrogen-ionizing photons ($f_\mathrm{gas}$).  Similar correlations
between specific SFR and ionization parameter (and/or line ratios such
as \editone{\oiii/\oii}) have been observed previously at low and high redshifts
\citep[e.g.,][]{nakajima14,kewley15,sand16a,bian16,kaas18,kashino19}.
In our CLEAR sample we see that this correlation persists even when
accounting for stellar mass and metallicity (see
Figures~\ref{fig:sSFR_q} and \ref{fig:stack_hi_lo}). Therefore, there
is strong evidence for a physical connection between specific SFR and
ionization.

A physical connection exists in the case where the \ion{H}{2} regions
are radiation bounded, where the ionization parameter can be written
as \citep{char01,nakajima14,stas15},
\begin{equation}\label{eqn:charlot01}
  q \approx \left(\ \frac{3Q\ n_\mathrm{H}\
      \epsilon^2\ \alpha_B^{2}  }{4\pi}\ \right)^{1/3}, 
\end{equation}
where $Q$ is the rate of Hydrogen ionizing photons, $\alpha_B$ is the
case-B Hydrogen recombination coefficient, $n_\mathrm{H}$ is the
number density of Hydrogen, and $\epsilon$ is the volume filling
factor of the gas.\footnote{The counter-intuitive relation, $q \sim
n_\mathrm{H}^{1/3}$, results from the fact that the size of the
\ion{H}{2} region depends on gas density.  In a radiation-bounded
nebula, the volume of the \ion{H}{2} region (defined by a
``St\"omgren'' sphere with volume $\sim R^3$, for radius $R$) is the
point where the rate of ionization ($Q$) is balanced by rate of
recombination (which scales as $n_e n_\mathrm{HII} \approx
n_\mathrm{H}^2$ for an ionized gas).  The radius therefore declines
with density as $R \sim n_\mathrm{H}^{-2/3}$.  The (volume-averaged)
ionization parameter, $q$, scales as the ionizing flux, $Q /(4\pi
R^2)$, divided by $n_\mathrm{H}$.  This leads to the relation where $q
\sim 1 / (R^{2} \times n_\mathrm{H})$.
%
%
Replacing $R$ by $n_\mathrm{H}^{-2/3}$ yields the relation that $q$
increases with gas density as $q \sim n_\mathrm{H}^{4/3} /
n_\mathrm{H} \sim n_\mathrm{H}^{1/3}$.   }     At present there is
little evidence for changes in $\epsilon$ in any environment
\citep[see,][]{brin08}, and we do not consider this further here.  The
ionization parameter therefore depends on $Q$ and $n_H$, \editone{which observations show are higher at
higher redshift (for the case of $n_H$; see e.g., \citealt{sand16a} and
\citealt{strom17}) or are expected to be higher at higher redshift (for
the case of $Q$, given the observed bluer colors and
lower metallicities of galaxies; see, e.g., \citealt{shivaei18}).}
This is a major driver of the
evolution in the MQR (Fig.~\ref{fig:MQR}).  The geometry of \ion{H}{2}
regions and star-forming nebula in distant galaxies may be more
complicated.  This could include nebulae that are either ``matter
bounded'' (also called ``density bounded''), where the region of
ionization (the Str\"omgren sphere) extends beyond the nebula (see,
e.g., \citealt{nakajima14}) and/or there denser clouds with a low
covering fraction \citep[e.g.,][]{naidu22}.  We consider these effects
in the discussion that follows.

\textit{Production rate of ionizing photons}. $Q$ depends on the age
of the stellar population, metallicity, and relative number of massive
stars (typically those with $\gsim 10~\msol$).  Therefore, to increase
$Q$ requires either an increase in the relative number of massive
stars and/or evolution in the properties of the stellar populations.
One obvious possibility is that there is a change in the IMF: either
an increase in the upper-mass cutoff, or high-mass slope. Currently
evidence for such a changes are  inconclusive
\citep[e.g.,][]{finkelstein11,narayanan13}, so we do not consider it
further.  However, it will remain important to consider for future
studies.

Several studies modeling the UV stellar continua and optical emission
line strengths have advocated for higher $Q$ values in some
star-forming galaxies at both high and low redshifts, particularly
when the metallicity of the stellar continuum is low \citep[$Z \sim
0.1~\zsol$;  e.g.,][]{topp20,berg21,oliv21}.  This may be exacerbated
by super-Solar $\alpha$/Fe ratios, which lead to harder ionizing
spectra (at fixed O/H).   \editone{There} is growing evidence for
increased $\alpha$/Fe in galaxies at $z\sim 2$
\citep[e.g.,][]{stei16a,strom18,shap19,sand20,topp20,runco21}. \editone{One
limitation of our work is that the MAPPINGS models do not currently
provide for variations in $\alpha$/Fe, and it will important to test
how this impacts the trends in our dataset. Nevertheless,} for this to
drive our observations, $\alpha$/Fe would need to vary with sSFR at
\textit{fixed} [O/H]\added{, which would itself be an important
discovery}.   Although our current dataset is insufficient at present,
this may be testable in the future by simultaneously modeling the
rest-UV continuum spectra and measurements of the optical emission
lines of galaxies (e.g., \citealt{topp20,oliv21}). 

\textit{Gas Density}.  There is considerable evidence that the gas
densities of galaxies at $z\sim 2$ are considerably higher than at low
redshift \citep[e.g.,][]{shirazi14,sand16a,acha19,runco21}.  This been
argued to contribute to the elevated emission-line ratios in
high-redshift galaxy samples \citep{brin08,nakajima14,kewley19b}.  For
changes in the gas density to account for our observations, these also
must correlate with the specific SFR.   One physical explanation for
such a correlation is an extension of the Kennicutt--Schmidt relation,
where the SFR density scales with gas density, $\rho_\mathrm{SFR} \sim
\rho_\mathrm{gas}^N$ \citep[see, e.g.,][]{bacc19}.  The exponent
scales from $N=1.4$ (for the classical Kennicutt-Schmidt relation) to
2 depending on timescales over which star-formation occurs (see,
\citealt{madore77,larson81,kenn07,krum14,elme15,bola17,bacc19}).
There are observations of galaxies at low redshift that indicate
higher O$_{32}$ ratios for galaxies with smaller sizes \citep[implying
higher densities, e.g.,][]{ji21}.  Regardless, this provides a
physical connection between the strength of H-recombination lines
(tracing the SFR density) and the ionization parameter (tracing gas
density in \ion{H}{2} nebula). 

A connection between specific SFR  and ionization parameter is
therefore expected, and borne out in our CLEAR dataset.   Testing if
the gas density drives this correlation will be
feasible using observations of emission lines whose ratios are
dependent on gas density.   For example, the ratio of the S$^+$
emission lines, [\ion{S}{2}] $\lambda 6716$/[\ion{S}{2}] $\lambda
6731$ varies by a factor of $\sim$2 as the gas density changes from
$n_e \simeq 10$ to 1000 cm$^{-3}$ \citep{ryden21}.  Currently,
\hst/WFC3 grism data have insufficient resolution to deblend \sii\ for
galaxies at $z \lsim 1.5$.  However, future spectroscopy at higher
spectral resolution would be able to test for density variations
directly. 

\textit{Ionizing Radiation Escape Fractions}.  Several studies of
low-redshift galaxies ($z\lsim 0.1$) show evidence that the escape
fraction H-ionizing photons, $f_\mathrm{esc}$, is correlated with the
ionization parameter, for example where $f_\mathrm{esc} \propto
(\mathrm{O}_{32})^2$ \citep{chis18,izotov18,ji21}.   This means that
the geometry of the \ion{H}{2} regions is likely an important factor,
where non-uniform covering fractions and/or ``matter-bounded'' nebulae
can lead to a higher escape fractions, $f_\mathrm{esc} > 0$
\citep[e.g.,][]{nakajima14,naidu22}.  At a \textit{fixed ionization
parameter}, a non-zero $f_\mathrm{esc}$ requires a higher number
density of ionizing photons relative to the gas density.  Using a
suite of simulations, \citet{giam05} showed that increasing the escape
fraction of Hydrogen-ionizing photons from  $f_\mathrm{esc}$=0 to 0.5
increases $\log q$ by as much as 1 dex.  This can be boosted in the
case of a ``density--'' (or ``matter--'') bounded nebula, where the
O$^{+}$ region in the \ion{H}{2}--region is truncated while the
O$^{++}$ region, located closer to the ionizing source, is not
\citep[e.g.,][]{nakajima14}.  This would lead to a potential
correlation between $f_\mathrm{esc}$ and \oiii/\oii\ ratio (see also
\citealt{brin08} and \citealt{kashino19}).  There is recent evidence
for this in the findings of \citet{naidu22}, who argue that high
\oiii/\oii\ line emission of $z\sim 2$ galaxies corresponds to a
short-lived phase where massive stars have cleared sightlines in the
nebular gas.  This produces a geometry with a smaller covering
fraction of dense gas, permitting higher $f_\mathrm{esc}$ (and higher
\oiii/\oii).    As pointed out by \citet[and others
above]{brin08}, such an increase in $f_\mathrm{esc}$ at higher
redshift is an important factor in interpreting the emission-line
ratios of high redshift galaxies.  There are an increasing number of
observations (either direct measurements or inferences) that the
escape fraction of H-ionizing photons is higher at high redshift
\citep{vanz16,vanz18,debarros16a,shap16,bian17,ji20,begley22}, where
inferences for reionization require $f_\mathrm{esc}\sim 0.1-0.2$
\citep[e.g.,][]{ouchi09,robe13,bouw15b,ishi18,fink19}.

The fact that we observe a correlation between ionization
parameter and specific SFR in our CLEAR galaxies then implies there
\editone{may} be a correlation between specific SFR and
$f_\mathrm{esc}$. \added{This prediction stems from the fact that the
ionization parameter and escape fraction are expected to correlate
\citep{nakajima14}, and we observe a correlation between specific SFR
and $\log q$ (Figure~\ref{fig:sSFR_q}) and between \hb\ EW and $\log q$
(Figure~\ref{fig:balmer_q}).  } \citet{brin08} speculated that such a
correlation could account for offsets in the emission-line ratios of
SDSS galaxies (see also \citealt{nakajima14} and \citealt{kashino19}).
\editone{Currently there are only weak
constraints on a correlation between specific SFR or ionization parameter
and} $f_\mathrm{esc}$ at high redshifts, though some recent
observations are suggestive (but sample sizes remain small, e.g.,
\citealt{bass19} and \citealt{naidu22}).  Future observations of the
escape fraction in high-$z$ galaxies like those in our CLEAR sample
(either with UV spectroscopy or imaging, see \citealt{siana10}),
probing a large range in specific SFR, \oiii/\oii\ ratio would provide
the data to test this directly.

We summarize this section by postulating that multiple effects likely
contribute to the correlation between specific SFR and ionization
parameter.   While the current dataset is insufficient to
differentiate these effects, we favor the explanation that both an
increase in the gas density ($n_H$) and the escape fraction of
H-ionizing fractions ($f_\mathrm{esc}$) drive the trend in specific
SFR and ionization parameter, as these have the strongest physical
bases.  This will be testable directly with future data. 
%

\section{Summary}\label{section:summary}

In this \textit{Paper}, we have used data from CLEAR, including deep
spectroscopy from the \hst/WFC3 IR grisms, combined with broad-band
photometry, to study the stellar populations, ionization and chemical
abundances in a sample of $\simeq$ 200 star-forming galaxies at
$z\sim~1.1-2.3$.   At these redshifts the grisms measure emission from
strong nebular lines in the rest-frame optical, including \oii\
$\lambda\lambda$3727, 3729, \oiii\ $\lambda\lambda$4959, 5008, and
\hb, which are sensitive to physical conditions in the galaxies'
star-forming regions (at $z \lsim 1.5$ the data also cover \ha+\nii\
$\lambda\lambda$6549, 6585 and \sii\ $\lambda\lambda$6718, 6732).
From the emission-line measurements, we derive constraints on the
oxygen abundances (\OH) and ionization parameters ($\log q$) of the
nebular gas using predictions from updated photoionization models
(MAPPINGS V, \citealt{kewley19b}).  Our findings can be summarized as
follows.
\begin{enumerate}
  \item  The CLEAR galaxies show evolution in the gas-phase
metallicity--mass relation (MZR) and ionization--mass relation (MQR)
as a function of redshift.    Galaxies with lower stellar masses have
lower gas-phase metallicities and higher ionization parameters.
Compared to low-redshift samples ($z\sim 0.2$) at fixed stellar mass,
$\log M_\ast/M_\odot = 9.4-9.8$, the CLEAR galaxies at $z=1.35$ (1.90)
have lower metallicity, $\Delta(\log Z)$ = 0.25 (0.35) dex, and higher
ionization, $\Delta(\log q)$ = 0.25 (0.35) dex.   We provided updated
analytic calibrations between O$_{32}$ and the ionization parameter
(Eq.~\ref{eqn:o32_q}), and R$_{23}$ and metallicity
(Eq.~\ref{eqn:r23_Z}). We further provide analytic fits for the MZR
and MQR at $z=1.3$ and $z=1.8$ (Eq.~\ref{eqn:MZR} and
Eq.~\ref{eqn:MQR}).   Our measurement of the MZR is consistent with other
derivations from the literature \citep[e.g.,][]{henry21,sand21}, and
motivates future studies of the systematics in calibrations of
metallicities from strong-line indicators.  

\item  We find evidence that the ionization parameter $q / c = U$, is
correlated with galaxy specific SFR,  where $q \sim
\mathrm{sSFR}^{0.4}$, derived from changes in the strength of galaxy
\hb\ EW, where alternatively $\Delta \log q = 1.2 \log
\mathrm{EW}(\hb)$ (see Eq.~\ref{eqn:sSFR_q}).    This persists for
galaxies at fixed mass and metallicity implying there is an underlying
physical connection (see Fig.~\ref{fig:balmer_q}). We consider multiple
physical effects for the origin of this relationship.  We conclude
that the higher gas ionization parameter is a consequence of
increasing gas density, $n_H$, (and/or variable gas geometry),
combined possibly with an increasing H-ionizing photon escape
fraction, $f_\mathrm{esc}$, and all of these must increase with
specific SFR.   

\end{enumerate}

Importantly, this work shows the capabilities that space-based
observations using near-IR slitless spectroscopy have for engaging in
these kinds of scientific studies.  This provides a complementary
picture to ground-based telescopes (where the capabilities from space
provide improvements in wavelength coverage and stable/uniform flux
sensitivity and calibration).   Future work will expand these types of
studies over vastly larger datasets (in the case of \textit{NGRST})
and wavelength space (in the case of \jwst).  
 
Lastly, the work here has important considerations for observations of
galaxies at even higher redshift.  There is mounting evidence that
galaxies near and into the EoR (e.g., $z \gsim 6$) have higher
specific SFRs \citep[e.g.,][]{salmon15,santini17}, Balmer emission and
\oiii\ emission-line ratios
\citep[e.g.,][]{smit15,robe16,matt17,stark17,reddy18,hutc19,endsley21}.
It seems likely therefore that the physical manifestation between
specific SFR and these emission line ratios will be similar in such
galaxies and those in our CLEAR data.  This will be directly testable
with forthcoming observations from \textit{JWST}.

\section*{$~$} 
  
  We thank our colleagues on the CLEAR team for their valuable
conversations and contributions.  We wish to acknowledge Yingjie
Cheng, Alison Coil, Alaina Henry, \added{Ryan Sanders, Alice Shapley}
and Allison Strom for helpful comments, feedback, and
\editone{suggestions} (and clarifications).  \added{We also thank the
anonymous referee for a helpful report, which greatly improved the
quality and clarity of this work.}   This work is based on data
obtained from the Hubble Space Telescope through program number
GO-14227.  Support for Program number GO-14227 was provided by NASA
through a grant from the Space Telescope Science Institute, which is
operated by the Association of Universities for Research in Astronomy,
Incorporated, under NASA contract NAS5-26555.  This work is supported
in part by the National Science Foundation through grant AST 1614668.
\editone{CP thanks Marsha and Ralph Schilling for generous support of
this research.} VEC acknowledges support from the NASA Headquarters
under the Future Investigators in NASA Earth and Space Science and
Technology (FINESST) award 19-ASTRO19-0122, as well as support from
the Hagler Institute for Advanced Study at Texas A\&M University.  IJ
acknowledges support from NASA under award number 80GSFC21M0002. This
work was supported in part by NASA contract NNG16PJ33C, the Studying
Cosmic Dawn with WFIRST Science Investigation Team.   The authors
acknowledge the Texas A\&M University Brazos HPC cluster and Texas
A\&M High Performance Research Computing Resources (HPRC,
\url{http://hprc.tamu.edu}) that contributed to the research reported
here.    This work benefited from generous support from the George
P. and Cynthia Woods Mitchell Institute for Fundamental Physics and
Astronomy at Texas A\&M University.

Funding for the Sloan Digital Sky  Survey IV has been provided by the
Alfred P. Sloan Foundation, the U.S.  Department of Energy Office of
Science, and the Participating  Institutions. 

SDSS-IV acknowledges support and  resources from the Center for High
Performance Computing  at the  University of Utah. The SDSS  website
is www.sdss.org.

SDSS-IV is managed by the  Astrophysical Research Consortium  for the
Participating Institutions  of the SDSS Collaboration including  the
Brazilian Participation Group,  the Carnegie Institution for Science,
Carnegie Mellon University, Center for  Astrophysics | Harvard \&
Smithsonian, the Chilean Participation  Group, the French
Participation Group,  Instituto de Astrof\'isica de  Canarias, The
Johns Hopkins  University, Kavli Institute for the  Physics and
Mathematics of the  Universe (IPMU) / University of  Tokyo, the Korean
Participation Group,  Lawrence Berkeley National Laboratory,  Leibniz
Institut f\"ur Astrophysik  Potsdam (AIP),  Max-Planck-Institut  f\"ur
Astronomie (MPIA Heidelberg),  Max-Planck-Institut f\"ur  Astrophysik
(MPA Garching),  Max-Planck-Institut f\"ur  Extraterrestrische Physik
(MPE),  National Astronomical Observatories of  China, New Mexico
State University,  New York University, University of  Notre Dame,
Observat\'ario  Nacional / MCTI, The Ohio State  University,
Pennsylvania State  University, Shanghai  Astronomical Observatory,
United  Kingdom Participation Group,  Universidad Nacional Aut\'onoma
de M\'exico, University of Arizona,  University of Colorado Boulder,
University of Oxford, University of  Portsmouth, University of Utah,
University of Virginia, University  of Washington, University of
Wisconsin, Vanderbilt University,  and Yale University.


\facilities{ \hst\ (NASA/ESA)}

\software{ AstroPy \citep{astropy13}, EAZY \citep{bram08}, \texttt{eazy-py} \citep{eazy-py}, \texttt{grizli} \citep{grizli},
  Interactive Data Language (IDL), 
  \texttt{IZI} \citep{blanc15},  \texttt{linmix}  \citep{kelly07},
  NumPy \citep{numpy11}, PANDAS \citep{reback2020pandas}, \ppxf\
  \citep{capp17}, SciPy \citep{scipy20}, Seaborn \citep{Waskom2021} }

\restartappendixnumbering

\begin{appendix}
  
\section{Modeling Gas-Phase Metallicity and Ionization with \oii,
  \oiii, \hb, \ha+\nii, and \sii}\label{section:appendix}

In addition to the lines used above in the analysis (i.e., \oii, \hb,
and \oiii), the WFC3 grism spectra cover \ha+\nii\ for galaxies in our
sample with $z \lsim 1.6$ and \sii\ to $z\lsim 1.5$ (where the \ha\
and \nii\  lines are blended at the resolution of the WFC3 grism data;
see Figures~\ref{fig:examplespec} and \ref{fig:allstack}).     For
those galaxies, we consider how including these emission lines, along
with \oii, \hb, and \oiii, impacts the constraints on the galaxy
gas-phase metallicities and ionization parameter.   

We selected the subsample of 87 galaxies that have spectroscopic
coverage of \ha+\nii\ (selected from the 196 galaxies in the full
sample, see Section~\ref{section:selection}).
We also include \sii\ emission if the line is present in the data
(for galaxies at $z < 1.5$).  We then repeated our analysis
using \izi\ with the same set of photoionization models (see
Section~\ref{section:izi}), but using the full set of emission line fluxes
(and uncertainties):  \oii, \hb, \oiii, and \ha+\nii\ (and \sii, if present).

\begin{figure*}
  \centering
  \includegraphics[width=0.4\textwidth]{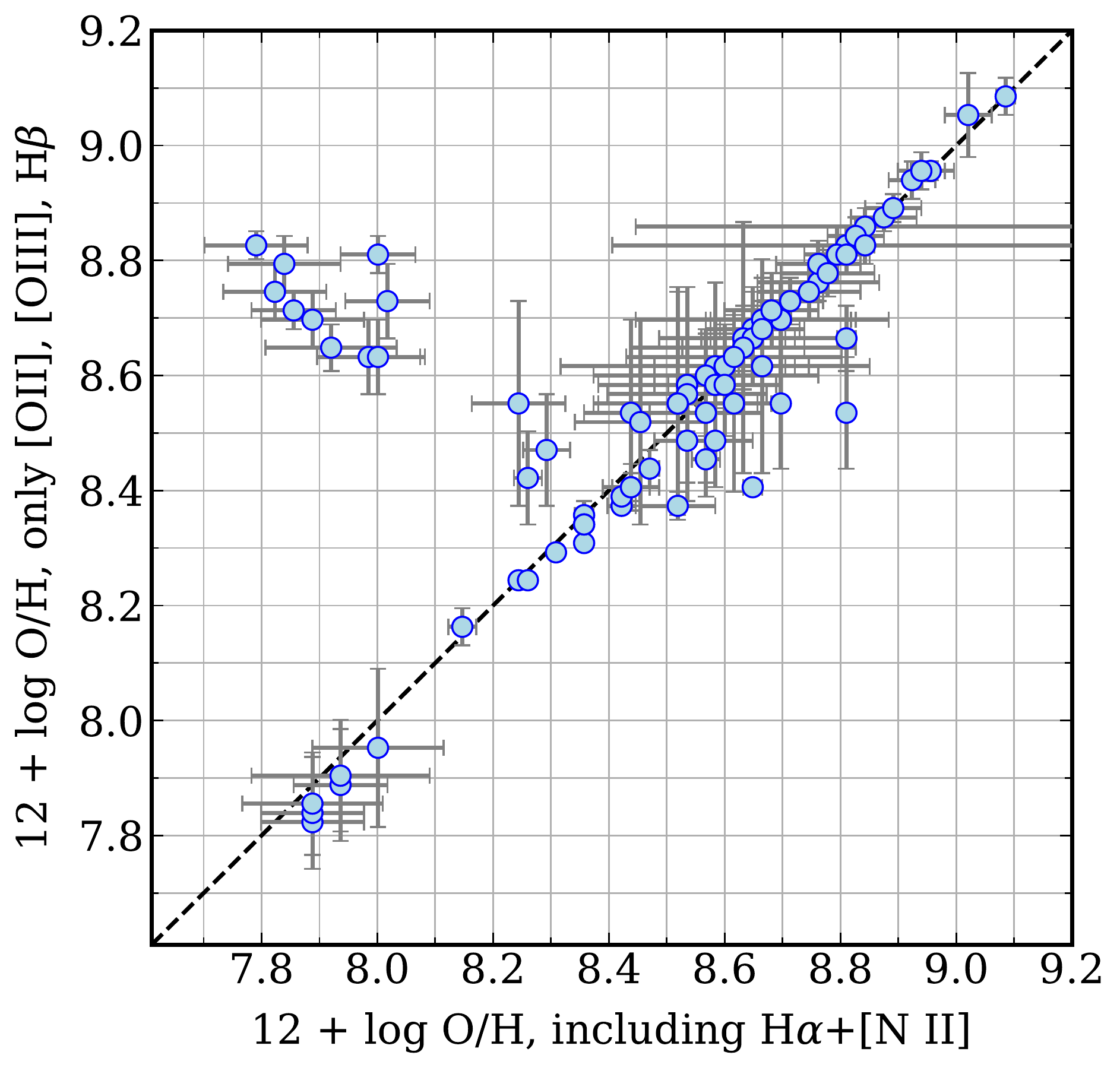}
  \includegraphics[width=0.4\textwidth]{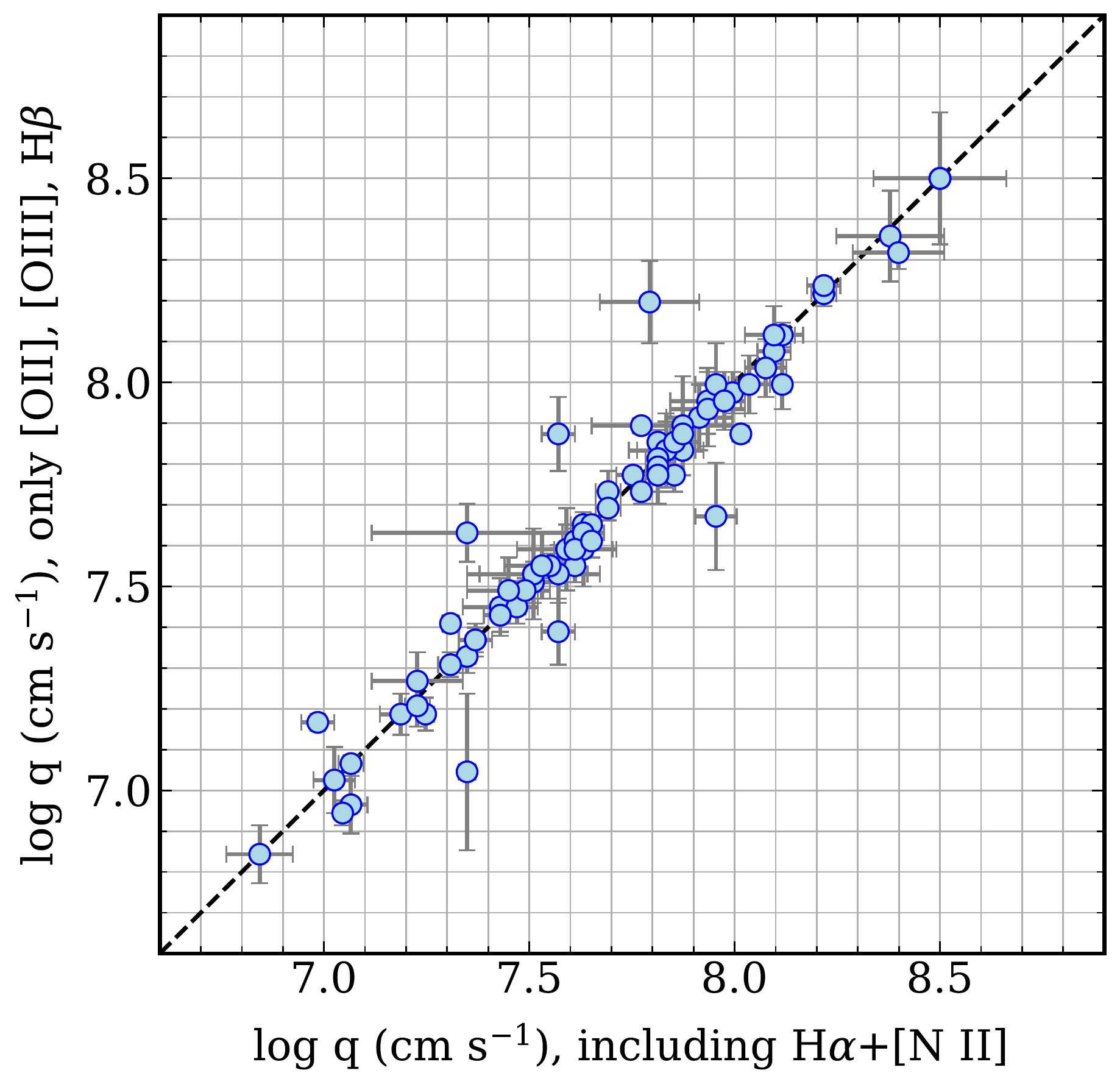}
    \caption{Comparison of gas-phase metallicities (12 + $\log$ O/H)
(left) and ionization parameters, $\log q$, (right) derived using
\oii, \oiii, \hb, \ha+\nii (which are blended at the resolution of the
WFC3 grism) and \sii\ (if present) (on the abscissa), versus fits
using only with \oii, \oiii, \hb\ (on the ordinate).  The data points
show the mode of the metallicity likelihood function, and the error
bars show the inter-68-percentile derived from the highest density
interval.  The dashed line shows the unity relation.   The majority of
data points have metallicities and ionization parameters that are
consistent between the two methods.  In a small number of cases (10
out of 87) the metallicities derived including \ha+\nii\ and \sii\
favor lower gas-phase metallicities.  This illustrates that for
objects on the lower-branch of the R$_{23}$ relation additional
information may be required to understand their gas phase
metallicities. } \label{fig:Ha_ZvZ}
  \end{figure*}
  
Figure~\ref{fig:Ha_ZvZ} compares the gas-phase metallicity ($12 +
\log\mathrm{O/H}$) and ionization parameters ($\log q$) between the
values derived using only \oii, \oiii, and \hb, to those that also
include \ha+\nii.    The majority of data points show that
the results for the ionization parameters from the two methods are
nearly unchanged (81/87 $=93\%$ fall on the unity relation within
their 68\% uncertainties).  The metallicities are also mostly
consistent between the two methods, where 73/87 $(=84\%)$  fall on the
unity relation within their 68\% uncertainties.  (In both cases the
measurements are not independent as they both use some of the
same emission lines).

\begin{figure*}
   \gridline{\fig{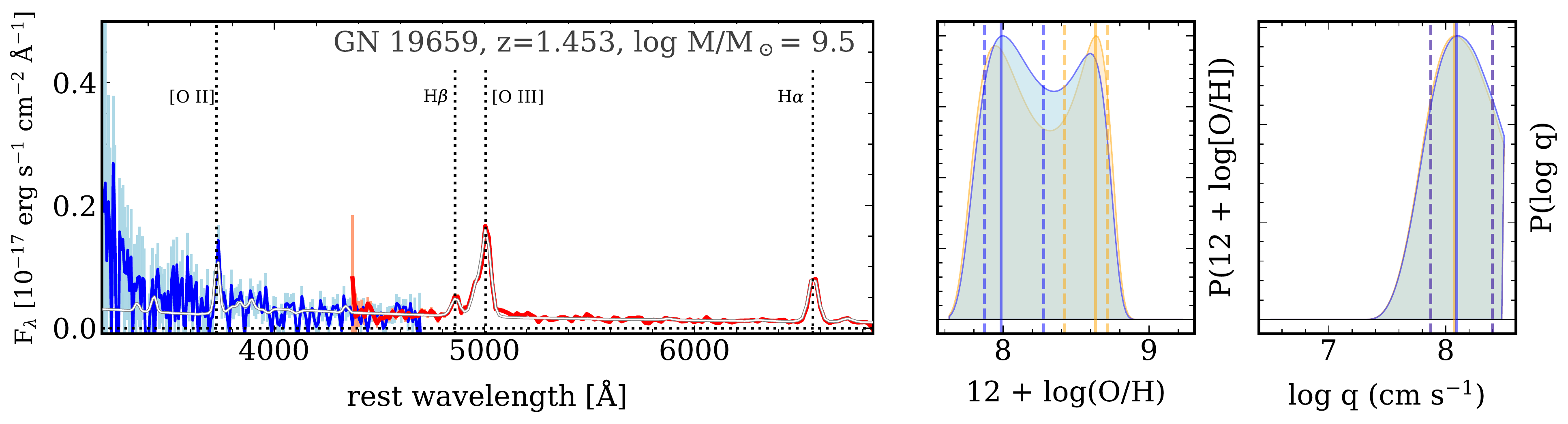}{0.45\textwidth}{}
     \fig{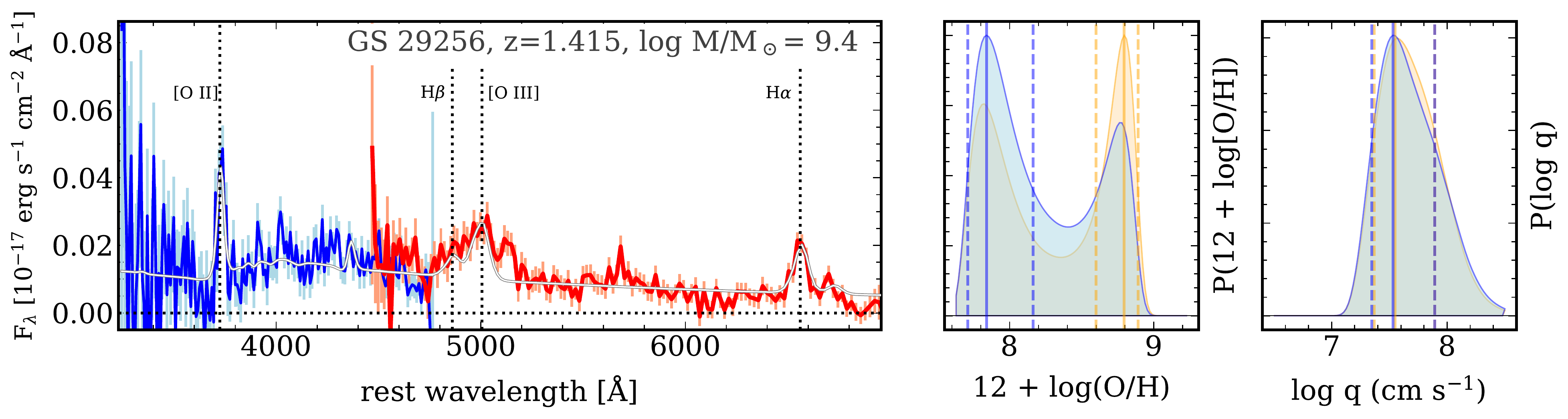}{0.45\textwidth}{}
  }
  \vspace{-24pt}
  \gridline{\fig{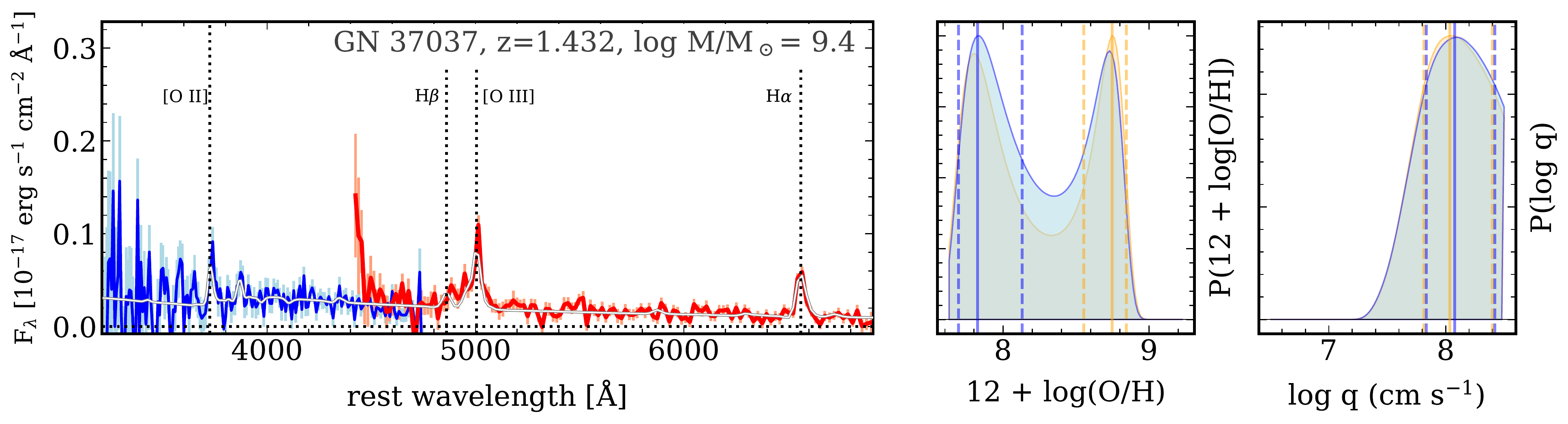}{0.45\textwidth}{}
    \fig{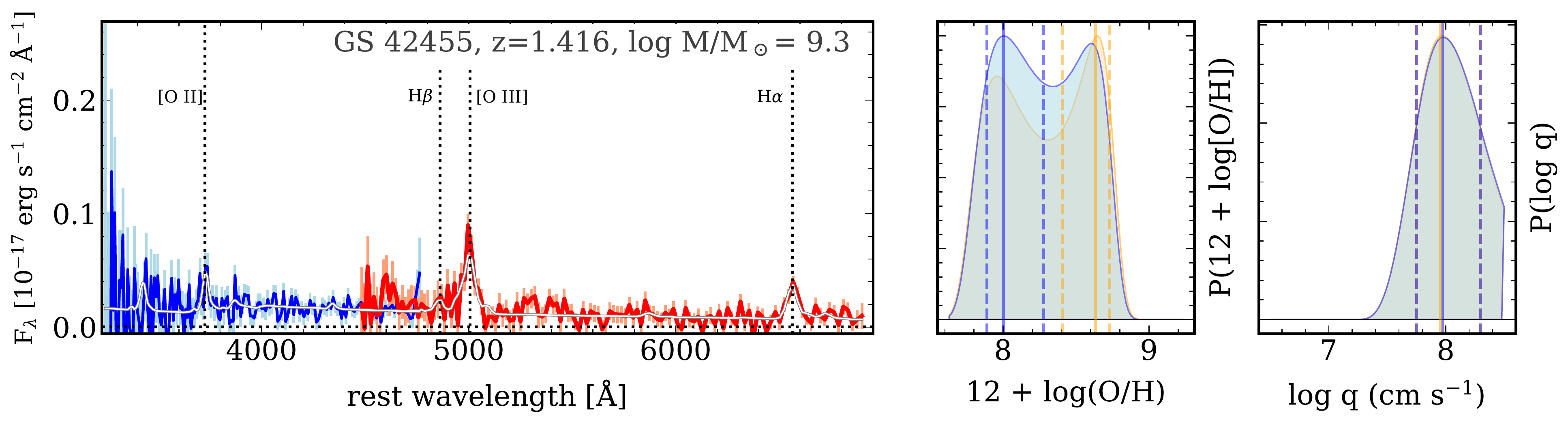}{0.45\textwidth}{}
  }
  \vspace{-24pt}
  \caption{Examples of galaxies in CLEAR at $1.1 < z < 1.5$ that have
gas-phase metallicities derived using \oii, \oiii, \hb, \ha+\nii, and
\sii\ that are substantially lower than those derived using \oii,
\oiii, and \hb\ alone.  Each row shows one galaxy in our sample (with
ID, spectroscopic redshift from the grism data, and stellar mass as
indicated).  The left panel in each row shows the WFC3 G102 (blue) and
G141 (red) 1D extracted spectra.  The dashed lines indicate strong
emission lines (\ha\ includes \nii, which is blended at the grism
resolution).   The middle and right panels show the posterior
likelihoods for the gas-phase metallicity, $P(12 +\log\mathrm{O/H})$,
and ionization parameter, $P(\log q)$, derived using only \oii, \oiii,
and \hb\ (shown in tan) compared to those derived using \oii, \oiii,
\hb, \ha+\nii, and \sii\ (shown in light blue).  The likelihood on the
ionization parameters are nearly identical showing these are strongly
constrained by \oii\ and \oiii.  The posterior likelihood for
metallicity for these galaxies shifts to the ``lower branch'' of the
relation when \ha+\nii\ and \sii\ are
included. \label{fig:spectra_izi_results_wha}}
  \end{figure*}

In a small number of cases (10 out of 87) the metallicities derived
including \ha+\nii\ and \sii\ favor significantly lower   gas-phase
metallicities.     Figure~\ref{fig:spectra_izi_results_wha} shows the
posterior likelihoods on the metallicity, $P(12+\log\mathrm{O/H})$ and
ionization parameter, $P(\log q)$, for some of these galaxies, comparing the
likelihoods derived using only \oii, \oiii, and \hb, and those that
include \ha+\nii.   A comparison of the
posterior likelihoods show that the constraints on ionization is
nearly unchanged.  This implies that the ionization is mostly driven
by measurements of the \oiii/\oii\ ratio, and that adding \ha+\nii\
(and \sii) provides little additional information.  However, inspection
of the metallicity posterior likelihood shows that adding \ha+\nii\
and \sii\ for some galaxies increases the probably density at lower
metallicities.  This is akin to stating that these adding \ha+\nii\
and \sii\ move the metallicity to the ``lower branch'' of the
R$_{23}$--metallicity relation (see, e.g., Figure~\ref{fig:r23_Z}) and
favors lower metallicity solutions as a result.  Visually inspecting
the spectra and posteriors in Figure~\ref{fig:spectra_izi_results_wha}
we see that in some cases the \ha\ is weakly detected (with lower SNR,
e.g., GS 47954 and GS 29256).  There are also additional effects, for
which our analysis does not account, such as the well-known
metallicity (and by proxy, stellar mass) dependence on the \nii/\ha\
ratio \citep[e.g.,][]{trem04,erb06b,nagao06,kewley19}, which may
complicate the fitting and interpretation.   This illustrates that
determining for objects on the lower-branch of the R$_{23}$ relation
additional information may require additional information to
understand their gas phase metallicities.   Said another way, with
CLEAR we are finding that galaxies at $1.1 < z < 2.3$ with stellar
masses around $\log M_\ast / M_\odot \sim 9.2-10$ have $R_{23}$ values
near or approaching the ``peak'' of the $R_{23}$--metallicity
distribution (see figure~\ref{fig:r23_Z}), which occurs at $R_{23}
\simeq $1.0 and $12 + \log(\mathrm{O/H}) \simeq 8.2$ (see
equation~\ref{eqn:r23_Z}).   It is around this inflection point that
additional information will be invaluable to diagnosis the
metallicities of galaxies.  
  
  \begin{figure*}
    \centering
 \includegraphics[width=0.75\textwidth]{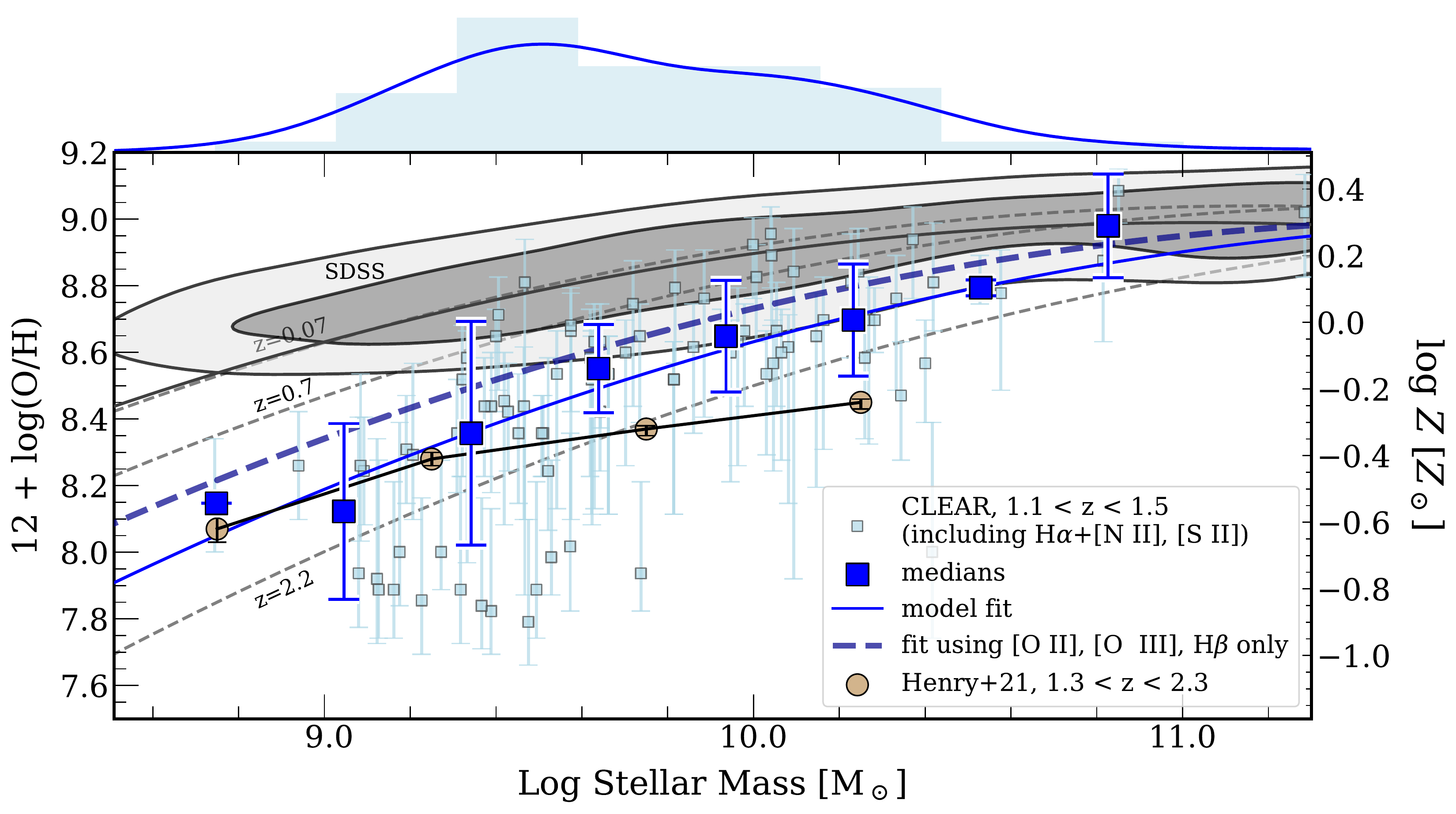}
\caption{  The stellar--mass, gas-phase metallicity relation (MZR) for
galaxies at $1.1 < z < 1.5$ from CLEAR with metallicities derived
using the combination of \oii, \oiii, \hb, \ha+\nii, and \sii.    The
small-blue data points show the results for individual galaxies.  The
large data points show medians in bins of stellar mass.   The shaded
region shows the distribution derived using the same combination of
emission lines from SDSS DR14.   The black solid, thin line shows our
fit to the SDSS galaxies.  The dotted lines show the MZR relation at
$z=0.07$, 0.7, and 2.2 from \citet{maio08}.  The histogram and curve
along the top of the panel shows the distribution of stellar masses
for the CLEAR galaxies in this subsample .   The solid, thick line
shows the fit to the CLEAR galaxies in this subsample; the dashed,
thick line shows the fit to these same CLEAR galaxies using the
metallicities derived only from \oii, \oiii, and \hb.   The large
circles show the relation derived by \citet{henry21} using all
available emission lines (including data for this sample) with a
different calibration of metallicity (see
text).    \label{fig:MZR_wHa}}
   \end{figure*}

Figure~\ref{fig:MZR_wHa} shows the impact on the stellar-mass,
gas-phase-metallicity (MZR) relation for galaxies in the CLEAR
subsample with $1.1 < z < 1.5$, that include metallicities derived
using the combination of \oii, \oiii, \hb, \ha+\nii, and
  \sii.   The figure compares the updated analytic fit
  (Eq.~\ref{eqn:MZR}) with $\log M_0 / M_\odot = 12.07 \pm 0.06$.
  Therefore, including the \ha+\nii\ lines shifts the MZR to
  lower metallicities at fixed stellar mass (compared to results using
  only \oii, \oiii, and \hb) of $\Delta\log Z = 0.15$~(0.10)~dex at a
  fixed stellar mass of $\log M_\ast / M_\odot = 9.0$ (10.0) at $1.1 <
  z < 1.5$, but the overall qualitative trend in the MZR is unchanged
  from the results above.  
  
  Figure~\ref{fig:MZR_wHa} also compares the results from CLEAR to
those from \citet{henry21}.  These authors combined results from
available WFC3 grism data for galaxies at $1.3 < z < 2.3$ that
includes CLEAR along with other datasets (WISPS, \citealt{atek10};
3DHST, \citealt{momc16}).   These authors derive gas-phase
metallicities calibrated against the relation from \citet{curti17}. As
illustrated in Figure~\ref{fig:r23_Z}, at lower values of metallicity
(i.e., $8.0 < 12 + \log\mathrm{O/H} < 8.2$) the \citeauthor{curti17}
relation is similar to our results (derived by fitting the
photoionization models from MAPPINGS V, \citealt{kewley19b}).
However, at higher metallicities (i.e., $12 + \log\mathrm{O/H} \gsim
8.4$) the \citeauthor{curti17} relation is offset by 0.2--0.3~dex
toward lower metallicities.  This is evidence in
Figure~\ref{fig:MZR_wHa} which shows the MZR we derive (including
\oii, \oiii, \hb, \ha+\nii) is consistent with that from
\citeauthor{henry21} for lower stellar masses/metallicities, but we
observe an offset to higher metallicity at fixed stellar mass for
higher masses/metallicities.  The magnitude of this offset consistent
with the systematic uncertainties in metallicity calibrations.  This
highlights the importance of including systematic uncertainties
arising from calibration when interpreting the absolute evolution of
the mass-metallicity relation. 
  
\section{On the Effects of Dust Attenuation
}\label{section:appendix_dust}

\added{ In the analysis above, we have implicitly assumed that the
dust attenuation of the nebular gas is equal to that of the stellar
continua (see Section~\ref{section:stacking}).  That is, we take,
$E(B-V)_\mathrm{nebular} = E(B-V)_\mathrm{continuum}$, where $E(B-V) =
A(B) - A(V)$ is the color excess.       The literature has found
varying relationships between $E(B-V)_\mathrm{gas}$ (sometimes called
$E(B-V)_\mathrm{nebular}$) and $E(B-V)_\mathrm{continuum}$ (sometimes
called $E(B-V)_\mathrm{stars}$).   \citet{calzetti01} discuss the
evidence that the nebular gas experiences roughly twice the dust
attenuation as the stars in the integrated emission for local
UV-luminous galaxies, with $E(B-V)_\mathrm{nebular} =
E(B-V)_\mathrm{continuum} / 0.44$.   However, galaxies at higher
redshifts, $z \sim 2$, show that the attenuation of the gas
is more consistent with that of the stars, where  $E(B-V)_\mathrm{nebular}
\approx E(B-V)_\mathrm{continuum}$ \citep{erb06a,reddy15}, at least
for galaxies with relatively low attenuation.  For example,
\citet{reddy15} found that the majority ($>50$\%) of objects in their
$H$-band selected sample of $z\sim 2$ galaxies (from MOSDEF) are
consistent with $E(B-V)_\mathrm{nebular} = E(B-V)_\mathrm{continuum}$
(within the 1$\sigma$ uncertainties).  \citeauthor{reddy15} also found
that the attenuation of the gas
increases with respect to that of the stars for galaxies with higher stellar
masses and SFRs, where the relation approaches $E(B-V)_\mathrm{nebular} =
E(B-V)_\mathrm{continuum}/0.44$ for galaxies with SFR $>$
20~$M_\odot$~yr$^{-1}$.   Nearly all galaxies in our CLEAR sample have
SFRs below this value: 95\% of the sample have SFRs in the range,
2.2 -- 19~$M_\odot yr^{-1}$ (see Figure~\ref{fig:sfms}).  Therefore we
we have assumed $E(B-V)_\mathrm{nebular} =
E(B-V)_\mathrm{stars}$. 
%
%
}

\begin{figure*}
      \centering
 \includegraphics[width=0.65\textwidth]{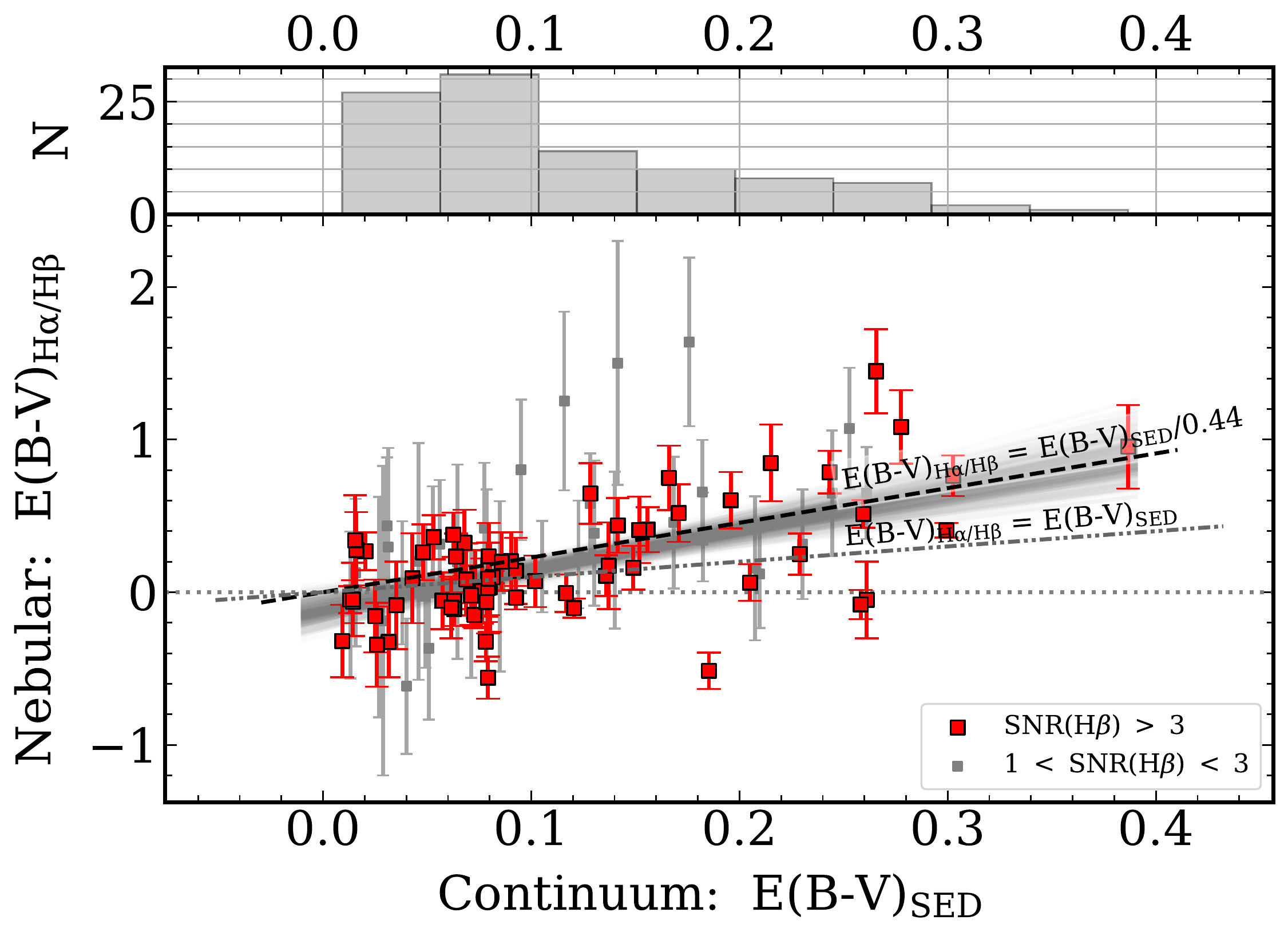}
\caption{  The relation between dust attenuation derived from the SED
fitting, $E(B-V)_\mathrm{SED}$ and that derived directly from the
\ha/\hb\ line ratio, $E(B-V)_\mathrm{H\alpha/H\beta}$.   This plot
shows data for CLEAR galaxies in our sample with $z < 1.5$ where both
\ha\ and \hb\ lines are detected.   The figure shows that considering
the whole sample the nebular gas (traced by the \ha/\hb\ ratio)
experiences more dust attenuation than the stars (traced by the
continuum, modeled by the SED fitting).  A linear fit (using
\texttt{linmix})to the full sample shows a best fit of
$E(B-V)_\mathrm{H\alpha/H\beta} = E(B-V)_\mathrm{SED}/(0.4 \pm 0.2$)
(indicated by the swath of gray lines, which show random draws from
the posterior of the fit).  However, the majority of the galaxies
($>66$\%) have $E(B-V)_\mathrm{SED} < 0.13$, for which the data are
consistent with $E(B-V)_\mathrm{H\alpha/H\beta} \approx
E(B-V)_\mathrm{SED}$.  We therefore adopt $E(B-V)_\mathrm{nebular} =
E(B-V)_\mathrm{SED}$, but we show in
Appendix~\ref{section:appendix_dust} that assuming higher dust
attenuation for the nebular gas would not impact substantially any of
our findings.\label{fig:dust_vs_dust}} 
\end{figure*}

\begin{figure*}
      \centering
 \includegraphics[width=0.7\textwidth]{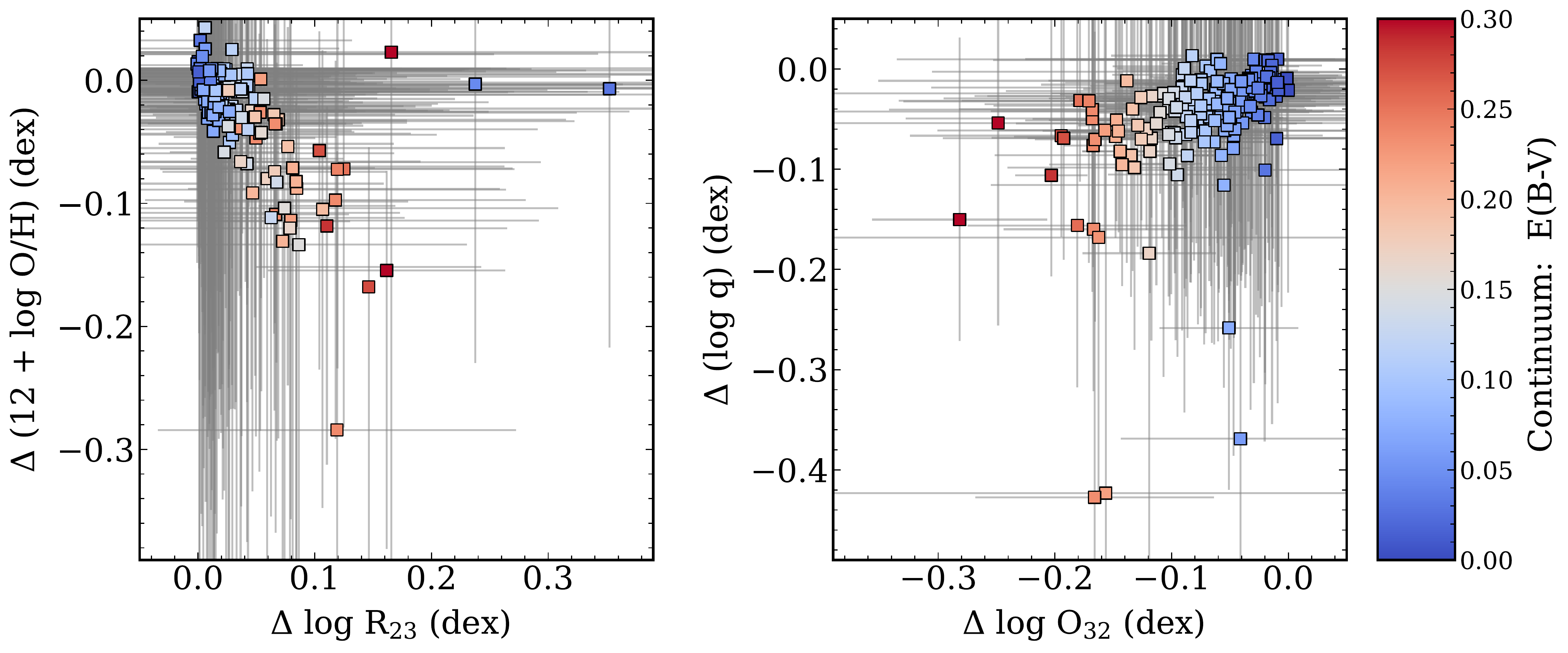}
 \includegraphics[width=0.7\textwidth]{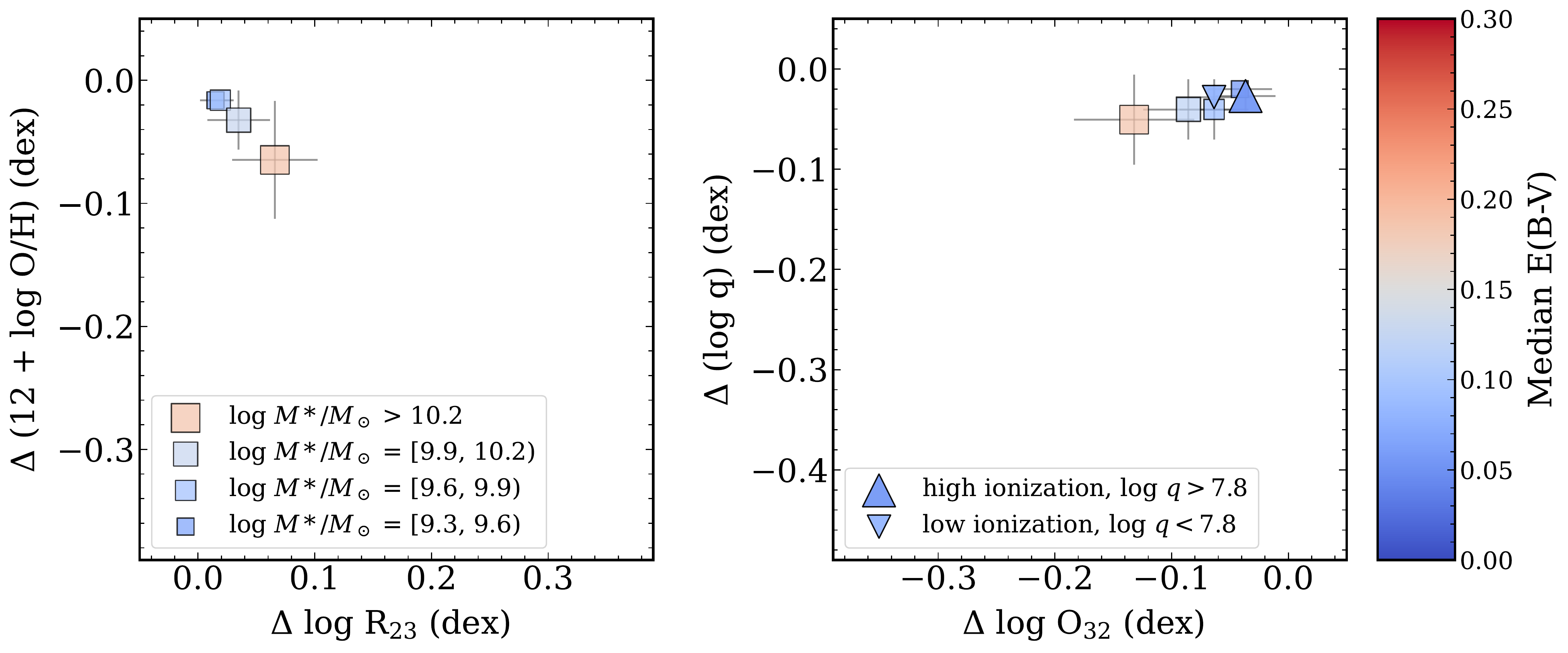}
\caption{ Change in R$_{23}$ and O$_{23}$ emission line ratios, and
the derived gas-phase metallicity ($\Delta \log \mathrm{O/H}$, and
ionization parameter, $Delta \log q$) when increasing the amount of
dust attenuation in the gas.  Each quantity is $\Delta x = x_2 - x_1$,
where $x_1$ is the quantity assuming equal dust attenuation in the gas
and stars, and $x_2$ is the quantify assuming more dust attenuation in
the gas (with $E(B-V)_\mathrm{nebular} = E(B-V)_\mathrm{SED}/0.44$).
The top panels show the effects on our full
sample, color-coded by $E(B-V)_\mathrm{SED}$. The bottom panels show
medians for different subsamples (as labeled in the legends).  For
most of the galaxies in our sample, the increase in R$_{23}$
corresponds to a decrease in metallicity.  Similarly the decrease in
O$_{32}$ corresponds to a decrease in ionization parameter.  However,
the observed effects are small, particularly in the medians for the
samples in the MZR, MQR, and when comparing the ionization parameters
of the samples of ``high'' and ``low'' ionization (see text and
Section~\ref{section:ZQR}).  \label{fig:dust_impact}}
  \end{figure*}

\added{ We are able to test the relation between
$E(B-V)_\mathrm{nebular}$ and $E(B-V)_\mathrm{stars}$ for galaxies in
our sample that are well detected in \hb\ (with S/N $>$1) and have
redshifts $z < 1.5$ such that \ha\ is also present in the data.   We
can then calculate $E(B-V)_\mathrm{nebular} =
E(B-V)_\mathrm{H\alpha/H\beta}$ directly assuming the lines have an
intrinsic ratio of \ha/\hb\ = 2.86 (for Case-B recombination,
following, e.g., \citealt{nelson16b}).  Figure~\ref{fig:dust_vs_dust}
compares this against  $E(B-V)_\mathrm{SED}$ derived from the SED
fitting (which we take as an estimate of $E(B-V)_\mathrm{stars}$).
The figure shows that for galaxies with low attenuation,
$E(B-V)_\mathrm{SED} \lsim 0.15$, there is little difference between
$E(B-V)_\mathrm{H\alpha/H\beta}$ and $E(B-V)_\mathrm{SED}$.   For
galaxies with higher attenuation, we see that the nebular attenuation
is higher than that derived from the SED fitting, approaching the
value $E(B-V)_\mathrm{nebular} = E(B-V)_\mathrm{stars}/0.44$
consistent with \citep{calzetti01,reddy15}.   Formally, we
parameterize the relation as $E(B-V)_\mathrm{H\alpha/H\beta} =
E(B-V)_\mathrm{SED} / \kappa$, where a fit to the entire sample in
Figure~\ref{fig:dust_vs_dust} gives $\kappa = 0.4 \pm 0.2$.  This is
consistent with \citet{calzetti01}.  However, most of our sample has
low over attenuation, with approximately two-thirds having
$E(B-V)_\mathrm{SED} < 0.13$.  For this sample, there is no evidence
to support a higher dust attenuation in the gas compared to the stars,
where we find $\kappa = 0.7 \pm 2.0$.  The CLEAR
sample with $E(B-V)_\mathrm{SED} < 0.13$ have median SFR of $\simeq
4$~$M_\odot$~yr$^{-1}$ (with an interquartile range of
2--6~$M_\odot$~yr$^{-1}$) compared to the sample CLEAR sample with
$E(B-V)_\mathrm{SED} > 0.15$ which has a median SFR =
9~$M_\odot$~yr$^{-1}$ (and interquartile range of
6--14~$M_\odot$~yr$^{-1}$).   Therefore, our findings for CLEAR seem
consistent with \citet{reddy15}.  Because the majority of our sample
has low color excess and lower SFRs, we therefore argue that
$E(B-V)_\mathrm{nebular} = E(B-V)_\mathrm{SED}$ is a reasonable
assumption. }

\added{ Nevertheless, we have conducted a study to determine what the
impact would be if we instead adopted a higher dust attenuation in the
nebular gas, using $E(B-V)_\mathrm{nebular} =
E(B-V)_\mathrm{SED}/0.44$.  We have repeated all the analyses assuming
this case (including recomputing the gas-phase metallicities and ionization
parameters).
%
%
Figure~\ref{fig:dust_impact} shows how the derived gas-phase
metallicities and ionization parameters are impacted by increasing the
dust attenuation, assuming $E(B-V)_\mathrm{nebular} =
E(B-V)_\mathrm{SED}/0.44$.   Each panel of the Figure shows the
quantities ($x$) as $\Delta x = x_2 - x_1$ where $x_1$ is the quantity
assuming $E(B-V)_\mathrm{nebular} =E(B-V)_\mathrm{SED}$ (equal dust
attenuation in the gas) and $x_2$ is the quantify assuming
$E(B-V)_\mathrm{nebular} = E(B-V)_\mathrm{SED}/0.44$ (more dust
attenuation in the gas).  The increased dust attenuation has more of an effect on the \oii\
flux, which causes the R$_{23}$ ratio to increase and the O$_{32}$
ratio to decrease.  The left panels show that on average the
R$_{23}$ values of galaxies \textit{decrease} by 0.05-0.1~dex,
which corresponds to an \textit{increase} in metallicity of $\Delta
\log \mathrm{O/H} = 0.02-0.06$~dex.  This small change would have a
minimal impact on the evolution of the MZR as these offsets are
significantly smaller than the evolution we measure (Figure~\ref{fig:MZR} and
Section~\ref{section:MZR}).  Similarly, the O$_{32}$ values on average
\textit{decrease} by 0.05--0.15~dex, which corresponds to a
\textit{decrease} in the ionization parameter by $\approx$0.05
dex. Again, this imposes minimal impact because this is much smaller
than the evolution we observe in the MQR
(Figure~\ref{fig:MQR} and Section~\ref{section:MQR}).  For the
subsamples of galaxies with ``high'' and ``low'' ionization (at fixed
stellar mass, see Figure~\ref{fig:balmer_q} and
Section~\ref{section:ZQR}), the change is also very minor.
Qualitatively, the reason for the minimal impact is because the dust
attenuation for these samples is already low.
Figure~\ref{fig:dust_impact} shows the ionization parameters would be
increased by $0.02-0.04$~dex if the gas experiences stronger dust
attenuation compared to that of the stellar continuum. Therefore, our
results are reasonably robust even if the nebular gas for our sample
is more attenuated than the derived from the SED fits. }

\end{appendix}

\bibstyle{aasjournal}
\bibliography{alpharefs}{}


\end{document}

%% file: defs.tex
\def\lesssim{\mathrel{\hbox{\rlap{\hbox{%
 \lower4pt\hbox{$\sim$}}}\hbox{$<$}}}}
\def\gtrsim{\mathrel{\hbox{\rlap{\hbox{%
 \lower4pt\hbox{$\sim$}}}\hbox{$>$}}}}
\def\I.{\kern.2em{\sc i}} 
\def\II.{\kern.2em{\sc ii}} 
\def\III.{\kern.2em{\sc iii}} 
\def\IV.{\kern.2em{\sc iv}} 
\newcommand{\hst}{\textit{HST}}

\newcommand{\gsim}{\gtrsim}
\newcommand{\lsim}{\lesssim}

\newcommand{\hbeta}{\hbox{H$\beta$}}

\newcommand{\hg}{\hbox{H$\gamma$}}

\newcommand{\hb}{\hbox{H$\beta$}}
\newcommand{\ha}{\hbox{H$\alpha$}}

\newcommand{\mstar}{\hbox{$M_\ast$}}

\newcommand{\msol}{\hbox{$M_\odot$}}
\newcommand{\zsol}{\hbox{$Z_\odot$}}

\newcommand\myRoman[1]{\@Roman{#1}\relax}%

\newcommand{\jwst}{\textit{JWST}}

\newcommand{\OH}{\hbox{$12 + \log(\mathrm{O/H})$}}

\definecolor{aggiemaroon}{HTML}{500000}

\newcommand{\neiii}{\hbox{[Ne\,{\sc iii}]}}
\newcommand{\oii}{\hbox{[O\,{\sc ii}]}}
\newcommand{\oiii}{\hbox{[O\,{\sc iii}]}}
\newcommand{\nii}{\hbox{[N\,{\sc ii}]}}
\newcommand{\sii}{\hbox{[S\,{\sc ii}]}}

\newcommand{\grizli}{\hbox{\texttt{grizli}}}

\newcommand{\izi}{\hbox{\texttt{IZI}}}
\newcommand{\ppxf}{\hbox{\texttt{pPXF}}}

\definecolor{aggiemaroon}{HTML}{500000}